\documentclass[amsmath,amssymb,showpacs]{revtex4}
\usepackage{amsmath,amsfonts,amssymb,latexsym,extarrows}
\newtheorem{thm}{Theorem}
\newtheorem{lemma}{Lemma}
\newtheorem{propo}{Proposition}
\newtheorem{rema}{Remark}

\renewcommand{\theequation}{\arabic{equation}}
\newcommand{\dis}{\displaystyle}

\newcommand{\bequ}{\begin{equation}}
\newcommand{\eequ}{\end{equation}}
\newcommand{\barr}{\begin{array}}
\newcommand{\earr}{\end{array}}
\newcommand{\bea}{\begin {eqnarray}}
\newcommand{\eea}{\end {eqnarray}}
\newcommand{\lb}{\label}

\newcommand{\qed}{\hfill \rule{2.25mm}{2.25mm}\vspace{.15cm}}

\renewcommand{\Im}{{\cal I}{\rm m}\:}
\renewcommand{\Re}{{\cal R}{\rm e}\:}

\begin{document}
% ====================================
\def \tr {\mathrm{Tr\,}}
\let\la=\lambda
\def \Z {\mathbb{Z}}
\def \Zt {\mathbb{Z}_o^4}
\def \R {\mathbb{R}}
\def \C {\mathbb{C}}
\def \La {\Lambda}
\def \ka {\kappa}
\def \vphi {\varphi}
\def \Zd {\Z ^d}
% ====================================
% **************************************************
% **************************************************
%\begin{frontmatter}
\title{On Yang-Mills Stability Bounds and Plaquette Field Generating Function}
\author{Paulo A. Faria da Veiga}\email{veiga@icmc.usp.br.  Orcid Registration Number: 0000-0003-0739-069X.}
\author{Michael O'Carroll}\email{michaelocarroll@gmail.com}
\affiliation{Departamento de Matem\'atica Aplicada e Estat\'{\i}stica - ICMC, USP-S\~ao Carlos,\\C.P. 668, 13560-970 S\~ao Carlos SP, Brazil}
% **************************************************
% **************************************************
% **************************************************
% **************************************************
% **************************************************
% **************************************************
% *******DATA*******DATA******DATA*****DATA*******DATA*****  DATA
\date{August 26, 2024.} %{\bf\large DRAFT VERSION}\vspace{.3cm}}
% **************************************************
% **************************************************
% **************************************************
% **************************************************
% **************************************************
% **************************************************
% **************************************************
% **************************************************
\begin{abstract}
We consider the gauge-invariant Yang-Mills quantum field theory in the imaginary-time functional integral Wilson formulation on the finite hypercubic lattice  $\Lambda\subset a\mathbb Z^d\subset\mathbb R^d$, $d = 2,3,4$, $a\in (0,1]$, with $L$ (even) sites on a side and with the gauge Lie groups $\mathcal G={\mathrm U(N),\mathrm{SU}(N)}$. The lattice provides an UV regularization of the continuum spacetime and Osterwalder-Seiler positivity is verified. To each $\Lambda$ bond $b$ there is assigned a unitary matrix gauge variable $U_b$ from an irrep of $\mathcal G$. The vector gauge potentials (gluons) are parameters in the Lie algebra of $\mathcal G$. The Wilson finite lattice partition function $Z_\Lambda(a)$ is used with an action $A_\Lambda(a)$ which is a sum of gauge-invariant plaquette (set of four bonds forming a minimal square) actions times $[a^{d-4}/g^2]$, $g^2\in(0,g_0^2]$, $0<g_0^2<\infty$. Each plaquette action has the product of four gauge variables; the partition function is the integral over the Boltzmann factor with a product over bonds of $\mathcal G$ Haar measures. Formally, in the UV limit $a\searrow 0$, the action gives the YM classical continuum action. For free and periodic b.c., and using scaled fields, defined with an $a$-dependent noncanonical scaling, we show thermodynamic and UV stable (TUV) stability bounds for a scaled partition function, with bound constants independent of $L$, $a$ and $g$. Passing to scaled fields does not alter the model energy-momentum spectrum, hence its particle contents, and can be interpreted as an a priori field strength renormalization, making the action more regular. With scaled fields, we can isolate the UV singularity of the finite lattice physical, unscaled free energy $f_\Lambda(a)=[\ln Z_\Lambda]/\Lambda_s$, where $\Lambda_s=L^d$ is the total number of lattice sites. This allows us to show the existence of, at least, the subsequential thermodynamic ($L\nearrow\infty$) and UV ($a\searrow 0$) limits of the scaled free energy. To obtain the TUV bounds, the Weyl integration formula is used in the gauge integral and the random matrix probability distributions of the CUE and GUE appear naturally. The lower bound on $Z_\Lambda(a)$ results from a global, new upper bound on $A_\Lambda(a)$, quadratic in the gluon fields. This is in contrast to what happens in the classical continuum action which has a quartic growth. We also define gauge-invariant physical and scaled plaquette fields. Using periodic b.c. and the multireflection method, the generating function of $r$ scaled plaquette correlations is bounded uniformly in $L$, $a$ and $g$. The bound is also independent of the location/orientation of the $r$ plaquette fields. Consequently, $r$-scaled plaquette field correlations are also bounded. Combining this result with the scaling relation shows that the physical two-plaquette correlation at coincident points has an $a^{-d}$ singular behavior, which is the same as that of correlation of the derivative of free scalar unscaled fields at coincident points.  Unlike other approaches, working with the Wilson model, there is no need to introduce any additional infrared regulator in the action for free and periodic boundary conditions. Using the free scalar field as a reference, we have a characterization of UV asymptotic freedom. Here and in other papers, various  Bose/Fermi models were treated with scaled fields. This method can be coupled with other methods to analyze the existence and properties of QFT models.\vspace{1mm}
\end{abstract}
%%%%%%%%%%%%%%%%%%%%%%%%%%%%%%%%%%%%%%%%%%%%%%%%%%%%%%%%%%%%%%%%%%%%
\maketitle
%%%%%%%%%%%%%%%%%%%%%%%%%%%%%%%%%%%%%%%%%%%%%%%%%%%%%%%%%%%%%%%%%%%%%%%%%%%%%%%%%%%%%%%%%%%%%%%%&&&&&&&&&&&&&&&&&&&&&&&%%%%%%%%%%%%%%%%%%%%%%%%%%%%%
\begin{center}{\em This article is dedicated to the memory of our friend Krzysztof Gawedzki who\\made important contributions to mathematics and mathematical physics.}\end{center}\vspace{2mm}%%%
%%%%%%%%%%%%%%%%%%%%%%%%%%%%%%%&&&&&&&&&&&&&&&&&&&&&&&%%%%%%%%%%%%%%%%%%%%%%%%%%%%%
\section{Introduction and Description of Results} \lb{intro}
%%%%%%%%%%%%%%%%%%%%%%%%%%%%%%&&&&&&&&&&&&&&&&&&&&&&&%%%%%%%%%%%%%%%%%%%%%%%%%%%%%
%%%%%%%%%%%%%%%%%%%%%%%%%%%%%%%%%%%%%%%%%%%%%%%%%%%%%%%%%%%%%%%%%%%%
To show the existence and properties of an interacting relativistic quantum field theory (QFT) in spacetime dimension $d\,=\,4$ is a fundamental problem in physics \cite{Wei,Banks,Gat,GJ,Dim2}. Many partial results have been obtained \cite{GJ,Riv,Summers,Sei}.  The quantum chromodynamics model (QCD) of interacting (anti)quarks and gauge, gluon fields is considered to be the best candidate for a four dimensional QFT model which rigorously exists. The so called Gap Problem of the Clay Foundation is related to the proof of two important features of QCD: first, one shall ensure the existence of the ultraviolet limit of QCD and, second, when considering the underlying physical quantum mechanical Hilbert space, we must prove  the so called gap problem, showing that the are no corresponding vectors for free quarks, antiquarks and gluons. The only physically admitted states are bound states of these fundamental particles of the model. This is the confinement property! The fact that the first vector state in the gluonic energy-momentum (E-M) sector spectrum has a mass ensures a finite range for the strong interactions, according to the Paley-Wiener theorem \cite{Stein}.

Unfortunately, up to now, we are not able to give a complete answer to these problems. This paper considers the UV limit of the pure Yang-Mills (YM) model which corresponds to discard the matter particles (quarks and antiquarks) from QCD and take only gluons and their nonlinear interaction into account.

We work on a spacetime lattice and use a special class of scaled fields, with a noncanonical scaling which preserves the decay of correlations (and then the E-M spectrum), we concentrate on obtaining finiteness properties for the model free energy, the generating functions of certain  gauge invariant plaquette field correlations for any value of the parameters. Of course, this finiteness properties do not lead directly to the construction of the UV limit but their proofs give some progress in this direction and we hope our methods can be coupled to more traditional methods in constructive quantum field theory to produce other results. We also discuss how to characterize UV asymptotic freedom on the lattice. Finally, we emphasize that our methods work also for models with Fermi (Grassmann) fields.

The action of QCD is a sum of an interacting Fermi-gauge field part and a pure-gauge, self-interacting YM field part.
The imaginary-time $d$-dimensional continuum spacetime classical smooth field Lagrangian or action of the local gauge-invariant YM model is given by \cite{Wei,GJ,Dim2}
\bequ\lb{actionTrF2}
\mathcal A_{\mathrm{classical}}\,=\,\sum_{\{\mu<\nu\}}\,\dis\int_{{\mathbb R}^d}\;{\mathrm Tr}[F_{\mu\nu}(x)]^2\,d^dx\,=\,\sum_{\{\mu<\nu\}}\:\;\dis\int_{{\mathbb R}^d}\;{\mathrm Tr}\left\{\partial_\mu A_\nu(x)-\partial_\nu A_\mu(x)+ig[A_\mu(x),A_\nu(x)]\right\}^2\,d^dx\,,
\eequ
where $\mu,\nu=0,1,\ldots,(d-1)$ (the label  $0$ denotes the time direction!), $A_\mu$ are the gauge fields or vector potentials, commonly called gluons, which are matrices in the Lie algebra of the gauge group $\mathcal G$, $F_{\mu\nu}$ is the second order antisymmetric field strength tensor, $g>0$ is the gauge field coupling and $[\,\cdot\,,\,\cdot\,]$ denotes the Lie algebra commutator.

Here, we will work in an Euclidean imaginary-time formulation of QFT. In this formulation, the model is defined by a partition function with a Boltzmann factor given by the exponential of minus the action and integration over all configurations of the fields is carried out.

One approximation is to work with a discretization of the Euclidean spacetime, replace the continuum fields by fields on a finite lattice $\Lambda\subset\mathbb R^d$ and make Riemann sum approximation to the integral in the classical action of Eq. (\ref{actionTrF2}). Carrying this out, infinities emerge due to the unbounded gauge (gluon) field integration and also the excess of gauge bond variables due to local gauge invariance.

Instead of this approach, Wilson (see e.g. Refs. \cite{Wil,MM}) proposed a finite valued finite lattice partition function. We can take  e.g. the lattice $\Lambda$ to be the finite hypercubic lattice with spacing $a$, $L\in\mathbb N$ ($L$ even) sites on a side and a total number of sites $\Lambda_s\,=\,L^d$. With this, we have $\Lambda\,\subset a\mathbb Z^d\,\subset \mathbb R^d$ and we take $a\in(0,1]$. The lattice volume $|\Lambda|\equiv \Lambda_s$ is taken to be the total number of sites $L^d$, instead of the volume $(aL)^d$ in $\mathbb R^d$, as it is usual in statistical mechanical lattice models \cite{Huang,GJ,Simon3,Galla}. 

In the Wilson finite lattice partition function, the gluon fields are parameters in the Lie algebra associated with the Lie gauge group $\mathcal G$. We choose $\mathcal G$ to be compact, e.g. we can take the groups of unitary matrices $\mathcal G=\mathrm U(N),\mathrm{SU}(N)$, $N\in\mathbb N$.

Let $b_\mu(x)\,=\,[x, x_\mu^+]$ denote a {\em positively directed lattice bond}, connecting the site $x$ to its neighboring lattice site $x_\mu^+\,=\, x\,+\,ae^\mu$, where $e^\mu$, $\mu=0,1,\ldots,(d-1)$, is the unitary vector of the $\mu$th spacetime direction. 

To each positively oriented bond (or, simply, bond), we assign a unitary matrix from an irreducible representation of the gauge group $\mathcal G$. These are our {\em bond variables}. The physical gluon fields are parameters in the Lie algebra of $\mathcal G$. The integration over the field configurations becomes a product of Haar integrals over the whole set of variables in gauge group $\mathcal G$. In this way, our gauge model is a random matrix model.

With the finite lattice Wilson partition function (see Eq. (\ref{partition}) below and Section \ref{sec2}), we get a regularization for the YM model and there are no more infinities. Besides, local gauge invariance is preserved and the property of Osterwalder-Seiler (OS) positivity, which allows the construction of an underlying quantum mechanical Hilbert space and then to prove the existence of a lattice QFT \cite{Sei}, is verified.

Moreover, it is expected that the control of the {\em thermodynamic limit} $\Lambda\nearrow a\mathbb Z^d$ and also the {\em continuum limit} $a\searrow 0$ of gauge field correlations lead to a continuum spacetime Euclidean QFT. Then, if the remaining OS axioms are verified, by the OS reconstruction theorem, we obtain a physically acceptable relativistic QFT in Minkowski space \cite{GJ,Dim2}. 

In this paper, we will concentrate on the pure-gauge, YM model. In an imaginary-time functional integral formulation, we adopt the above defined hypercubic lattice ultraviolet (UV) regularization $\Lambda\subset a\mathbb Z^d\subset\mathbb R^d$, $d=2,3,4$, $a\in(0,1]$. 

The starting point is the Wilson plaquette action partition function given below in Eq. (\ref{partition}). For small gauge coupling $0\,<\,g\,\ll\, 1$, stability bounds (see Ref. \cite{Rue}) for the corresponding model partition function have been proved in the seminal work of Balaban (see \cite{Bal,Bal2} and Refs. therein), using renormalization group (RG) methods and the heavy machinery of multiscale analysis. Applying RG methods in the continuum spacetime and using momentum slices, the UV limit of the YM model, in $d=4$ and with an additional infrared cutoff, was treated in Ref. \cite{MRS}. Using softer methods, in Ref. \cite{Ash}, the $d=2$ YM model was solved exactly.  It is expected that partition function stability bounds of Refs. \cite{Bal,Bal2} lead to bounds on YM gauge field correlations. 

Recently, the scaling limit of the YM-Higgs model was analyzed in \cite{Chat} for the gauge group ${\mathrm SU}(2)$. More recently (see e.g. Ref. \cite{GH}), a completely different approach considers the construction of Euclidean QFT using stochastic quantization and SPDE´s (Stochastic Differential Equations). For the scalar $\phi^4_3$ model in ${\mathbb R}^3$, some of the important OS axioms were shown and it is to be stressed that one of the virtues of this method is that the results are valid for any value of the couplings. Using this approach, the YM model with a general compact Lie gauge group in three dimensions was considered in Ref. \cite{Hairer}. 

Indeed, in the context of the RG, considering models which are small perturbations of the free field, the generating function of gauge field correlations and then correlations can be obtained through a formula which involves the effective actions generated by applying the RG transformations to the partition function (see e.g. \cite{MB}). However, unfortunately, in the case of gauge fields, this question, as well as the incorporation of fermionic quark/antiquark fields and the verification of the whole set of Osterwalder-Schrader-Seiler (OSS) axioms \cite{GJ,Dim2}, have never been completely analyzed up to now. This includes the physically interesting $d=4$ case. In $d=4$, after showing the mathematical existence of QCD, and checking the Osterwalder-Schrader-Seiler (OSS) axioms hold, an analysis of the low-lying E-M spectrum is expected to lead to the confinement of quarks/antiquarks and to the solution of the YM gap problem.
 
Recently, in the related Refs. \cite{M,MP,MP2}, we introduced an a priori QFT renormalization procedure on the lattice, with lattice spacing $a\in(0,1]$. This alternative procedure is based on a scaling of the physical (original) fields, like a wavefunction renormalization.  The $a$-dependent scaling is noncanonical. Besides, it preserves OSS positivity and the particle spectrum since it does not alter the decay rates of correlations. The new fields are called {\em scaled fields}. In the scaled fields, the original, physical action is more regular. The scaling is a smoothing process and may render some models finite and other are smoothed by it and present less or milder singularities. Our scaling transformation can then be viewed as a partial renormalization.

For the free field, the scaling removes the infinities from the free energy and correlations. Correlations are finite {\sl even at coincident points}. No smearing with test functions is needed. %Moreover, series expansions in the scaled coupling parameter converge up to and including the critical value, similarly to what happens in usual random walk expansions \cite{FFS}. 
Moreover, for the scaled field partition function and correlations, series expansions in the scaled coupling parameter converge absolutely, up to and including the critical value. The coefficients of these series are given explicitly. The random walk expansion associated with scaled field correlations also converge in this region \cite{FFS}. These free field results are detailed in Appendix A.

In the unpublished papers \cite{YM,GYM}, using scaled gauge fields, a simple proof of TUV stability bounds is given by a direct analysis of the finite lattice physical Wilson partition function $Z(\Lambda,a)$, with free boundary conditions (b.c.) in configuration space, starting with the model in the finite hypercubic lattice  $\Lambda\subset a\mathbb Z^d$. The gauge group is taken to be $\mathcal G = \mathrm U(N),\:\mathrm{SU}(N)$, with respective dimensions \bequ\lb{dims}\delta_N\,=\, N^2\quad ;\quad \delta_N\,=\,N^2-1\,,\eequ
but our methods extend to any other compact Lie group. Recall $e^\mu$, $\mu=0,1,\ldots,(d-1)$, is the lattice unit vector in the spacetime direction $\mu$. For each lattice oriented bond $b\,\equiv\,[x,x_\mu^+\,\equiv\,x\,+\,a\,e^\mu]$, there is a bond variable, i.e. a unitary matrix of the gauge group, $U_b\in \mathcal G$. There are a total of $\Lambda_b\,=\,dL^{d-1}(L-1)$ bonds in the lattice $\Lambda$.

By the exponential map, the gauge fields $A_b$ are elements of the Lie algebra of $\mathcal G$, and we write \bequ\lb{param}U_b\,=\,\exp(igaA_b)\,.\eequ 
With this convenient parametrization, the gauge fields are the usual physical gauge potentials. The $N\times N$ gluon field matrix $A_b$, in a suitable basis, is given by $$A_b=\sum_{c=1}^{\delta_N}\,A_b^c\,\theta_c\,,$$
and has components which are the ordinary gluon fields.
Here, for $\alpha= 1,2,\ldots,\delta_N$, we take the self-adjoint $\theta_\alpha$ to form a basis for the self-adjoint matrices of $\mathcal G\,=\,\mathrm U(N),\,\mathrm{SU}(N)$. They are the Lie algebra generators and are normalized accordingly to the trace condition $\mathrm{Tr} \theta_\alpha \theta_\beta=\delta_{\alpha\beta}$, with a Kronecker delta. With all this, if $b_\mu(x)=[x,x_\mu^+\equiv x+ae^\mu]$, then $A_b$ is the usual gauge field $A_\mu(x)$.

A lattice plaquette $p$ is a set of four bonds forming a minimal lattice square. The bonds connect four neighbor lattice sites. If $p\,\equiv\, p_{\mu\nu}(x)$ is a plaquette in the $\mu<\nu$ coordinate plane ($\mu,\nu=0,1,\ldots,(d-1)$), and if $x$ denotes the lower left corner of the plaquette, then the sites of $\Lambda$ taking part in $p$ are 
\bequ\lb{pmunu}{\mathrm vertices\;\,of\;\,}p\,=\,p_{\mu\nu}(x) \,:\;\;\;\; \;\;x\;,\;x_\mu^+\;,\;x_\mu^++ae^\nu\;,\;x_\nu^+\,.\eequ
Each plaquette $p\in\Lambda$ is associated with a positive Wilson action $A_p$ involving the trace of $U_p$, which is given by the ordered product of the four bond variables comprising the four consecutive sides of the plaquette. For the plaquette $p\,=\,p_{\mu\nu}(x)$, and using the physical parametrization of Eq. (\ref{param}), we have
\bequ\lb{extendedplaq}\barr{lll} 
U_p(x)&=& \exp\left[iag A_\mu(x) \right]\;\exp\left[iag A_\nu(x_\mu^+) \right]\;\left\{\exp\left[ iag A_\nu(x)\right]\;\exp\left[ iag A_\mu(x_\nu^+)\right]\right\}^\dagger\vspace{3mm}\\
&=&\exp\left[iag A_\mu(x) \right]\;\exp\left[iag A_\nu(x_\mu^+) \right]\;\exp\left[- iag A_\mu(x_\nu^+)\right]\;\exp\left[- iag A_\nu(x)\right]\,.
\earr
\eequ
\begin{rema}
We warn the reader that, as usual, we use the same notation $g$ both for a gauge group element, $g\in\mathcal G$, and the gauge coupling parameter $g>0$, which usually appears as $g^2$. There should be no confusion! Also, as the case of the unitary gauge group $\mathcal G=\mathrm U(N)$ is a little simpler than $\mathcal G={\mathrm SU}(N)$, in the text below, we will concentrate on $\mathrm U(N)$. The analysis for $\mathcal G=\mathrm {SU}(N)$ can be recovered from $\mathcal G=\mathrm U(N)$ with some minor modifications \cite{Simon2}.
\end{rema}

Now, taking $\mathcal G=\mathrm U(N)$, we give a schematic description of the model partition function and the stability bounds. Precise definitions are given in sections below. Mainly, we refer the reader to Section II where most of the needed definitions can be found.

The original, physical model partition function $Z(\Lambda,a)\,\equiv\,Z_\Lambda(a)$ is an integral over the Boltzmann factor (exponential of minus the action), with a product measure of $\mathrm U(N)$ ($\mathrm {SU}(N)$) Haar measures \cite{Bump,Simon2,Far} $$d\mu(U)\,=\,\prod_{b\in\Lambda}\,d\mu(U_b)\,,$$ one normalized measure for each bond. It reads
\bequ\lb{partition}
Z_\Lambda(a)\,=\,\dis\int \;\exp\left[-\dfrac{a^{d-4}}{g^2}\,\dis\sum_{p\in\Lambda}\,A_p(U_p)\right]\:d\mu(U)\,\equiv\,\dis\int \;\exp\left[-\dfrac{a^{d-4}}{g^2}\,\dis\sum_{p\in\Lambda}\,2\,\Re\,\tr (1\,-\,U_p)\right]\:d\mu(U)\,.\eequ

The Wilson plaquette action $A_p\,\geq\,0$ is given by \bequ\lb{AAp}A_p(U_p)\,=\,2\,\Re\,\tr\,(1-U_p)\,=\,\|U_p\,-\,1\|^2_{H-S}\,,\eequ and each plaquette action has a prefactor $[a^{d-4}/g^2]$, where we take the squared gauge coupling parameter $g^2\in(0,g_0^2]$, $0<g_0<\infty$. Here, $\|\,M\,\|_{H-S}\,\equiv\,\left[\tr\,M^*M\right]^{1/2}$ denotes the Hilbert-Schmidt norm (see Section II for the last equality). 
\begin{rema}
Although we have the physical parametrization of the gauge group bond variable $U_b$ in Eq. (\ref{extendedplaq}), the value of the partition function representation of Eq. (\ref{partition}) is independent of the parametrization. Sometimes, we denote this Wilson partition function as $Z^w_\Lambda(a)$. 
\end{rema}
\begin{rema}\lb{apparent}
As we see below, the apparent singularity at $g=0$, due to the prefactor $(1/g^2)$ in Eq. (\ref{partition}), is removed by adopting the physical parametrization $U_b\,=\,\exp[igaA_b]$.
\end{rema}

All our results hold for all $g^2$ in the range $g^2\in(0,g_0^2]$. Therefore, contrary to what happens in other works (see, e.g., \cite{Bal,Bal2}), our statements are not restricted to small enough $g^2$.
\begin{rema}
The adjoint of the positive oriented bond variable of Eq. (\ref{extendedplaq}) can be interpreted as associated with the negatively oriented bond. For the above plaquette $p=p_{\mu\nu}(x)$, the plaquette action $A_p$ can be interpreted as an ordered product of group variables going around the perimeter of the plaquette in a counterclockwise fashion, with a bond variable for a positively oriented traverse and its adjoint for a negatively oriented traverse.
\end{rema}

Formally, in Ref. \cite{Gat}, using the Baker-Campbell-Hausdorff formula [see Ref. \cite{Far} and Eq. (\ref{BCH}) below], it is shown that, for small lattice spacing $a$, the Wilson plaquette action $\left\{\left(a^{d-4}/g^2\right)\,\sum_{p\in\Lambda}\,A_p(U_p)\right\}$, as explained in Eq. (\ref{actionTrF2}) is the Riemann sum approximation to the usual classical smooth field continuum YM action\bequ\lb{TrF2}
{\mathcal A}\,=\,\sum_{\{\mu<\nu\}}\,\dis\int_{[-La,La]^d}\;{\mathrm Tr}[F_{\mu\nu}(x)]^2\,d^dx\,\simeq\,\sum_{\{\mu<\nu\}}\:\;\sum_{x\in\Lambda}\;a^d\, {\mathrm Tr}\left\{F_{\mu\nu}^a\,\equiv\,\partial^a_\mu A_\nu(x)-\partial^a_\nu A_\mu(x)+ig[A_\mu(x),A_\nu(x)]\right\}^2\,.
\eequ
Here, we have the finite difference derivatives $\partial^a_\mu A_\nu(x)\,=\,a^{-1}\;[A_\nu(x+ae^\mu)\,-\,A_\nu(x)]$ and
used the notation $ \{\mu<\nu\}\,\equiv\{\mu,\nu=0,...,(d-1)\,/\,\mu<\nu\}$.

Associated with the classical statistical mechanical model partition function $Z_\Lambda(a)$ of Eq. (\ref{partition}), there is a lattice QFT.  The Osterwalder-Seiler construction provides, via a Feynman-Kac formula, a quantum mechanical Hilbert space, self-adjoint mutually commuting spatial momentum operators and a positive energy operator. A key property in the construction is Osterwalder-Seiler reflection positivity, which is ensured here by choosing $L\in\mathbb N$ to be even! (see e.g. Ref. \cite{Sei}).

In principle, considering this lattice QFT and neglecting all the possible internal degrees of freedom in a more general case (spin, isospin, etc), there are $\Lambda_s$ sites in the lattice $\Lambda$, and our system has $\Lambda_b\delta_N$ degrees of freedom, where we recall $\delta_N$ is the gauge group dimension [see Eq. (\ref{dims})]. However, as we explain more precisely below,  there is a copy of the gauge group $\mathcal G$ attached to each spacetime finite lattice site $x\in\Lambda$, and due to local gauge invariance of the plaquette actions $A_p$  in Eq. (\ref{partition}), when considering the total number $\Lambda_b$ of bonds and then the whole set of gauge variables in the lattice $\Lambda$, there is an excess of variables.

By a gauge fixing procedure, the extraneous gauge variables can be eliminated. Here, we sometimes fix what we call the {\em enhanced temporal gauge}. In this gauge, the temporal bond variables in $\Lambda$ are set to the identity (leading to a trivial gauge group integration), as well as certain specified bond variables on the boundary $\partial\Lambda$ of $\Lambda$. The gauged away bond variables involve bonds which {\em do not} form a lattice loop. This guarantees that the value of partition function is unchanged \cite{GJ}. Also, the maximal number of relevant variables is given by
\bequ
\lb{lambdar}\Lambda_r\,\approx (d-1)L^{d-1}\,=\,[(d-1)\Lambda_s/L]\,,
\eequ 
which is roughly the number of non-temporal (spatial) bonds.  These are the {\em retained bond variables} or simply, retained bonds. The {\em effective number of degrees of freedom in our model} is then $[\delta_N\Lambda_r]$.
\begin{rema}
We warn the reader that we use abusively the name volume for the quantity $\Lambda_r$. Below, we talk about the volumetric free energy using $\Lambda_r$ instead of the lattice volume $\Lambda_s=L^d$. We are not using the physical volume $(aL)^d$ in $\mathbb R^d$. By doing so, as there is a finite proportionality between them, we will be neglecting a finite additive ($d$-dependent) constant value of the free energy.
\end{rema}

In the context of a generic lattice model with partition function $\mathcal Z_\Lambda$, when $\Lambda_f$ degrees of freedom are present, a stability bound is a lower/upper bound on $\mathcal Z_\Lambda$ of the form
\bequ\lb{bdunscaled}
e^{c_\ell\Lambda_f}\,\leq\,\mathcal Z_\Lambda\,\leq\,e^{c_u\Lambda_f}\,,
\eequ
for some finite real `constants' $c_\ell$ and $c_u$.

Usually, in statistical mechanical models, to control the thermodynamic limit of the free energy, the only requirement is that $c_\ell$ and $c_u$ are uniform in the `volume' $\Lambda_f$. Here, in our lattice YM model, besides this condition, since we are also interested in taking the UV (continuum) limit $a\searrow 0$ afterwards, we do require that $c_\ell$ and $c_u$ are also uniform in the lattice spacing $a$. This is why we call our stability bound a {\em thermodynamic and UV stable stability bound}, or simply {\em TUV  bound}, for short.

In addition to the physical, {\em unscaled} gluon fields which, in order to improve clarity, we sometimes denote by $A_b^u$, with a superscript $u$, we will also be using local, {\em scaled gluon fields} $A_\mu(x)$ (without any superscript, or with a superscript $s$) which are related to $A_\mu^u(x)$ by the $a$-dependent scaling transformation
\bequ\lb{scalingA}
A_\mu(x)\,=\,a^{(d-2)/2}\,A^u_\mu(x)\,.
\eequ
This a priori scaling corresponds to a partial renormalization and can interpreted as a field strength renormalization.

Associated with the scaled fields are the {\em scaled free energy}, {\em scaled generating functionals} and {\em scaled correlations}. As it will be made clear below, the {\em scaled field quantities have good UV regularity properties} in the lattice spacing $a$ and the gauge coupling $g$. For instance, the scaled free energy and correlations are bounded uniformly in $a\in(0,1]$ for any finite gauge coupling $g^2$.

Before embarking on the scaling transformations and specific results for the YM model free energy and correlations, we give a general picture of the use of our scaled fields and its consequences. For simplicity, we consider the free lattice scalar field. This case is analyzed in more detail in Appendix A. It is important to stress that the scaling used in the analysis of the free scalar field case is also employed for each color component $A_\mu^c(x)\in\mathbb R$, $x\in\Lambda\subset a\mathbb Z^d$, of our gauge fields in our YM model.

Considering the same hypercubic lattice defined before, we denote the $\Lambda$ lattice physical (or unscaled) scalar field at a lattice site $x\in\Lambda$ by $\phi^u(x)$. Up to boundary conditions, the physical, {\sl free} unscaled lattice scalar model action is given by
\bequ\lb{freeboson}
A^u_\Lambda(\phi^u)\,=\,\dfrac12\, \kappa_u^2 a^{d-2}\,\sum_{x,\mu}\,\left[ \phi^u(x_\mu^+)\,-\,\phi^u(x)\right]^2\,+\,\dfrac12\,m_u^2a^d\,\sum_{x}\,[\phi^u(x)]^2\,.
\eequ
Here, $\kappa_u^2,\,m_u\,>\,0$ and $\sum_{x,\mu}$ sums over the finite lattice sites $x\in\Lambda\subset a\mathbb Z^d$ and $\mu\,=\,0,1,\ldots, (d-1)$.
% is the unscaled hopping parameter, $m_u$ is the unscaled field mass, and we recall that $\mu=0,1,\ldots,d$ is a spacetime direction and $x_\mu^+\,=\,x\,+\,ae^\mu$, for the unit vector $e^\mu$ of the direction $\mu$.

As usual, the finite lattice unscaled partition function is
\bequ\lb{partu1}
Z^u_\Lambda(a)\,=\,\dis\int\,e^{-A^u_\Lambda(\phi^u)}\,D\phi^u\,,
\eequ
where $D\phi^u$ is the $\Lambda$ product of single site Lebesgue measures $d\phi^u(x)/\sqrt{2\pi}$. Recalling that lattice $\Lambda$ has $L\in\mathbb N$, $L$ even, sites on a side, and its volume is $\Lambda_s\,=\,L^d$, we can take the thermodynamic limit of the physical, unscaled free energy as
$$
f^u(a)\,=\,\lim_{L\nearrow\infty}\,\dfrac1{L^d}\,\ln Z^u_\Lambda(a)\,.
$$

The scalar model correlations {\em at coincident points} are singular in the continuum limit $a\searrow 0$.

We now introduce $a$-dependent scaled scalar fields by making the change of variables
\bequ\lb{cvs}
\phi(x)\,=\, s(a)\,\phi^u(x)\qquad;\qquad s(a)\,=\,a^{(d-2)/2}\,t\quad,\quad t\,=\,\left(2d\kappa_u^2\,+\,m_u^2a^2\right)^{1/2}\,.
\eequ

The scaled scalar field action is
$$
A_\Lambda(\phi)\,=\,A_\Lambda^u(\phi^u\,=\,[s(a)]^{-1}\,\phi)\,=\,-\, \kappa^2\sum_{x,\mu}\,\phi(x)\phi(x_\mu^+)\,+\,\dfrac12\,\sum_x\,\phi^2(x)\,,
$$
where $\kappa^2$ is the scaled hopping parameter $\kappa^2\,=\,\left(2d\,+\,m_u^2a^2/\kappa_u^2\right)^{-1}$.

The finite lattice scaled free field partition function is 
\bequ\lb{Zff}
Z_\Lambda(a)\,=\,\dis\int\,e^{-A_\Lambda(\phi)}\,D\phi\,=\,s^{\Lambda_s}\,Z^u_\Lambda(a) \,,
\eequ
where we recall that $\Lambda_s\,=\,L^d$ is the number of sites in $\Lambda$ and the scaled measure $D\phi$ is defined a product measure, similarly to the unscaled measure $D\phi^u$ in Eq. (\ref{partu1}). 

The scaled scalar model partition function $Z_\Lambda(a)$ obeys the TUV stability bound
$$
e^{c_\ell\Lambda_s}\,\leq\,Z_\Lambda(a)\,\leq\,e^{c_u\Lambda_s}\,,
$$
with real finite constants $c_\ell$ and $c_u$ independent of $\Lambda_s$ and the lattice spacing $a$.

The important point is that, in terms of scaled fields, all singularities disappear, and the scaled free energy is finite, for $d\,=\,3,4$, and is given by
$$
f(a)\,=\,\lim_{L\nearrow\infty}\,\dfrac1{L^d}\,\ln Z_\Lambda(a)\,.
$$
We observe that scaled field correlations, to be defined below, also remain {\em finite}  as $a\searrow 0$, {\em even at coincident points}, for $d\,=\,3,4$.

Furthermore, in the thermodynamic limit $L\nearrow\infty$, the unscaled free energy is related to the scaled one by
\bequ\lb{freerelation}
f(a)\,=\,\ln s(a)\,+\,f^u(a)\,,
\eequ
meaning that we have isolated completely the $a\searrow 0$ singularity of $f^u(a)$.

Regarding correlations, the physical, unscaled two-point correlation
$$
\langle \phi^u(x)\, \phi^u(y)\rangle^u\,\equiv\, \dfrac 1{Z^u_\Lambda}\,\dis\int\,\phi^u(x)\, \phi^u(y)\;e^{-A_\Lambda^u(\phi^u)}\;D\phi^u\,,
$$
using the change of variables of Eq. (\ref{cvs}), is related to the corresponding scaled two-point correlation (similarly defined without the $u$ superscripts) by
$$
\langle \phi^u(x)\, \phi^u(y)\rangle^u\,=\,\dfrac1{s^2(a)}\,\langle \phi(x)\, \phi(y)\rangle\,.
$$
Both correlations have the same asymptotic decay rate at large distances. More generally, the scaling transformation does not affect the Osterwalder-Schrader positivity property \cite{GJ,Dim2} nor does it alter the decay rates of correlations. Thus, the E-M spectrum is unchanged.

As is known, the unscaled correlations are singular in the continuum limit $a\searrow 0$. However, the scaled free scalar model correlations remain finite
as $a\searrow 0$, even at coincident points, for $d=3,4$. For example, for the unscaled two-point correlation at coincident points we have the following singular behavior
$$
\langle [\phi^u(x)]^2\rangle\,\simeq\,\dfrac1{s^2(a)}\,\simeq\,a^{2-d}\,.
$$
This behavior can be taken as a characterization of UV asymptotic freedom on the lattice.

In the YM case, we consider only gauge-invariant correlations which behave as derivative fields (think of the electromagnetic field for the abelian gauge group $\mathrm U(1)$). For scalar free fields the unscaled finite lattice derivative field correlation is
$$
\left\langle \partial^a_\mu\phi^u(x)\, \partial^a_\mu\phi^u(y)\right\rangle^u\,\equiv \,\left\langle\dfrac1a\, \left[\phi^u(x_\mu^+)\,-\,\phi^u(x)\right]\, \dfrac1a\,\left[\phi^u(y_\mu^+)\,-\,\phi^u(y)\right]\right\rangle^u\,,
$$
with derivatives replaced by finite differences. This correlation behaves as $s^{-2}(a)\,a^{-2}\,\simeq\,a^{-d}$, for coincident points. This behavior can also be taken as a characterization of UV asymptotic freedom on the lattice for derivative field correlations.

We remind the reader that, even for a finite lattice with spacing $a\in(0,1]$, the partition function may not exist due to zero modes in the action. This is e.g. the case for periodic b.c. where the action has a zero mode for $m_u\,=\,0$. This problem is usually eliminated by adding to the action a small mass term (an infrared regulator). Then the partition function exists and obeys TUV stability bounds. Then, the regulator may be removed after taking the thermodynamic limit. Alternatively, for example, if free b.c. is used, there is no zero mode, the finite lattice partition function exists and agrees with that in the infrared regulator procedure. In this paper, we deal with the Wilson YM model and no infrared regulator is needed for free and periodic b.c..

We now return to the YM model and describe how we apply the scaling transformation.
For the abelian $\mathrm U(1)$ case we can define the scaling transformation by a similar change of variables of Eq. (\ref{scalingA}), namely
$$
A_\mu(x)\,=\, a^{(d-2)/2}\,A_\mu^{u}(x)\,.
$$ 
In terms of the gauge fields $A_\mu(x)$, the action becomes regular both in $g$ and $a\in(0,1]$, and the $\mathrm U(1)$ Haar measure is proportional to the Lebesgue measure. 

In the nonabelian case, the implementation and the effect of the scaling is more complicated than the scalar free field case. The scaling is performed in each color component $A_\mu^{u,c}$ of the physical or unscaled gluon field $A_\mu^u$ by defining
\bequ\lb{nonabscaling}
A_\mu^c(x)\,=\, a^{(d-2)/2}\,A_\mu^{u,c}(x)\,,
\eequ
so that the bond variable $U_b$ is parametrized as $U_b\,=\,\exp[iga^{-(d-4)/2}\,A_b]$. In doing this, the action of each plaquette Boltzmann factor becomes also regular. However, the Haar measure is not proportional to the Lebesgue and does not transform simply by a multiplicative factor. This complexity does not lead to a multiplicative scaling for the partition function and correlations. Below, we show how we deal with this problem. The Haar measure for unitary groups is obtained in Refs. \cite{Nelson,Holland,Marinov,Euler}.

Before we get to this point, in order to be able to compare our method with the renormalization method developed in Ref. \cite{GH}, we illustrate how our scaling leads to the absence of singularities in the scalar $\phi^4_3$ model. From Refs. \cite{BFSo,GH,Hairer}, the unscaled action is the free action $A^u_\Lambda(\phi^u)$ of Eq. (\ref{freeboson}) plus a local quartic potential. Namely, for spacetime dimension $d=3$, we have (for $\lambda\,>\,0$)
\bequ\lb{boson43}\barr{lll}
A_{4,\Lambda}^u(\phi^u)&=&\dfrac12\, \kappa_u^2 a\,\sum_{x,\mu}\,\left[ \phi^u(x_\mu^+)\,-\,\phi^u(x)\right]^2\,+\,\dfrac12\,\left( m_u^2\,-\,\dfrac{c\lambda}a\right)\;a^3\,\sum_{x}\,[\phi^u(x)]^2\,+\,\lambda\,a^3\, \sum_{x}\,[\phi^u(x)]^4\vspace{2mm}\\
&\equiv&A^u_\Lambda(\phi^u) \,+\,A_{I,\Lambda}^u(\phi^u)\,.\earr
\eequ
In $d=3$, the sums are over $x\in\Lambda\subset a\mathbb Z^3$ and $\mu\,=\,0,1,2$. Also, we have  $c\,\equiv\,c(a,\lambda)\,=\,c_1(a)\,+\,c_2(a)\,\lambda$, where $c_1(a)\,=\,{\mathcal O}(1)$ and $c_2(a)\,=\,{\mathcal O}(|\ln a|)$. The form of $[\lambda c(a,\lambda)/a]$ is a renormalization which removes the UV divergences and keeps the physical mass finite in the continuum limit $a\searrow 0$. In $\lambda$ perturbation, we take care of the ${\mathcal O}(\lambda)$ tadpole contribution and the so called `rising sun' ${\mathcal O}(\lambda^2)$ contribution. For simplicity, we suppress the $a,\,\lambda$ dependence of $c(a,\lambda)$ and put $c\equiv c(a,\lambda)$. 

As before, in terms of scaled fields, $$\phi(x)\,=\,s\,\phi^u(x)\quad,\quad s= a^{(d-2)/2}\,t\quad;\quad t\,=\, [2d\kappa_u^2\,+\,(m_ua)^2]^{1/2}\,,$$ the scaled $\phi^4_ 3$ action is
\bequ\lb{scboson43}\barr{lll}
A_{4,\Lambda}(\phi^u)&=&-\, \kappa^2\,\sum_{x,\mu}\, \phi(x)\,\phi(x_\mu^+)\,+\, \dfrac12\,\sum_{x}\,[\phi(x)]^2\,-\,\dfrac{c\lambda a^{2}}{s^2}\,\sum_{x}\,[\phi(x)]^2 \,+\,\dfrac{\lambda a^3}{s^4}\, \sum_{x}\,[\phi(x)]^4\vspace{2mm}\\
&\equiv&A_\Lambda(\phi) \,+\,A_{I,\Lambda}(\phi)\,,\earr
\eequ
where $A_\Lambda(\phi)$ is the scaled free action. The key fact is that $A^I(\phi)$ has a minimum at $\phi^2(x)\,=\,c/2$, where it assumes the value $(-\,\lambda a c^2\,\Lambda_s/4)$.

We let $Z_\Lambda^{\mathrm{fr}}$ denote also the scaled free scalar field partition function of Eq. (\ref{Zff}). Then, $Z_\Lambda^{\mathrm{fr}}$ satisfies the TUV stability bound (recalling $\Lambda_s\,=\,L^d$, $d=3$)
$$
e^{c^{\mathrm fr}_\ell\,\Lambda_ s}\,\leq\,Z_\Lambda^{\mathrm{fr}}\,\leq\,e^{c^{\mathrm fr}_u\,\Lambda_ s}\,,
$$
where
$c_u^{\mathrm fr}\,=\,[(1\,-\,L^{-1})\,\ln\sqrt{2}]$ (see Appendix C) and $c_\ell^{\mathrm fr}\,=\,0$ (by Jensen's inequality).

Now, we let $Z_{4,\Lambda}$ denote the partition function for the $\phi^4_3$ model. We have 
$$
Z_{4,\Lambda}\,=\,\int\,e^{-A_{4,\Lambda}(\phi)}\,d\phi\,\equiv\,Z^{\mathrm fr}_\Lambda\,\int\,e^{-A_{I,\Lambda}(\phi)}\,d\mu^{\mathrm fr}(\phi)\,,
$$
with, using Eq. (\ref{scboson43}), $d\mu^{\mathrm fr}(\phi)\,=\,\int\,e^{-A_\Lambda(\phi)}\,d\phi/Z^{\mathrm fr}_\Lambda$ is a probability measure.

For the upper bound, reinstating the $\lambda\,,\,a$ dependence in $c$, taking the minimum value of $A_{I,\Lambda}(\phi)$ given above, we have
$$
Z_{4,\Lambda}\,\leq\, \exp\left[ a\,\lambda\,c^2(\lambda,a)\,\Lambda_s \right]
\:\exp\left[c_u^{\mathrm fr}\,\Lambda_s \right]\,\equiv\,e^{c_u\,\Lambda_s}\,.$$

Now, by Jensen's inequality and using the lower bound $Z_\Lambda^{\mathrm fr}\geq 1$, we have the lower bound
$$
Z_{4,\Lambda}\,\geq\,e^{-\,\int\,A_{I,\Lambda}(\phi)\,d\mu^{\mathrm fr}(\phi)}\,\equiv\,e^{c_\ell\,\Lambda_s}\,,
$$
with, here, $c_\ell\,=\,\left[(c\lambda a/t^2)\,C_0\,-\,(3\lambda a /t^4)\,C_0^2\right]$ and $C_0$ is the coincident point two-point correlation for the free scaled field given in Eq. (\ref{2pfscalar}) and bounded in Eq. (\ref{bbb}).

Writing the upper and lower bound together, we have the TUV stability bound
\bequ\lb{phi43bd}
e^{c_\ell\,\Lambda_ s}\,\leq\,Z_{4,\Lambda}\,\leq\,e^{c_u\,\Lambda_ s}\,.
\eequ
From the values of $c_\ell$, $c_u$ and $c$ given above and using, for $C_0$ Remark \ref{unifcs} below, we see that the constants in Eq. (\ref{phi43bd}) are uniform in $a\in(0,1]$ so that the bound in Eq. (\ref{phi43bd}) is a TUV stability bound.

Let us now come back to the gauge field case. We consider the abelian case of the gauge group $\mathrm U(1)$ case in more detail. For the special case of $\mathrm U(1)$, both the pure-gauge action and the coupling with Bose and Fermi fields were treated in Refs. \cite{Driver,BalabanHiggs,King,Dim}. The starting point for all these papers is a quadratic action for the electromagnetic potentials. In these papers, in order to remove the null space of the quadratic form and define the model partition function, a gauge fixing is required at the onset. This is {\em not} what we do here. Instead, we make a rigorous connection between the Wilson partition function and the Wilson plaquette action which is {\em not} quadratic in the fields.

To see the effect of the scaling transformation on the $\mathrm U(1)$ Haar measure, we parametrize the bond variable $U_b$ with the physical potential as $U_b\,=\,\exp(iagA_b^u)$. The measure is
$\dis\prod_b\,\left(\dfrac{ag}{2\pi}\,dA_b^u\right)$. In terms of scaled fields $A_b$, the measure is 
$\dis\prod_b\,\left[\left(\dfrac{g^2}{a^{d-4}}\right)^{1/2}\,\,dA_b/(2\pi)\right]$. Hence, for $\mathcal G\,=\,\mathrm U(1)$, the unscaled and scaled gauge finite lattice partition functions are related by
$$
Z^u_\Lambda(a)\,\equiv\,Z^w_\Lambda(a)\,=\,\left(\dfrac{g^2}{a^{d-4}}\right)^{\Lambda_r/2}\,Z^s_\Lambda(a)\,,
$$
where we recall that $\Lambda_r$ is the number of retained gauge variables after the enhanced temporal gauge is fixed (see Eq. (\ref{lambdar})).

Here, $Z^s_\Lambda(a)$ is the {\em scaled partition function}, expressed with an action written in terms of scaled fields and with the measure $\dis\prod_b\,dA_b/(2\pi)$, $|A_b|\,<\,\pi$.

Also, $Z^s_\Lambda(a)$ obeys TUV stability bounds ($\delta_N\,=\,1$ is the $\mathrm U(1)$ group dimension)
\bequ\lb{TUVu1}
e^{c_\ell\delta_N\Lambda_r}\,\leq\, Z^s_\Lambda(a)\,\leq\,  e^{c_u\delta_N\Lambda_r}\,.
\eequ

Motivated by the scaling relation above, for the abelian case, in the nonabelian YM case of the gauge groups $\mathcal G\,=\,\mathrm U(N),\;\mathrm{SU}(N)$, by defining a scaled field partition function by
\bequ\lb{Zbound}
Z^s_\Lambda(a)\,=\,\left(\dfrac{a^{d-4}}{g^2}\right)^{\delta_N\Lambda_r/2}\;Z^u_\Lambda(a)\,,
\eequ
with one factor of $\left(\dfrac{a^{d-4}}{g^2}\right)^{1/2}$ for each of the $\delta_N\Lambda_r$ effective degrees of freedom, $Z^s_\Lambda(a)$ obeys TUV stability bounds.

%After the above explanation about scaled fields and coming back to our introduction and description of results, in Refs. \cite{M,MP,MP2}, we found that the use of scaled fields, with a noncanonical scaling, depending on the lattice spacing $a$, {\em reduces the UV singularity of QFT models}. In the particular case of the real scalar free fields, we found that both the free energy and correlations (even at coincident points!) are bounded uniformly in $a\in(0,1]$, such that they are finite in the continuum limit $a\searrow 0$. This is detailed in Appendix A. \vspace{1cm}For this reason, we scale each component $A_b^c$ of the gluon field, in a similar way as we did for the scalar field. The original/physical quantities as fields, partition function,... defined above are sometimes denoted by a superscript $u$, e.g $A_b^u$ and $A_b^{u,c}$, and we call them {\em unscaled}. The scaled quantities are denoted without a superscript. 

Here, we consider two types of boundary conditions (b.c.) on the lattice $\Lambda$. Namely, our TUV bounds on the scaled partition function $Z^{s,B}_{\Lambda}\equiv Z^{s,B}(\Lambda,a)$ hold for both free and periodic b.c. The index $B$ is left blank, for free b.c., and $B=P$, for periodic b.c. The adoption of periodic b.c extends the TUV bound result of Refs. \cite{YM} and is convenient for us to use the multiple reflection method \cite{GJ} to analyze a generating function and correlations. The partition function $Z^{s,B}_{\Lambda}$ is related to $Z^u_\Lambda(a)$ as in Eq. (\ref{Zbound}) obeys the the same TUV stability bound of Eq. (\ref{TUVu1}) with a corrected value of $\delta_N\,=\,N^2,\;(N^2-1)$. Namely, we have 
\bequ\lb{TUVnorm}
e^{c_\ell\delta_N\Lambda_r}\,\leq\,Z^{s,B}_{\Lambda}\,\leq \, e^{c_u\delta_N\Lambda_r}\,,
\eequ
where we recall that the finite constants $c_\ell$ and $c_u$ are both uniform in $\Lambda_r$ and $a\in(0,1]$.

Using Eq. (\ref{Zbound}), the proof of the bound of Eq. (\ref{TUVnorm}) is one of our main results and is stated in Theorem \ref{thm1}.

The TUV stability bound on the scaled partition function [see Eq. (\ref{TUVnorm})] arises from an interesting factorization structure of the bounds for $Z^{u,B}_\Lambda$ and hence $Z^{s,B}_\Lambda$. For $Z^{u,B}_\Lambda$, the upper and lower bounds factorize as
$$
z_\ell^{\Lambda_r}\,\leq\,Z^{u,B}_\Lambda\,\leq\,z_u^{\Lambda_r}
$$
where each factor is a {\em single plaquette partition function of one bond variable}. Expressions for the `constants' $z_\ell$ and $z_u$ involve probability distributions of the circular unitary random matrix and the Gaussian unitary ensembles, CUE \cite{Metha,Deift}, and are analyzed in Theorem \ref{thm2}. This analysis leads to the TUV stability bound of Eq. (\ref{TUVnorm}). 
 \begin{rema}
A key point in our method, as it can be checked in the expressions and bounds on $z_\ell$ and $z_u$ given below, is that our lower and upper bounds on the unscaled partition functions do exhibit the same multiplicative singular factor. This is what allows us to define a scaled partition function by extracting the singular factor multiplicatively.
\end{rema}

The random matrix ensemble CUE arises naturally in the above context. To go further, and extract the singular behavior of $z_\ell$ and $z_u$, the Gaussian unitary random matrix ensemble GUE \cite{Metha,Deift} also shows up. We notice that both $z_\ell$ and $z_u$ are given by integrals with {\em class function} integrands. We emphasize that this property was not present for the integrand of the unscaled partition function $Z^u_\Lambda$ but does hold for the bounds and we recall that a {\em class function} $f(U)$ on the gauge group $\mathcal G$ is constant over each group conjugacy class, i.e. $f(U)$ satisfies the property $f(U)= f(VUV^{-1})$, for all $V\in\mathcal G$.

We now explain how these random matrix ensembles appear in our bounds.

Our upper and lower stability bounds have an interesting factorization structure. In continuum scalar field models such a factorization is achieved by imposing Dirichlet decoupling on the covariance of the free field Gaussian measure (see Chap. 9 of \cite{GJ} and \cite{GRS,GRS2,Simon4}). In the continuum, in the exponents of the stability bounds, we typically have the volume in $\mathbb R^d$. We show in Appendix C, how this factorization is accomplished in lattice scalar field models where each factor is a partition function of single bond `transfer matrix'. Here, the exponent in the TUV stability bound is the number $\Lambda_b\,=\,d(L-1)L^{d-1}$ of lattice bonds in $\Lambda$.

In the YM model, the factorization involves products of single-plaquette, single-bond variable partition functions. A new, global quadratic upper bound in the gluon fields, for the positive Wilson plaquette action, is proved in Lemma \ref{lema1}. This upper bound gives a lower bound on the partition function. This bound gives rise to the factorized lower bound on the partition function. For the upper bound, since each plaquette action is positive (it is  a Hilbert-Schmidt norm!), we simply set some plaquettes actions to zero. We denote by $z_u$ ($z_\ell$) the single-bond Haar integral partition functions describing  the single-plaquette partition function for the upper (lower) stability bound on the partition function with periodic b.c. 

By the spectral theorem, as $U$ is unitary, there exists a unitary $V$ which diagonalizes $U_b$, i.e. $V^{-1}U_b V=\mathrm{diag} (e^{i\lambda_1},\ldots,e^{i\lambda_N})$,  $\lambda_j\in(-\pi,\pi]$.  The $\lambda_j$ are called the {\em angular eigenvalues} of $U$. Recalling Eq. (\ref{scalingA}), the fundamental relation between scaled gluon fields and the angular eigenvalues is given by the equality (see Lemma \ref{lemma0})
\bequ\lb{quadrado}
\sum_{j=1}^N\; \lambda_j^2\,=\, a^2g^2\;\sum_{c=1}^{\delta_N}\, \left|A_b^{u,c}\right|^2\,=\,\dfrac{g^2}{a^{d-4}}\;\sum_{c=1}^{\delta_N}\, \left|A_b^{c}\right|^2\,,\eequ
which is used to deal with the above quadratic bound on the Wilson action. It is immediate for $\mathcal G=\mathrm U(1)$.

With this, for $\lambda=(\lambda_1,\ldots,\lambda_N)$ and $d\lambda=d\lambda_1\dots d\lambda_N$, by the Weyl integration formula \cite{Weyl,Bump,Simon2,Far}
\bequ\lb{weyll}
\dis\int_{{\mathrm U}(N)}\;f(U)\;d\sigma(U)\,=\,\dfrac1{N!}\,\dis\int_{(-\pi,\pi]^N}\;f\left({\rm diag}(\lambda_1,\lambda_2,\ldots,\lambda_N)\right)\;\dis\prod_{k,j=1,\ldots,N; k<j}\,|e^{i\lambda_j}-e^{i\lambda_k}|^2\;\dfrac {d\lambda}{(2\pi)^N}\,,
\eequ
the $N^2$-dimensional Haar integration over the $N\times N$ matrix unitary gauge group $\mathcal G$ is reduced to an $N$-dimensional integration over the angular eigenvalues of $U$. In Eq. (\ref{weyll}), the measure corresponds to the probability density of the circular unitary ensemble (CUE) and in the bounds on $z_u$ and $z_\ell$ the probability density for the Gaussian unitary ensemble (GUE). Random matrix theory appears in a natural way in our context (see Refs. \cite{Metha,Deift}). 

From the Weyl formula of Eq. (\ref{weyll}), we see that, even in the case of class functions, the measure $$\prod_{k,j=1,\ldots,N;\; k<j}\:|e^{i\lambda_j}-e^{i\lambda_k}|^2\;d\lambda$$ does not obey a multiplicative scaling relation under a change of variables transformation. However, for small enough $\lambda_k$, scaling $\lambda_k$ by $(s\lambda_k)$, this measure scales with a factor $s^{\delta_N}$. 

Going further, the importance of the TUV bound of Eq. (\ref{TUVnorm}), for the scaled partition function $Z^s_\Lambda(a)\,\equiv\,Z^s(\Lambda,a)$, is that it ensures us that the finite lattice scaled YM free energy
\bequ\lb{free}
f^s_\Lambda(a)\,\equiv\,f^s(\Lambda,a)\,=\,\dfrac1{\Lambda}\,\ln Z^s_\Lambda(a)
\eequ satisfies the bound \bequ\lb{bdf}c_\ell\,\leq\,f^s_\Lambda(a)\,\leq\,c_u\,.\eequ
However, recalling that the finite constants $c_\ell$ and $c_u$ are both uniform in $\Lambda_r$ and $a$, by the Bolzano-Weierstrass theorem \cite{Rudin}, this shows that the thermodynamic limit
$$f^s(a)\,=\,\lim_{\Lambda\nearrow a\mathbb Z^d}\, f^s(\Lambda,a)\,$$ exists at least in the subsequential sense. Subsequently, by the same reason, the continuum limit exists, at least in the subsequential sense, and defines the bounded function $$f^s\,=\,\lim_{a\searrow 0}\,f^s(a)\,,$$ of the model parameters.

From the above discussion, we see that our scaling transformation, used to define the scaled gluon fields, allowed us to subtract the exact singularity of the physical or unscaled free energy $$f^u_\Lambda(a)\equiv f^u(\Lambda,a)\,=\,(\ln Z^u_\Lambda(a))/\Lambda_r\,.$$ In fact, we see that the scaled and unscaled free energy are related by
\bequ\lb{relf}
f^s_\Lambda(a)\,=\,f^u_\Lambda(a)\,+\,\delta_N\;\ln \dfrac{a^{(d-4)/2}}{g}\,,
\eequ
which is analogous to Eq. (\ref{freerelation}) for the scaled scalar free field model.
%\begin{rema}
%As shown in Appendix A, for a scalar field model, a similar relation to Eq. (\ref{relf}) occurs. There, for the scaled and unscaled finite lattice free energies, we obtain
%\bequ\lb{relation1}
%f_\Lambda(a)\,=\, \ln s\,+\,f^u_\Lambda(a)\quad,\quad s\equiv s(a)\,=\, \left(2d\kappa_u^2a^{d-2}\,+\,m_u^2a^d \right)^{1/2}\,,
%\eequ
%where $\kappa_u$ is the unscaled hopping parameter, $m_u$ is the unscaled field mass and $s$ is the scaling factor used to define scaled scalar fields in Eq. (\ref{cvs}), i.e. $\phi(x)\,=\,s(a)\,\phi^u(x)$.
%\end{rema}

It is to be remarked that, as a first step towards showing the existence (the continuum and thermodynamic limits) of the YM models and QCD, here and in Ref. \cite{YM}, we only concentrated on the finiteness of the scaled free energy per degree of freedom. Next, in Ref. \cite{GYM}, the finiteness analysis was extended to the generating function of gauge-invariant plaquette fields and their correlations. There, we did not invest in analyzing the properties these quantities satisfy and if there is possibly more than the physical models of interest. It is also worth noticing that the techniques and methods used in Ref. \cite{YM}, combined with the results of Refs. \cite{M,MP,MP2} were used to prove the existence of a scaled free energy for a bosonic lattice QCD model, with the (anti)quark fields replaced with spin zero, multicomponent complex or real scalar fields. This is the content of Ref. \cite{bQCD}. With these results in mind, we mention that our methods and techniques can eventually be coupled with new and more traditional methods, such as explicit renormalization and multiscale analysis, to make progress towards the complete construction of QFT models, e.g. to show the existence of the continuum and the thermodynamic limits of correlations for the $\mathrm{SU}(3)$ YM model and QCD, and also other interesting models which are still not fully understood.

We now turn to generating functions and correlations. The folklore tells us that the TUV bound of Eq. (\ref{TUVnorm}) is enough to bound the generating function and correlations. For the case of the scaled free field (see Appendix A), the generating function for powers of the field at a single point $x$ is
$$
\langle e^{J\phi(x)}    \rangle\,=\, \exp \frac12\left[ J^2\,\langle [\phi(x)]^2  \rangle  \right]\,,
$$
where $J$ is a constant source strength, and the expectations $\langle [\phi(x)]^r \rangle$, $r=1,2,\ldots$ are Gaussians and are also bounded. So as not to think that these boundedness properties only hold for the scaled scalar free field, in Appendix B, we prove that they also hold for another scaled field model, which we call {\em truncated model} and which is a good candidate for a continuum QFT model that exists and in perturbation theory it is nontrivial (non-Gaussian) in $d=4$. This is important because of the recent triviality results for $\phi^4_4$ by Aizenman and Dumenil-Copin \cite{AizDC}.

The model partition function, denoted here by $Z^t_\Lambda$, is obtained from the scaled free field partition function (see Eqs. (\ref{scscaledpart}) and (\ref{scact}) of Appendix A) by replacing the bond factor $\exp[\kappa^2\,\phi(x)\phi(x^+_\mu)]$, where $\kappa^2>0$ is the (squared) hopping parameter and $x^+_\mu=x+ae^\mu$, by the truncation $\left[ 1\,+\,\alpha \,\phi(x)\phi(x^+_\mu)\right]$. We call this model the {\em truncated model}. The partition function $Z^t_\Lambda$ obeys the TUV stability bound, recalling that $\Lambda_s\,=\,L^d$,
\bequ\lb{tsb}
e^{c_\ell\Lambda_s}\,\leq\,Z^t_\Lambda\,\leq\,e^{c_u\Lambda_s}\,,
\eequ
with finite constants $c_\ell$ and $c_u$ independent of $\Lambda_s$ and $a\in(0,1]$. The bounds depend on $\Lambda_s$ in the exponent and not the  physical volume $(aL)^d$ in $\mathbb R^d$. 

We also prove that the generating function $\langle e^{J\phi(x)} \rangle$ obeys the bound
$$\langle e^{J\phi(x)} \rangle\,\leq\,e^{c_u-c_\ell}\, e^{J^2}\,,$$ with the same constants $c_\ell$ and $c_u$ of Eq. (\ref{tsb}). Applying Cauchy estimates, this bound leads to the bound on the coincident point correlation with $r=1,2,...$ fields $\phi$ 
$$
\langle  \phi^r(x)\rangle\,\leq\,e\,e^{c_u-c_\ell}\, r!\:\,.
$$
The point is that the TUV stability bound using the lattice number of sites $\Lambda_s=L^d$ is {\it enough} to bound the generating function and correlations.

Concerning the YM model, inspired by Ref. \cite{Schor}, we define a  gauge-invariant physical, unscaled plaquette field as follows. Consider the plaquette $p\,=\,p_{\mu\nu}(x)$, $\mu<\nu$, in the $\mu\nu$ coordinate plane. The unscaled plaquette field associated with $p$ is given by, recalling that $U_p\,=\,\exp\{ iX_p \}$, 
\bequ\lb{plaqfield}\barr{lll}
\tr {\mathcal F}^u_{p}(x)&=&\dfrac 1 {a^2g}\,\Im\tr(U_p-1)\,=\,\dfrac {i}{2a^2g}\,\tr\,[U^\dagger_p\,-\,U_p]\,=\,%\vspace{2mm}\\&=&
 \dfrac1{a^2g}\,\tr (\sin X_p)\,.\earr
\eequ

The reason for the above $[1/(a^2g)]$ multiplicative factor in Eq. (\ref{plaqfield}) is that, if one uses the physical parametrization for $U_b\,=\,\exp(igaA_b)$, then we obtain, for $0<a\ll 1$, that $\tr \mathcal F^u_p\,\simeq\,\tr F_{\mu\nu}^a$, where
\bequ\lb{fmunua}F_{\mu\nu}^a\,=\,\partial^a_\mu A_\nu(x)-\partial^a_\nu A_\mu(x)+ig[A_\mu(x),A_\nu(x)]\,,\eequ
with a commutator in the Lie algebra of $\mathcal G\,=\, \mathrm U(N),\:\mathrm{SU}(N)$. 

Similarly to what we did for the gauge fields and Wilson action, we define the scaled plaquette field ${\mathcal F}^s_{p}(x)$ by
\bequ\lb{scaledplaq}
\tr\,{\mathcal F}^s_{p}(x)\,=\,  a^{d/2}\;\tr\,{\mathcal F}^u_{p}(x)\,=\,\dfrac{ a^{(d-4)/2}}{g}\;\Im \tr \left(U_p\,-\,1\right)\,.
\eequ

As before, we emphasize that, in the above equations, $g=0$ is only an apparent singularity follows from Lemma \ref{lema1} (see below).

To analyze the plaquette field generating functions, we use this scaled plaquette field. The scaled field plaquette correlations are proved to be bounded, uniformly in $a\in(0,1]$. These bounds imply bounds on the singular behavior, when $a\searrow 0$, of the physical, unscaled plaquette correlations. For example, the bound implies that the physical, unscaled plaquette-plaquette correlation has a singularity of at most $a^{-d}$, when $a\searrow 0$.
\begin{rema}
It is important to remark that the exponential decay of physical, unscaled plaquette field correlations is the same as that of scaled plaquette field correlations. Hence, the associated E-M spectrum is also the same. Also, in the abelian case, it is known that local polynomials in the plaquette fields form a dense set in the subspace of gauge-invariant vectors in the associated model quantum mechanical Hilbert space $\mathcal H$. Whether or not these local fields form a dense set in the nonabelian case requires further investigation. We know it generates the low-lying E-M massive glueball spectrum. Possibly, more general loop variables are needed to go up in the spectrum \cite{Sei}. 
\end{rema}

Rather than bounding the scaled plaquette field correlations directly, we bound the generating function of $r-$scaled ($r\in\mathbb N$) plaquette field correlations. To do this, we use periodic b.c. and the multireflection method (see Ref. \cite{GJ}). Based on the work of Ref. \cite{Schor}, we define a scaled generating function for the correlation of $r\in\mathbb N$ gauge-invariant scaled plaquette fields as
\bequ\lb{scaledgener}
\barr{lll}
\left\langle \exp\left\{ \sum_{j=1}^r\,J_j\,\tr\,{\mathcal F}^s_{p_j}(x_j)\right\}  \right\rangle&=&\dfrac 1{Z^P_\Lambda}\,\dis\int\,\exp\left\{ \sum_{j=1}^r\,J_j\,\tr\,{\mathcal F}^s_{p_j}(x_j)\,-\,\dfrac{a^{d-4}}{g^2}\,\dis\sum_{p\in\Lambda}\,A_p(U_p)\right\}\,d\mu(U)\vspace{2mm}\\
&\equiv& G_{r,\Lambda}(J^{(r),x})\,\equiv G_{r}(\Lambda,a,J^{(r),x})\,,
\earr
\eequ
where $J^{(r)}$ denotes the whole set of $r$ source strengths $\{J_1,\ldots,J_r\}$ and $x$ the set of $r$ external points $\{x_1,\ldots,x_r\}$. We note that the denominator is the $J^{(r)}=0$ value of the numerator integral, with periodic b.c., which is the periodic b.c. partition function $Z^P_\Lambda$. Also, here $p_j$ is a short notation for $p_{\mu_j\nu_j}(x_j)$.

Of course, the scaled plaquette fields correlation are given by the source derivatives $(\partial/\partial J_j)$ at $J^{(r)}=0$, i.e. setting all the sources to zero. Namely, for $y_E\,=\,\{y_1,\ldots,y_r\}$, we have
\bequ\lb{scaledcorrel}
C_{r,\Lambda}(y_E)\,\equiv\,C_{r}(\Lambda,a,y_E)\,=\,\left.\dfrac{\partial^r}{\partial J_1(y_1)\ldots\partial J_r(y_r)}\,G_{r,\Lambda,a}(J^{(r)})\,\right|_{J^{(r)}=0}\,.\eequ
 
In Theorem \ref{thm4} below, we prove that the scaled generating function $G_{r,\Lambda}(J^{(r),x})$ is absolutely bounded, with a bound that is independent of $L$, $a$, $g$, and the location and orientation of the $r$ {\em external} plaquette fields. This bound leads to the existence of a sequential or subsequential thermodynamic limit $G_{r,a}(J^{(r)})$ and then to a continuum $a\searrow0$ (at least subsequential) limit $G_{r}(J^{(r)})$.

The generating function bound also has an interesting structure. The bound has only a product of single-plaquette, single bond-variable partition function $z_u(J)$ with a source strength field $J$ in the numerator;  in the denominator only a product of $z_\ell$ (the same as in the preceding case!) occurs. In the bound for $z_u(J)$  the probability density for the Gaussian symplectic ensemble (GSE) appears (see \cite{Deift}). The generating function  is jointly analytic, entire function in the source strengths $J_1$, ..., $J_r$ of the $r$ plaquette fields. The $r$-plaquette field correlations admit a Cauchy integral representation and are bounded by Cauchy bounds. In particular, the coincident point plaquette-plaquette physical field correlation is bounded by $\mathrm{const}\,a^{-d}$. The $a^{-d}$ factor at small $a$ behavior is the same as that of the physical or unscaled real derivative scalar free field two-point correlation (the physical free field correlation has a singular behavior $a^{-(d-2)}$). For the free field, these singular behaviors are a measure of ultraviolet asymptotic freedom, in the context of the lattice approximation to a continuum QFT.

In this way, we conclude that the singular behavior of the plaquette correlations is bounded by the singular behavior of the free derivative scalar field correlations in $d=2,3,4$. For the physically relevant $d=4$ case, we can say more. The behavior of the coincident plaquette, physical plaquette-plaquette correlation is exactly $a^{-d}\,h(g)$, for some function $h(g)$ which is bounded.

For the free physical scalar field, locally scaled field correlations are bounded uniformly in $a\in(0,1]$, such as no smearing of the fields is needed to achieve boundedness. The two-point correlation of physical fields for coincident points has an $a^{2-d}$ singular behavior, for $d=3,4$. If we consider correlations of physical derivative scalar fields, then the singular behavior is different. The two-point correlation of physical derivative scalar fields, at coincident points, has an $a^{-d}$, $a\searrow 0$ singularity, for $d=2,3,4$; for the massless case the exact value is $a^{-d}/d$.

For the free field, the relation between the correlations of physical fields or physical derivative fields and their scaled counterparts is developed Appendix A.

In this paper, first, we give detailed and much simplified proofs of the Theorems of the unpublished Refs. \cite{YM,GYM}, using free boundary conditions. Moreover, we are able to incorporate the case of periodic boundary conditions in the present analysis. Similar finiteness results are shown to hold for the generating function of gauge-invariant plaquette fields \cite{Schor} and their correlations. For doing this, we adopt periodic b.c. and apply the multireflection method.

Besides, in order the allow the reader see how our methods work in a simpler case, we discuss the special case of the abelian gauge group ${\mathrm U}(1)$. For this group, the Haar measure is much simpler as compared with ${\mathrm U}(N>1)$ \cite{Nelson,Holland,Marinov,Euler} and computations can be carried out more explicitly and transparently, and we emphasize that the independence of our bounds on $a\in(0,1]$ is already manifest in this case and the reader can better appreciate why this holds true.   

For both, free and periodic b.c., our TUV stability bounds on the scaled partition functions (defined by extracting the $a\searrow 0$ singularity) lead to at least the existence of the subsequential thermodynamic and ultraviolet limits of the corresponding   scaled free energies per effective degree of freedom. The existence of these subsequential continuum limits apply to any gauge model with the same Wilson action and free/periodic b.c.

We now emphasize that our method does not intend to be a substitute to other powerful methods such as the RG multiscale formalism. However, it corresponds to a simple way to obtain TUV and generating function bounds, as well as it allows to a simple characterization of UV asymptotic freedom in the context of a lattice field theory. Our results hold for the whole one-parameter family of models depending on the gauge coupling $g$, such that $g^2\in(0,g_0^2]$, $0<g_0^2<\infty$. As pointed out before, the type of singularity we met for the plaquette-plaquette correlation is typical of UV asymptotic models, as it is expected on physical grounds. Concerning the mass, whether or not our YM models have a mass is not known for now. However, we can think that applying our methods with the ones developed in Ref. \cite{Schor} can be used to attack this question. Recall that in \cite{Schor}, a gauge-invariant glueball state was shown to be present in the E-M spectrum, with mass gap of $-8\ln\beta$ ($0\,<\,\beta=1/g^2\,\ll\,1$). This state is isolated from above and below (going down to the vacuum state!) in the spectrum and shows it determines a mass gap. The needed spectral methods, based on the analysis of correlations and the Bethe-Salpeter operator kernel, are the same developed by Ref. \cite{Sp} and employed e.g. in Refs. \cite{BS,BS2} and in many other cases of strongly coupled lattice QCD. We still have a long way to go to get more substantial progress regarding YM and QCD models.

We give a brief overview of our main results. In Lemma \ref{lema1}, we give a  representation for the Wilson plaquette action and for plaquette fields. This representation is used to prove an upper bound on the action which is quadratic in the gluon fields. In turn, this bound is used to obtain a lower bound on the Wilson partition function in Theorem 1.

In Theorem 1, we obtain a factorized stability bound on the Wilson partition function. The positivity of each plaquette action in the Wilson action plays a role in the proof of the upper stability bound and its factorization. Each factor is a partition function of a single plaquette action with only a single bond variable. The stability bound is good enough for the existence of the thermodynamic limit of the free energy. However, the bound is not uniform in the lattice spacing $a\in(0,1]$. So, it is not enough to show, subsequently, the existence of the continuum $a\searrow 0$ limit. Upper and lower bounds on the single plaquette partition function are obtained in Theorem 2.

We define a scaled partition function and, using the results of Theorems 1 and 2, in Theorem 3, we prove that it satisfies upper and lower stability bound, which are uniform in $a\in(0,1]$ and leads to a scaled free energy which does have a subsequential continuum limit, at least. To obtain TUV stability bound, it is crucial to scale the physical gluon field $A^u_b$, as it appears in the physical parametrization $U_b\,=\,\exp[igaA^u_b]$ and use the scaled gluon field $A_b\,=\,a^{(d-2)/2}\,A^u_b$

Also, we consider the generating function for correlations of $r\,=\,0,1,2,\ldots$, plaquette fields defined in Eqs. (\ref{plaqfield}) and (\ref{scaledplaq}. In Theorem 4, we prove that this generating function is bounded uniformly in the number of lattice sites, $a\in(0,1]$ and $g^2\,\in\,(0,g_0^2<\infty]$. The exponent of the exponential bound is the sum of the square of the $r$ source strengths. The bound is also independent of the location and orientation of the $r$ plaquette fields. Using analyticity in the sources and Cauchy bounds these results imply that the $r$-plaquette correlations are also bounded uniformly in $a\in(0,1]$ and $g^2\,\in\,(0,g_0^2<\infty]$. In particular, the bound is UV regular, meaning the bound holds if some of the positions of the plaquette fields coincide.

The paper is organized as follows. In Section \ref{sec2}, we define the model with the Wilson action for periodic and free b.c. In Section \ref{sec3}, we define and treat an approximate model. In the approximate model, we set to zero, in the Wilson action, plaquette actions corresponding to interior horizontal plaquettes (i.e. plaquettes orthogonal to the time direction), plus some specified plaquettes on the boundary $\partial \Lambda$ of $\Lambda$. Next, by a judicious integration procedure, we carry out
all the remaining gauge bond variable integrations. In each integration, a factor is extracted which is a plaquette partition function depending only on a single bond variable. In this way, we obtain explicit and exact results for the approximate model partition function, free energy and plaquette correlations,   as well as their continuum limits, in subsections \ref{sec3A}, \ref{sec3B} and \ref{sec3C}. For the complete model, TUV stability bounds and bounds for the generating functions for the gauge-invariant plaquette correlations are given in sections \ref{sec4} and \ref{sec5}, as our four main theorems. These theorems are proved in section \ref{sec6}. Section \ref{sec7} is devoted to some concluding remarks. We provide three appendices aiming at giving the reader the opportunity to understand essential steps in our proofs for the YM model in a context of simpler model cases. In Appendix A, we develop the relation between original physical, unscaled quantities and their scaled counterparts for the free field. The use of scaled fields removes the UV divergences of the free energy and correlations, even at coincident points. Comparing the $a\searrow 0$ behavior of the free scalar case with the physical field coincident-point plaquette-plaquette correlation gives us a characterization of ultraviolet asymptotic freedom.

As a bonus, the scaled field free energy and correlations admit absolutely convergent power series in the scaled coupling (hopping) parameter up to and including the critical point. The same holds for the random walk expansion associated with the scaled field free energy and scaled field correlations.

In contrast with the scalar model analyzed in Ref. \cite{AizDC}, in Appendix B, we treat a lattice scalar field model, which is non-Gaussian in perturbation theory in $d=3,4$. We prove that it obeys TUV stability bounds with the exponent of the exponential proportional to the number of lattice sites $\Lambda_s\,=\,L^d$ ({\em not} the volume $(aL)^d$ in $\mathbb R^d$). The proportionality constants are uniform in $L\in\mathbb N$ (even) and $a\in(0,1]$. This bound is sufficient to bound the scaled field free energy and correlations, even at coincident points, uniformly in $a\in(0,1]$.

Finally, we note that we have obtained local factorized bounds of the YM partition function in the text. In Appendix C, we show how local factorized bounds are obtained in the case of Bose fields.
%========================================================================================
%========================================================================================
%========================================================================================
%%%%%%%%%%%%%%%%%%%%%%%%%%%%%%%%%%%%%%%%%%%%%%%%%%%%%%%%%%%%%%%%%%%%
\section{The Wilson YM Model}\lb{sec2}
This section is devoted to the definition of our lattice YM model. 
\begin{rema}We warn the reader that all definitions given above (lattice, sites, plaquettes, partition function,...) are assumed to hold here and below in our text. In order to simplify the notation, sometimes, we will drop the $u$ and $s$ superscripts to differ the same quantity in its unscaled and scaled versions, respectively. Instead, we will assign two different letters for them, making clear in the text which is which. Also, our notation will avoid to write the $a$-dependence explicitly. Finally, as it was observed above, we concentrate on the case of the gauge group $\mathcal G\,=\,\mathrm U(N)$. The case of the gauge group $\mathcal G\,=\,\mathrm{SU}(N) $ is obtained from this one with minor adaptations.\end{rema}

We describe the partition function of the free and periodic b.c. scaled YM models and their gauge invariance properties. The superscript $P$ will denote periodic b.c. quantities.\vspace{3mm}

\noindent {\em Free b.c. Bonds}: Recalling that a lattice site is $x\,=\,(x^0,x^1,\ldots,x^{d-1})$, where $0$ labels the time direction, $e^\mu$, $\mu=0,1,\ldots,(d-1)$ is the unit vector of the $\mu$-th direction and $b_\mu(x)$ is the lattice bond with initial point $x$ and terminal point $x_+^\mu\equiv x + ae^\mu \in\Lambda$, the number of free b.c. bonds in $\Lambda$ is $\Lambda_b=d(L-1)L^{d-1}$. Sometimes, we refer to the bonds in the time direction $x^0$ as {\em vertical} bonds. The other (spatial) bonds are called {\em horizontal}. \vspace{3mm}

\noindent {\em Periodic b.c. Bonds}:  In addition to free b.c. bonds, we have additional or extra bonds in our periodic lattice.  An extra bond has initial point at the extreme right lattice site and terminal point at the extreme left lattice site, in each coordinate direction.  If $\Lambda_e$ denotes the number of extra bonds, we have $\Lambda_e=dL^{d-1}$.  The total number bonds in $\Lambda$ with periodic b.c. (henceforth called {\em periodic bonds}) is $\Lambda_b^P=\Lambda_b+\Lambda_e$.\vspace{3mm}

\noindent {\em Free b.c. Plaquettes} : The free b.c. plaquettes are the plaquettes $p_{\mu\nu}(x)$, in the $\mu\nu$-plane, with $\mu<\nu$. Its vertices are the lattice sites $x$, $x^\mu_+\,\equiv\,x + ae^\mu$, $x^\mu_+\,+\,ae^\nu$, $x^\nu_+\,\equiv\,x + ae^\nu$ of $\Lambda$. \vspace{3mm}

\noindent {\em Periodic b.c. Plaquettes}:  In addition to the   free b.c. lattice plaquettes, there are also extra plaquettes formed with at least with one periodic b.c. bond. The periodic b.c. plaquettes are comprised of all plaquettes that can be formed from the totality of periodic b.c. bonds. We denote the total number of free (periodic) plaquettes by $\Lambda_p$ ($\Lambda_p^P$).   We have, $\Lambda_p=\Lambda_r$, for $d=2$; $\Lambda_p\simeq 3L^3,6L^4$, respectively, for $d=3,4$.  $\Lambda_p^P$ is given by $\Lambda_p$ plus the number of boundary plaquettes.\vspace{3mm}

Recalling that $a\in(0,1]$ and $g^2\in(0,g^2_0)$, $0<g^2_0<\infty$,   and letting $B\,=\, {\rm blank\: or}\: P$, to denote the free and periodic b.c., respectively, we represent the model partition function, with $B$-type b.c., by Eq. (\ref{partition}). Namely, we have
\bequ\lb{partitionB}Z^B_\Lambda(a)\,=\,\dis\int \;\exp\left[-\dfrac{a^{d-4}}{g^2}\,\dis\sum_{p\in\Lambda}\,A_p(U_p)\right]\:d\mu(U)\,\equiv\,\dis\int \;\exp\left[-\dfrac{a^{d-4}}{g^2}\,\dis\sum_{p\in\Lambda}\,2\,\Re\,\tr (1\,-\,U_p)\right]\:d\mu(U)\,,\eequ
with $d\mu(U)\,=\,\prod_{b\in\Lambda}\,d\mu(U_b)$. The Wilson plaquette action $A_p(U_p)$ is defined according Eq. (\ref{AAp}) and is given by  \bequ\lb{WA}A_p(U_p)\,=\,2\,\Re\,\tr\,(1-U_p)\,=\,\|U_p\,-\,1\|^2_{H-S}\,=\, \tr\left(2-U_p-U_p^\dagger\right)\,=\,2\tr(1–\cos X_p)\,.\eequ Here, $\|\,\cdot\,\|_{H-S}$ means the Hilbert-Schmidt norm and the dagger sign denotes the adjoint. The last equality is proved below, before Lemma \ref{lemma0}.  For $p\,=\,p_{\mu\nu}(x)$, the plaquette with vertices as in Eq. (\ref{pmunu}), and bonds $b_1\,=\,[x\,,\,x^\mu_+]$, $b_2\,=\,[x^\mu_+\,,\,x^\mu_+\,+\,ae^\nu]$, $b_3\,=\,[x_+^\nu\,,\,x^\nu_+\,+\,ae^\mu]$ and $b_4\,=\,[x, x^\nu_+]$, we have
\bequ\lb{Upp}
U_p\,=\,U_{b_1}\,U_{b_2}\,U^\dagger_{b_3}\,U^\dagger_{b_4}\,=\,U_{b_1}\,U_{b_2}\,\left[U_{b_4}\,U_{b_3}\right]^\dagger\,=\,e^{iX_p}.
\eequ
The last equality for $A_p(U_p)$ above uses the fact $U_p$ is a unitary variable, as given in Eq. (\ref{Upp}). All the above definitions are independent of the gauge variable parametrization.

If we adopt the physical parametrization of Eq. (\ref{AAp}), i.e. $U_b\,=\,\exp[igaA_b]$, with $A_p\,\rightarrow\,A_p^u$, denoting the {\em unscaled Wilson action}, we obtain
\bequ\lb{part}
Z^{u,B}_{\Lambda,a}\,=\,\int\, \exp\left[-\,\dfrac{a^{d-4}}{g^2}\,A^{u,B}\right]\,dg^B\,.
\eequ
Here, we assigned a gauge bond variable  for each bond $b$ which is a unitary matrix $U\in \mathcal G\,=\,\mathrm U(N)$. The measure $dg^B$ is the product over bonds $b$ of the single-bond $\mathcal G$ Haar measures $d\sigma(U)$. $dg^B$ is expressed in terms of the physical, unscaled gauge potentials $A_b^u$, according to Eq. (\ref{param}). For $p$ denoting any fixed plaquette, the model unscaled action is
\bequ\lb{action}A^{u,B}= \sum_p  \,A_p^u\,,\eequ
where the sum $\sum_p$ is over plaquettes in the lattice $\Lambda$ with the b.c. of type $B$. Obviously, the plaquette actions $A^u_p$, and then the total action $A^{u,B}$, are {\em pointwise positive}. This is an important property which we will use when deriving our bounds below.

To define $A^u_p$, we first recall some important facts about unitary matrices and their representation in terms of elements of the Lie algebra of self-adjoint matrices associated with the gauge group $\mathcal G$.

For an $N\times N$ matrix $M$, recall the Hilbert-Schmidt norm is $\|M\|_{H-S}= [Tr(M^\dagger M)]^{1/2}$.  Let $M_1$ and $M_2$ be $N\times N$ matrices.  Then $(M_1,M_2)\,\equiv\,Tr(M_1^\dagger M_2)$ is a sesquilinear inner product.  We also have the following properties, which are summarized in the next lemma.
\begin{lemma}\lb{lemma0}
\begin{enumerate}
\item Let $X$ be a self-adjoint matrix. Define $\exp(iX)$ by the Taylor series expansion of the exponential.  Then $\exp(iX)$ is unitary.
\item Given a unitary $N\times N$ matrix $U$, by the spectral theorem, there exists a unitary $V$ such that $V^{-1}UV= \mathrm{diag}(e^{i\lambda_1}, \ldots,e^{i\lambda_N})$, $\lambda_j\in(-\pi,\pi]$. The $\lambda_j$ are the angular eigenvalues of $U$.  Define $X=V^{-1}\mathrm{diag}(\lambda_1,\ldots,\lambda_N)V$. Then, $X$ is self-adjoint, $U= \exp(iX)$, and the exponential map is onto (see \cite{Bump}).
\item For $\alpha= 1,2,\ldots,N$, let the self-adjoint $\theta_\alpha$ form a basis for the self-adjoint matrices (the   $\mathrm U(N)$ Lie algebra generators), with the normalization condition $\mathrm{Tr} \theta_\alpha \theta_\beta=\delta_{\alpha\beta}$, with a Kronecker delta. Then, with $X$ being an $N\times N$ self-adjoint matrix, $X$ has the representation $X=\sum_{1\leq \alpha\leq N^2} \;x_\alpha\theta_\alpha$, with $x_\alpha\,=\,\tr X\theta_\alpha$, for $x_\alpha$ real.
\item For $U$ and $X$ related as in item $2$, we have the important equalities:
$$
\| X\|^2_{H-S}\,=\,\mathrm{Tr}\left(X^\dagger X\right)\,=\,\sum_{1\leq\alpha\leq N^2}|x_\alpha|^2=|x|^2\,=\,\sum_{1\leq  j\leq N}\,\lambda_j^2\leq N\pi^2 \quad ,\quad \,\lambda_j\in(-\pi,\pi]\,.
$$
Thus, the exponential map is onto, for $|x|\leq N^{1/2}\pi$.
\item In particular, if $X_b\,=\,agA_b^u$, the equality of the last item takes the form \bequ\lb{item5}\| X\|^2_{H-S}\,=\,\mathrm{Tr}\left(X^\dagger X\right)\,=\,\sum_{1\leq\alpha\leq N^2}|x_\alpha|^2\,=\,\sum_{j=1}^N\; \lambda_j^2\,=\, a^2g^2\;\sum_{c=1}^{\delta_N}\, \left|A_b^{u,c}\right|^2\,=\,\dfrac{g^2}{a^{d-4}}\;\sum_{c=1}^{\delta_N}\, \left|A_b^{c}\right|^2\,.\eequ
where we used the scaling relation given in Eq. (\ref{scalingA}), namely, \bequ\lb{Ascaled}A_b\,=\,a^{(d-2)/2} \,A_b^u\,.\eequ Eq. (\ref{item5}) uses the content of Eq. (\ref{quadrado}).
\end{enumerate}
\end{lemma}

The unscaled gauge field $A^u_b$ has the representation $A^u_b=\sum_{1\leq \alpha\leq \delta_N} A_{b}^{u,\alpha}\, \theta_\alpha$, and we refer to $A_{b}^{u,\alpha}$ as the physical, unscaled color or gauge components of $A^u_b$. For $p=p_{\mu\nu}(x)$, the plaquette variable $U_p$, is given by Eq. (\ref{extendedplaq}), with $A_\mu(x)\,\rightarrow\,A^u_\mu(x)$. Likewise, adopting the physical parametrization for the $U_b$ in $U_p$ and with $U_p= e^{iX_p}$, the positive unscaled plaquette action $A^u_p$ for the plaquette $p$ is defined as in below Eq. (\ref{partition}), by
\bequ\lb{Ap}
A^u_p\,=\,\|U_p-1\|^2_{H-S}\,=\,2\,\Re\tr(1-U_p)\,=\,2\tr(1–\cos X_p)\,=\, \tr\left(2-U_p-U_p^\dagger\right)\,.
\eequ

For concreteness, we give the case of the gauge group $\mathcal G\,=\,\mathrm U(2)$, as an example. Here, $X\,=\,\sum_{\alpha=1,\ldots,4}\,x_\alpha\,\theta_\alpha$, with $\tr \theta_\alpha\theta_\beta\,=\, \delta_{\alpha\beta}$ (with a Kronecker delta), and, for $\sigma_j$, for $j=1,2,3,4$ being the three $2\times 2$ traceless and hermitian Pauli spin matrices $\sigma_1,\,\sigma_2,\,\sigma_3$, and $\sigma_4\,=\,I$ is the $2\times 2$ identity. For $\mathrm U(2)$, we can take $\theta_j=\sigma_j/\sqrt{2}$.

This completes the description of the model.

Using the Baker-Campbell-Hausdorff formula \cite{Far}
for non-commuting operators $X$ and $Y$, 
\bequ\barr{l}\lb{BCH}
e^X\,e^Y=e^Z\quad;\quad
Z=X+Y+{\frac {1}{2}}[X,Y]+{\frac {1}{12}}[X,[X,Y]]-{\frac {1}{12}}[Y,[X,Y]]+\cdots \,,\earr
\eequ
formally, it is shown in Ref. \cite{Gat}, for small lattice spacing $a>0$, that 
$$U_p\,=\,\exp\left[ia^2g\,F^a_{\mu\nu}(x)\,+\,R\right]\quad,\quad  R\,=\,\mathcal O(a^3)\,,$$
where $F_{\mu\nu}^a(x)$ is the finite lattice unscaled nonabelian strength field tensor given in Eq. (\ref{fmunua}). Namely, we have $$F_{\mu\nu}^a(x)\,=\,\partial^a_\mu A^u_\nu(x)\,–\,\partial^a_\nu A^u_\mu(x)\,+\,ig[A^u_\mu(x),A^u_\nu(x)]\,,$$ with finite difference derivatives \bequ\lb{partiala}\partial^a_\mu A^u_\nu(x)\,=\,a^{-1}\,\left[A^u_\nu(x+ae_\mu)\,–\,A^u_\nu(x)\right]\,,\eequ  

Also, it is shown that the unscaled plaquette action satisfies, for small $a$,
$A^u_p\,\simeq\,a^4g^2\,\tr\left[F^a_{\mu\nu}(x)\right]^2$. Each term in $\left[F^a_{\mu\nu}(x)\right]$ is self-adjoint. Hence, the square is a self-adjoint and positive matrix, and its trace is positive.  The quantity $\{\,[a^{d-4}/g^2]\,\sum_p\,A_p\,\}$ is the Riemann sum approximation to the classical smooth field continuum YM action $\{\,\int\,\,\tr(F^a_{\mu\nu})^2(x)\,d^dx\,\}$. When $\Lambda\nearrow a\mathbb Z^d$ and after $a\searrow 0$, formally, and the finite difference derivatives become ordinary partial derivatives.

We now discuss gauge invariance and gauge fixing in detail. We take the global gauge group $\mathcal G_{\mathrm{global}}$ as the product of a $\mathcal G\,=\,\mathrm U(N),\;\mathrm{SU}(N)$ group at each lattice site. Namely, $$\mathcal G_{\mathrm{global}}\,=\,\prod_{x\in\Lambda}\, {\mathcal G}_x\,,$$  where an element of  ${\mathcal G}_x$ is an element of ${\mathcal G}$. It transforms the bond variables $U_{b}\,\equiv\,U_{x,x_\mu^+}$ and its adjoint $U^\dagger_{b}$ to 
\bequ\lb{gt}U_{x,x^+_\mu}\,\rightarrow\,U_x\,U_{x,x^+_\mu}\,U^\dagger_{x^+_\mu}\qquad;\qquad U^\dagger_{x,x^+_\mu}\,\rightarrow\,U_{x^+_\mu}\,U^\dagger_{x,x^+_\mu}\,U^\dagger_{x}\,,\eequ respectively. It is easy to see that $\tr U_p$ is invariant under these transformation. From this, it follows that the plaquette action $A_p(U_p)$ and the total action $\sum_p\,A_p(U_p)$ are also invariant under this {\em local gauge transformation}. 

Due to the local gauge invariance of the action $A^u_p$, and so also $A^{u,B}\,=\,\sum_p\,A^u_p$, there is an excess of gauge variables in the definition of the partition function given of Eq. (\ref{part}). By a gauge fixing procedure \cite{GJ}, we eliminate gauge variables by setting them,  in the action, equal to the identity and performing the trivial gauge bond variable integration. In this process of gauging away some of the gauge group bond variables, the value of the partition function is {\em unchanged}, as long as the gauged away bonds do {\em not} form a closed loop in $\Lambda$ (see \cite{GJ}).

We work with the {\em enhanced temporal gauge}. This gauge will be fixed to prove some of our main results.

In the enhanced temporal gauge, the temporal bond variables in $\Lambda$  are set to the identity, as well as certain specified bond variables on the boundary $\partial\Lambda$ of $\Lambda$. The number $\Lambda_r$ of retained bonds (see Eq. (\ref{lambdar})) is, for free b.c., $\Lambda_r=[(L-1)^2]$, $[(2L+1)(L-1)^2]$, $[(3L^3-L^2-L-1)(L-1)]$, respectively, for
$d=2,3,4$. Clearly, $\Lambda_r\simeq (d-1) L^d$, for sufficiently large $L$, and $\Lambda_r\nearrow\infty$ as $\Lambda\nearrow a \mathbb Z^d$. For periodic b.c., the same bond variables are gauged away; the number of non-gauged away bond variables is then $(\Lambda_r\,+\,\Lambda_e)$, where we recall that $\Lambda_e$ is the number of extra bonds to implement periodic b.c.

The precise definition of gauged away bonds, for free b.c., is as follows (see page 4 of \cite{bQCD} for more details). We label the sites of the $\mu$-th lattice coordinate by $1,2,\ldots,L$. The enhanced temporal gauge is defined by setting in $\Lambda$ the following bond variables to $1$. First, for any $d =2,3,4$, we gauge away all temporal bond variables in $\Lambda$ by setting $g_{b_0(x)}=1$. For $d=2$, take also $g_{b_1(x^0=1,x^1)}=1$. For $d = 3$, set also $g_{b_1(x^0=1,x^1,x^2)}=1$ and $g_{b_2(x^0=1,x^1=1,x^2)}=1$. Similarly, for $d=4$, set also to $1$ all $g_{b_1(x^0=1,x^1,x^2,x^3)}$, $g_{b_2(x^0=1,x^1=1,x^2,x^3)}$ and $g_{b_3(x^0=1,x^1=1,x^2=1,x^3)}$. For $d=2$ the gauged away bond variables form a comb with the teeth along the temporal direction, and the open end at the maximum
value of $x^0$. For $d=3$, the gauged away bonds can be visualized as forming a scrub brush with bristles along the $x^0$ direction and the grip forming a comb. For any $d$, all gauged away bond variables are associated with bonds in the hypercubic lattice $\Lambda$ which form a maximal tree. Hence, by adding any other bond to this set, we form a closed loop.

In the next section, we consider the case of an approximate model and obtain TUV stability bounds. This analysis is intend to allow the reader to get an overall view on how the TUV bounds hold, showing a factorization structure.
%%%%%%%%%%%%%%%%%%%%%%%%%%%%%%%%%%%%%%%%%%%%%%%%%%%%%%%%%%%%%%%%%
%%%%%%%%%%%%%%%%%%%%%%%%%%%%%%%%%%%%%%%%%%%%%%%%%%%%%%%%%%%%%%%%%
\section{TUV Stability and Plaquette Field Correlations for an Approximate Model}\lb{sec3}
%%%%%%%%%%%%%%%%%%%%%%%%%%%%%%%%%%%%%%%%%%%%%%%%%%%%%%%%%%%%%%%%%
%%%%%%%%%%%%%%%%%%%%%%%%%%%%%%%%%%%%%%%%%%%%%%%%%%%%%%%%%%%%%%%%%
With the definitions given in section \ref{sec2} for our YM model in mind, in this section, we restrict our attention to a simplified lattice YM model. Of particular interest are the contents of the basic Eqs. (\ref{partitionB}), (\ref{WA}),  (\ref{Upp}), (\ref{part}), (\ref{action}) and (\ref{Ap}), which will be used repeatedly here, as well as in the next sections. 

Before attacking the complete nonabelian mode, in order to allow the reader to understand better the main points of our proofs, we consider a {\em simplified} YM model, with free b.c. In the Wilson action, we set to zero the actions   corresponding to internal horizontal plaquettes (i.e., those plaquettes which are orthogonal to the time direction), plus certain specified plaquettes on the boundary   $\partial \Lambda$ of the lattice $\Lambda$. We refer to this model as the {\em  approximate model}.

For the approximate model, the free energy, plaquette field  [check Eq. (\ref{plaqfield})] correlations and their thermodynamic limits,   as well as their continuum limits, are obtained explicitly and exactly. The bounds obeyed in the approximate model are a good guide for the model without approximation.

In subsection \ref{sec3A}, we define the approximate model and treat stability. In subsection \ref{sec3B}, we obtain plaquette field correlations considering the gauge group $\mathrm U(1)$. The plaquette field correlation results are extended to $\mathrm U(N\geq 2)$ in subsection \ref{sec3C}.

The complete, non-approximate model is treated in the ensuing sections. For $d=2$, the results obtained for the complete model and the approximate model coincide.

The physical gauge-invariant plaquette field plaquette-plaquette correlation is most singular for coincident points. The ultraviolet limit $a\searrow 0$ singular behavior is $({\rm const}/a^d)$. The same behavior occurs for the coincident-point derivative field correlations in the case of the real, massless scalar free field, as shown in Appendix A.

Of course, the abelian $\mathcal G\,=\,\mathrm U(1)$ case and, for the model without approximation, the formal $g\searrow 0$ limit gives us the lattice free electromagnetic field with a quadratic action. (See Remark \ref{rema2} in subsection \ref{sec3A}). The plaquette-plaquette field correlations can be obtained exactly in a momentum space representation and the coincident point plaquette-plaquette field correlation is equal to $\{4/[d(d-1)a^d]\}$.

Using our scaled field method, we obtain TUV stability bounds and bounds on the scaled free energy and also the boundedness of two-point plaquette scaled field correlation. For the abelian gauge group $\mathcal G={\mathrm U}(1)$, the Haar measure is simpler, formulas are more familiar and the analysis becomes more transparent.

For $d=2$, the results for the two-point plaquette field correlation are exact. For $d=3,4$, the results are also exact for the approximate model. This seemingly gross approximation gives the correct picture for bounds for the complete YM model with the nonabelian gauge group $\mathcal G={\mathrm U}(N>1)$.
%%%%%%%%%%%%%%%%%%%%%%%%%%%%%%%%%%%%%%%%%%%%%%%%%%%%%%%%%%%%%%%%%%%%
%%%%%%%%%%%%%%%%%%%%%%%%%%%%%%%%%%%%%%%%%%%%%%%%%%%%%%%%%%%%%%%%%%%%
\subsection {Approximate Model: TUV Stability}\lb{sec3A}
%%%%%%%%%%%%%%%%%%%%%%%%%%%%%%%%%%%%%%%%%%%%%%%%%%%%%%%%%%%%%%%%%%%%
%%%%%%%%%%%%%%%%%%%%%%%%%%%%%%%%%%%%%%%%%%%%%%%%%%%%%%%%%%%%%%%%%%%%
Using the physical parametrization and starting from the free b.c. partition function of Eq. (\ref{part}), we set
$$
U_b\,=e^{i\theta_b}\,=\,e^{iagA^u_b}\,,
$$
where we recall $A^u_b$ is the physical, unscaled gauge potential. For the plaquette $p\,=\,p_{\mu\nu}(x)$, set   
$$\theta_{\mu\nu}(x)\,=\,\theta_\mu(x)\,+\,\theta_\nu(x_\mu^+)\--\,\theta_\nu(x)\,-\,\theta_\mu(x_\nu^+)\,,$$
where $\theta_\mu(x)\equiv \theta_{b=b_\mu(x)}$.

Then, the finite lattice free b.c. unscaled partition function reads
$$
Z^u_{\Lambda}(a)\,=\,\dis\int_{|\theta_b|\leq\pi}\,\exp\left[-\,\dfrac{a^{d-4}}{g^2}\,\sum_{x,\mu<\nu}\,2\left[1\,-\,\cos\left(\theta_{\mu\nu}(x)\right) \right] \right]\;\prod_{b\in\Lambda}\,\dfrac{d\theta_b}{2\pi}\,.
$$

In terms of the fields $A^u_b$, setting $A_{\mu\nu}(x)\,\equiv\,a\,F^a_{\mu\nu}(x)$, where $F^a_{\mu\nu}(x)$, given in Eq. (\ref{fmunua}), is the usual field strength antisymmetric second order tensor, defined with finite difference derivatives. We have
$$
Z^u_{\Lambda}(a)\,=\,\left(\dfrac{ag}{2\pi}\right)^{\Lambda_r}\;\dis\int_{|A_b|\leq\pi/(ag)}\,\exp\left\{-\,\dfrac{a^{d-4}}{g^2}\,\sum_{x,\mu<\nu}\,2\left[1\,-\,\cos\left(a^2gF^a_{\mu\nu}(x) \right)\right] \right\}\;\prod_b\,dA_b\,,
$$

Now, we transform to the locally scaled fields $\chi_b$ defined by
\bequ\lb{chi}
\chi_b\,=\,a^{(d-2)/2}\,A^u_b\,.
\eequ
In terms of these fields, the free b.c. unscaled partition function is
$$
Z^u_{\Lambda}(a)\,=\,\left(\dfrac{g}{2\pi}\,a^{(4-d)/2}\right)^{\Lambda_r}\;\dis\int_{|\chi_b|\leq(\pi/g)\,a^{(d-4)/2}}\,\exp\left\{-\,\dfrac{a^{d-4}}{g^2}\,\sum_{x,\mu<\nu}\,2\left[1\,-\,\cos\left(ga^{(4-d)/2}\chi_{\mu\nu}(x) \right)\right] \right\}\;\prod_b\,d\chi_b\,,
$$
\begin{rema}\lb{rema1}
We note that, instead of the above simple expression, in the nonabelian case $\mathrm U(N>1)$, the Haar measure presents also a weight function factor besides the   product of Lebesgue measures \cite{Nelson,Holland,Marinov,Euler}.
\end{rema}
\begin{rema}\lb{rema2}
	In the $A_b$ variables, the Boltzmann factor, for $a\searrow 0$, is approximately $$
	\exp\left\{-\,a^{d}\,\sum_{x,\mu<\nu}\,\left[F^a_{\mu\nu}(x)\right]^2 \right\}\,,
	$$
for $d=2,3$, and for $d=4$ and $g\searrow 0$. In both cases, the action approximates the continuum model action.

In the $ \chi_b$ variables, the Boltzmann factor, for $a\searrow 0$, is approximately 
$$
\exp\left\{-\,\sum_{x,\mu<\nu}\,\left[ \chi_{\mu\nu}(x)\right]^2 \right\}\,,
$$ for $d=2,3$ and the same holds for $d=4$ and $g\searrow 0$. Here, in both cases, the action is independent of the lattice spacing $a$. In the above quadratic approximation of the action, the model can be solved explicitly by diagonalizing the corresponding quadratic form.
\end{rema}

We now define more precisely and analyze our approximate model. We also outline the bond gauge integration procedure. This is done for each value of the spacetime dimension $d=2,3,4$. For simplicity, we identify coordinates of a lattice site in each lattice direction, $\mu=0,1,2,\ldots,(d-1)$, with the labels $1,2,\ldots,L$. We have:
\begin{itemize}
	\item $d=4$: For $x^0\,=\,L, L-1,\ldots,2$, set the plaquette actions to zero in the planes parallel to the $\mu\nu\,=\,12,13,23$ coordinate planes. For $x^0\,=\,1$, $x^3\,=\,L,\ldots,2$, set the plaquette actions to zero in the coordinate planes parallel to the $12$-plane;
	\item $d=3$: For $x^0\,=\,L,L-1,\ldots,2$, set to zero the plaquette actions in the planes parallel to the $12$-plane;
	\item $d=2$: maintain all the plaquette actions.
\end{itemize}
\begin{rema} \lb{rema3new}
We remark that a simpler approximate model can be defined by setting to zero {\em all} horizontal plaquette actions. Such a model can also be solved exactly and the same results given here, for our approximate model, also hold. Boundary effects disappear in the thermodynamic limit. In our approximate model, fewer plaquette actions are discarded.
\end{rema}

\noindent{\em Simplified, Approximate Model for the Abelian Gauge Group $\mathcal G=\mathrm U(1)$}:\vspace{1mm}

With these definitions, we now perform the bond integration. For ease of visualization, we carry it out explicitly for $d=3$.

For $d=3$, integrate over successive planes of horizontal bonds starting at the coordinate $x^0\,=\,L$ and ending at $x^0\,=\,2$. For the $x^0\,=\,1$ horizontal plane, integrate over successive lines of horizontal bonds in the coordinate direction two, starting at $x^1\,=\,L$ and ending at $x^1\,=\,2$. For each horizontal bond variable integration, the bond variable appears in only one plaquette.

The simplification that occurs in our original model is that, in the approximate model, we can carry out all bond integrations. Besides, for each integration, we can extract a single plaquette partition function of a single bond variable.

We emphasize that, for $d\,=\,2$, the model was solved without any approximation in Ref. \cite{Ash}.

After integration, each integral depends, in principle, on the other bond variables of the plaquette which are present in the plaquette variable $U_p$ [see Eq. (\ref{extendedplaq})]. However, as in Ref. \cite{Ash}, for $d=2$, by a change of variables, the integral is independent of the other variables and their integrals are trivially done. Here, we are using the simplest case of the {\em left and right invariance of the gauge group Haar measure} (see e.g Refs. \cite{Simon2,Far,Bump} and Eq. (\ref{lrinvariance}) below). In this way, a factor is extracted from the partition function which corresponds to the scaled partition function of a single plaquette of a single bond variable.

After the bond integration procedure is completed, we obtain
\bequ\lb{Zz}
Z^u_{\Lambda}(a)\,=\,\left[ \dfrac{g}{2\pi\,a^{(d-4)/2}} \right]^{\Lambda_r}\;z^{\Lambda_r}\,,
\eequ
where $z$ is the scaled single bond partition function. Namely, we have
$$
z\,=\,\dis\int_{|X|\leq(\pi/g)\,a^{(d-4)/2}}\,\exp\left\{-\,\dfrac{a^{d-4}}{g^2}\,\sum_{x,\mu<\nu}\,2\left[1\,-\,\cos\left(ga^{(4-d)/2}\,X \right)\right] \right\}\,dX\,.
$$

Using the following elementary trigonometric inequalities in the above integrand (see e.g. Ref. \cite{Simon3} for a proof of the second one)
\bequ\lb{trig}\barr{lll}
1\,-\,\cos u&\leq&u^2/2\quad,\quad u\in\mathbb R\vspace{1.2mm}\\
1\,-\,\cos u&\geq&\dfrac{2u^2}{\pi^2}\quad,\quad u\in(-\pi,\pi]\,,
\earr\eequ
we obtain the   upper and lower bounds 
\bequ\lb{appupper}
z\,\leq\,\dis\int_{|X|\leq(\pi/g)\,a^{(d-4)/2}}\,\exp\left[-\,\dfrac{4}{\pi^2}\,X^2\right]\,dX\,\equiv z_u\,,
\eequ
and
\bequ\lb{applower}
z\,\geq\,\dis\int_{|X|\leq(\pi/g)\,a^{(d-4)/2}}\,e^{-X^2}\,dX\,\geq\,\dis\int_{|X|\leq(\pi/g_0)}\,e^{-X^2}\,dX\,\equiv\,\tilde z_\ell\,>\,0\,,
\eequ 
for all $a\in(0,1]$ and $0\,<\,g^2\,\leq\,g^2_0\,<\,\infty$.

We now define the scaled free b.c. partition   function $Z^s_{\Lambda}(a)$, by extracting the $a\searrow 0$ singularity in Eq. (\ref{Zz}). It reads
\bequ\lb{Znorm}
Z^s_{\Lambda}(a)\,=\,\left[ \dfrac{g}{2\pi\,a^{(d-4)/2}} \right]^{-\Lambda_r}\,Z^u_{\Lambda}(a)\,=\,z^{\Lambda_r}\,.
\eequ

In this way, in terms of $Z^s_{\Lambda}(a)$, we obtain the TUV stability bound
$$
0\,<\,\tilde z_\ell^{\Lambda_r}\, \leq\,Z^s_{\Lambda}(a)\,\leq z_u^{\Lambda_r}\,,
$$
so that, defining the scaled free energy per effective degree of freedom in the finite $d$-dimensional hypercubic lattice $\Lambda$ by 
\bequ\lb{fn}
f^s_{\Lambda}(a)\,\equiv\,\dfrac1{\Lambda_r}\,\ln Z^s_{\Lambda}(a)\,=\,\ln z\,,
\eequ
the TUV bounds ensure, the thermodynamic limit $\Lambda\nearrow a\mathbb Z^d$ and the continuum limit $a\searrow 0$ exist (here, not only the subsequential limits as below!) and we obtain  
$$\barr{lll}
f^s&\equiv&\lim_{a\searrow 0}\,\lim_{\Lambda\nearrow a\mathbb Z^d}\,f^s_{\Lambda}(a)\vspace{1.2mm}\\
&=&\lim_{a\searrow 0} \ln z\vspace{2mm}\\
&=&\left\{\barr{l}\dis\int_{\mathbb R}\, e^{-X^2}\,dX\,=\,\sqrt{\pi/2}\:,\:d=2,3, \vspace{3.2mm}\\ \dis\int_{|X|\leq \pi/g}\, e^{-2g^{-2}[1-\cos(gX)]}\,dX\:,\:d=4\,.\earr\right.\earr
$$
Besides, for $d=4$, we have $\lim_{g\searrow 0}f^n\,=\,\sqrt{\pi/2}$.\vspace{6mm}

\noindent{\em Simplified Approximate Model with Gauge Group $\mathcal G=\mathrm U(N)$}: \vspace{1mm}

  Still considering the approximate model, here we extend our TUV bounds to the more general nonabelian $\mathcal G\,=\,\mathrm U(N)$ case. Using the same bond integration procedure as in the above $\mathrm U(1)$ case, the simplified model, free b.c. unscaled partition function with the gauge group ${\mathrm U}(N)$ also factorizes as
$$
Z^u_{\Lambda,a}\,=\,z^{\Lambda_r}\,,
$$
where
\bequ\lb{zz}
z\,=\,\dis\int_{{\mathrm U}(N)}\;\exp\left[-\,\dfrac {a^{d-4}}{g^2}\,\tr \left(2-U-U^\dagger\right)\right]\;d\sigma(U)\,.
\eequ
  Here, $z$ is the partition function of a single plaquette with the single bond variable $U$.

We explain how the factorization occurs, and we use the left and right invariance of the single bond Haar measure $d\sigma(U)$. We recall the invariance property (see e.g. \cite{Bump,Simon2,Far}): let $f(U)$ be a function of the bond variable $U\in{\mathrm U}(N)$ and let $W\in{\mathrm U}(N)$. Then,
\bequ\lb{lrinvariance}
\int_{{\mathrm U}(N)}\,f(U)\,d\sigma(U)\,=\, \int_{{\mathrm U}(N)}\,f(WU)\,d\sigma(U)\,=\, \int_{{\mathrm U}(N)}\,f(UW)\,d\sigma(U)\,. 
\eequ

Returning to the bond integration procedure, let $U_1,U_2,U_3,U_4$ be the plaquette $p$ bond variables and $U_p\,=\,U_1U_2U_3U_4$, as in Eq. (\ref{extendedplaq}). Consider the integration over $U_1$, where, in the partition function $Z^u_{\Lambda}(a)$,  $U_1$ only appears in the plaquette $p$. The integral over the bond variable $U_1$ is
$$
\int_{{\mathrm U}(N)}\,\exp\left\{-\dfrac{a^{(d-4)}}{g^2}\,\tr\left( 2-U_p-U^\dagger_p\right)\right\}\,d\sigma(U_1)\,.
$$
By the Haar measure left and right invariance (take $W\,=\,U_2U_3U_4$ and $U=U_1$ above!), the integral is just the single bond partition function $z$, and is independent of the other bond variables. In this way, we extract the factors $z$ from the partition function $Z^u_{\Lambda}(a)$.

To continue our analysis, we note that the above integrand is a class function on $\mathcal G$. For the ${\mathrm U}(N)$ group integral of a class function, the $N^2$  dimensional integral over the $n\times n$ matrix group reduces to an $N$ dimensional integral over the angular eigenvalues of $U$, according to the Weyl integration formula \cite{Weyl,Bump,Simon2,Far}. 

The angular eigenvalues are defined as follows. If the eigenvalues of the unitary matrix $U$ are denoted by $\{e^{i\lambda_1},\ldots,e^{i\lambda_N}\}$, with $\lambda_j\in(-\pi,\pi]$, $j=1,\ldots,N$, then $\lambda\,\equiv\,(\la_1,\ldots,\lambda_N)$ are called the angular eigenvalues of $U$. The Weyl integration formula reads
\bequ\lb{weyl1}
\dis\int_{{\mathrm U}(N)}\;f(U)\;d\sigma(U)\,=\,\dfrac1{N!}\,\dis\int_{(-\pi,\pi]^N}\;f(\lambda)\,\rho(\lambda)\,\dfrac {d\lambda}{(2\pi)^N}\,,
\eequ
where $d\lambda=d\lambda_1\ldots d\lambda_N$ is a product measure of Lebesgue measures, and the weight function or density $\rho(\lambda)$ arises from a squared Vandermonde determinant. It is given by
\bequ\lb{rho}
\rho(\lambda)\,=\,\prod_{1\leq j<k\leq N}\,\left|e^{i\lambda_j}-e^{i\lambda_k}\right|^2\,=\,\prod_{1\leq j<k\leq N}\,\left\{2\left[1\,-\,\cos\left(\lambda_j-\lambda_k \right) \right]\right\}\,.
\eequ

In this way, applying the Weyl integration formula to $z$ of Eq. (\ref{zz}), we obtain
$$
z\,=\,\dfrac1{N!(2\pi)^N}\,\dis\int_{(-\pi,\pi]^N}\;\exp\left[-\,\dfrac {2a^{d-4}}{g^2}\,\sum_{j=1,\ldots,N}\,\left(1\,-\,\cos \lambda_j\right)\right]\;\rho(\lambda)\,d\lambda\,.
$$

Next, we use Eq. (\ref{trig}) to give bounds on $(1-\cos u)$ and the density bound
$$
\left(\dfrac4{\pi^2} \right)^{N(N-1)/2}\,\hat\rho(\lambda)\,\leq\,\rho(\lambda)\,\leq\,\hat\rho(\lambda)\,,
$$
where  $\hat\rho(\lambda)\,=\,\prod_{1\leq j<k\leq N}\,\left|{\lambda_j}-{\lambda_k}\right|^2$. The lower bound holds for all $|\lambda_j|\,\leq\pi/2$ and there is no restriction for the upper bound. Besides, we make use of the changes of variables, with $y\,=\,(y_1,\ldots,y_N)$,
$$
y\,=\,\left(\dfrac{a^{d-4}}{g^2} \right)^{1/2} \, \lambda\qquad;\qquad y\,=\,\left(\dfrac{a^{d-4}}{g^2} \right)^{1/2} \,\dfrac 2\pi\,\lambda\,,
$$
respectively, in the lower and upper bounds. Doing this, we obtain the following bound on $z$
$$\barr{l}
\left(\dfrac2{\pi}\right)^{N(N-1)}\;\dfrac1{N!(2\pi)^N}\left(\dfrac{g^2}{a^{d-4}}\right)^{N^2/2}\,\dis\int_{\mathcal L}\exp\left[-\,
%\dfrac {2a^{d-4}}{g^2}
\sum_{1\leq j\leq N}\,y_j^2\right]\hat\rho(y)dy\vspace{3mm}\\\qquad\quad \leq\,z\leq\,\left(\dfrac{\pi}2\right)^{N^2}\;\dfrac1{N!(2\pi)^N}\;\left(\dfrac{g^2}{a^{d-4}}\right)^{N^2/2}\,\dis\int_{\mathcal U}\exp\left[-\,
%\dfrac {2a^{d-4}}{g^2}\,
\sum_{1\leq j\leq N}\,y_j^2\right]\,\hat\rho(y)dy\,.\earr
$$
where we have the integration domains $\mathcal L\,=\,\{y\,:\,{|y_k|\leq(\pi/2)\,(a^{d-4}/g^2)^{1/2}}\}$ and $\mathcal U\,=\,\{y\,:\,{|y_k|\leq\,2\,(a^{d-4}/g^2)^{1/2}}\}$. We easily recognize the above integrands as being proportional to the well known (see e.g. Refs. \cite{Metha,Deift}) Gaussian Unitary Ensemble (GUE) probability density in $\mathbb R^N$ of random matrix theory.

Extracting the $a\searrow 0$ singularity and defining the normalized $\mathrm U(N)$ approximate model finite lattice scaled partition with free b.c. by
\bequ\lb{Zappscaled}
Z^s_{\Lambda}(a)\,=\,\left(\dfrac{a^{d-4}}{g^2} \right)^{N^2\Lambda_r/2}\,Z^u_{\Lambda,a}\,,
\eequ
then $Z^s_{\Lambda,a}$ obeys the TUV bound
\bequ\lb{ZappTUV}
z_\ell^{\Lambda_r}
\,\leq\,Z^s_{\Lambda}(a)\,=\,z_s^{\Lambda_r}\,\leq z_u^{\Lambda_r}\,,\eequ
with
\bequ\lb{zs}
z_s\,=\,\dfrac1{N!(2\pi)^N}\,\left(\dfrac{a^{d-4}}{g^2}\right)^{N^2/2}\,\dis\int_{(-\pi,\pi]^N}\;\exp\left[-\,\dfrac {2a^{d-4}}{g^2}\,\sum_{j=1,\ldots,N}\,\left(1\,-\,\cos \lambda_j\right)\right]\;\rho(\lambda)\,d\lambda\,.
\eequ
In Eq. (\ref{ZappTUV}), we have $$\barr{l}z_\ell\,=\:(\frac2{\pi})^{N(N-1)}\;G\left((a^{d-4}/g^2)^{1/2}\,\pi/2\right)\vspace{2mm}\\ z_u\,=\:(\frac{\pi}2)^{N^2}\; G\left((a^{d-4}/g^2)^{1/2}\,2\right)\,,\earr$$ where, up to a normalization (see \cite{Metha,bQCD}), $G$ is the probability in the GUE given by
$$
G(u)\,=\,\dfrac1{N!\,(2\pi)^N}\,\dis\int_{|y_k|<u}\,\exp\left[-\,\sum_{1\leq j\leq N}\,y_j^2\right]\hat\rho(\lambda)d\lambda\,\leq\,G(\infty)\,.
$$

We now define a scaled finite lattice free energy by 
$$
f^s_{\Lambda}(a)\,=\,\dfrac1{\Lambda_r}\,\ln Z^s_{\Lambda}(a)\,.
$$
Hence, the above TUV bounds ensure the existence of the thermodynamic and continuum limits of the scaled free energy given by, with $G(\infty)\,\equiv\,\lim_{u\nearrow\infty}\,G(u)$,
$$
f^s\,=\,\lim_{a\searrow 0}\,\lim_{\Lambda\nearrow a\mathbb Z^d}\;f^s_{\Lambda}(a)\,=\,\left\{\barr{l}\ln G(\infty)\,,\quad d=2,3\vspace{2mm}\\
\ln \left\{
\dfrac1{N!(2\pi)^N}\,\dis\int_{|y_k|\leq \pi/g}\,\exp\left[-\,2g^{-2}\,\sum_{1\leq j\leq N}\,\left(1-\cos (gy_j)\right)\right]\,\hat\rho(y)\,dy\right\},\quad d=4\,,\earr   \right.
$$
at least in the subsequential sense. Furthermore, for $d=4$, we get $\lim_{g\searrow 0}\,f^s\,=\,\ln G(\infty)$.
%%%%%%%%%%%%%%%%%%%%%%%%%%%%%%%%%%%%%%%%%%%%%%%%%%%%%%%%%%%%%%%%%%%%
%%%%%%%%%%%%%%%%%%%%%%%%%%%%%%%%%%%%%%%%%%%%%%%%%%%%%%%%%%%%%%%%%%%%
\subsection{Approximate Model: Plaquette Field Correlations for ${\mathrm U}(1)$}
%%%%%%%%%%%%%%%%%%%%%%%%%%%%%%%%%%%%%%%%%%%%%%%%%%%%%%%%%%%%%%%%%%%%
%%%%%%%%%%%%%%%%%%%%%%%%%%%%%%%%%%%%%%%%%%%%%%%%%%%%%%%%%%%%%%%%%%%%
\lb{sec3B}
Here, first we take the gauge group to be   ${{\mathcal G}=\mathrm U}(1)$. As shown below, in this simple abelian group case, we are able to compute the plaquette-plaquette correlation exactly for the approximate model and for vertical plaquettes (plaquettes with two vertical bonds).   This computation allows us to show the boundedness of the scaled field plaquette-plaquette   correlation. In the next subsection, we consider the nonabelian gauge groups ${\rm U}(N\geq 2)$.

For the plaquette $p=p_{\mu\nu}(x)$, we define the physical unscaled gauge-invariant plaquette field as in Eq. (\ref{plaqfield}), using the unscaled fields $A^u_b$. In the abelian case, with the physical parametrization $U_b\,=\,e^{iagA^u_b}$, it reduces to
\bequ\lb{upfield}
{\cal F}^u_p(U_p)\,=\,\dfrac 1{a^2g}\,\sin\left[a^2gF^{a}_{\mu\nu} (x)\right]\,
\eequ
where $A^u_p\,=\,a\,F^{a}_{\mu\nu}$, for the abelian version of the field tensor $F^a_{\mu\nu}$ given in Eq. (\ref{fmunua}), i.e. without the commutator term.

Next, considering a sufficiently small lattice spacing $a$, we show this plaquette field leads to the expected physical correlation, i.e.
$$
{\cal F}^u_p(U_p)\,\simeq\,F^{a}_{\mu\nu}\,=\,\partial^a_\mu A^u_\nu\,-\,\partial^a_\nu A^u_\mu\quad,\quad 0\,<\,a\,\ll\,1\,.
$$
Then, the gauge-invariant unscaled plaquette-plaquette correlation is defined by
\bequ\lb{2plaqU1}
\barr{lll}\!\langle {\cal F}^u_{\mu\nu}(x){\cal F}^u_{\rho\sigma}(y)\rangle\!\!&=&\!\!
\dfrac1{\cal N}\,\dis\int_{|A^u_b|<\pi/(ag)}\,\!\left\{\left[\dfrac 1{a^2g}\, \,\!\!\sin\left(a^2\,g\,F^{a}_{\mu\nu}(x)\right)\right]\,\left[\dfrac1{a^2g}\, \sin\left(a^2g\,F^{a}_{\rho\sigma}(y)\right)\right]\right\}\vspace{2mm}\\
&&\!\!\times\;\exp\left\{ -\,\dfrac{a^{d-4}}{g^2}\;\sum_{z,\mu<\nu}\,2\left[1\,-\,\cos \left(a^2\,g\, F^{a}_{\mu\nu}(z) \right)\right]\right\}\;\prod_b dA^u_b\,.
\earr\eequ
As seen above, we emphasize that in the abelian case we can deal easily with the $\mathcal G$ Haar measure and express it in terms of the unscaled fields $A^u_b$.

For small $a$, the right-hand-side of Eq. (\ref{2plaqU1}) becomes
$$
\dfrac1{\cal N}\,\dis\int_{|A^u_b|\,\leq\,(\pi/ag)}\,F^{a}_{\mu\nu}(x)\,F^{a}_{\rho\sigma}(y)\;\exp\left\{ -\,a^d\;\sum_{z,\mu<\nu}\,\left[ F^{a}_{\mu\nu}(z)\right]^2\right\}\;\prod_b dA^u_b\,.
$$
Note that the above action is the Riemann sum approximation to the smooth field classical continuum action $\sum_{\mu<\nu}\,\int_{[-La,La]^d}\,d^dx\,\left[F^a_{\mu\nu}(x)) \right]^2$, where the field strength antisymmetric tensor in the abelian case is $F^a_{\mu\nu}(x)\,=\,\partial_\mu A^u_\nu(x)\,-\,\partial_\nu A^u_\mu(x)$. Hence, we obtain the lattice approximation to the unscaled plaquette-plaquette correlation.

Now, for $a\in(0,1]$, we define a  ${\rm U}(1)$ gauge-invariant scaled plaquette-plaquette correlation by
\bequ\lb{2plaque}
\barr{lll}\langle {\cal F}^s_{\mu\nu}(x)\,{\cal F}^s_{\rho\sigma}(y)\rangle&=&\dfrac1{\cal N}\,\dis\int_{|A^u_b|<\pi/(ag)}\,\left[\left(\dfrac{a^{d-4}}{g^2} \right)^{1/2}\,\sin[a^2gF_{\mu\nu}^{a}(x)]\right]\;\left[\left(\dfrac{a^{d-4}}{g^2} \right)^{1/2}\,\sin[a^2gF_{\rho\sigma}^{a}(y)]\right]\vspace{2mm}\\&&\times\;\exp\left\{ -\,\dfrac{a^{d-4}}{g^2}\;\sum_{z,\mu<\nu}\,2\left(1\,-\,\cos (a^2gF_{\mu\nu}^{a}(z)) \right)\right\}\;\prod_b dA_b^u\vspace{2mm}\\&=&a^d\;\langle {\cal F}^u_{\mu\nu}(x) {\cal F}^u_{\rho\sigma}(y) \rangle\,,
\earr\eequ
where we observe the change of coefficients on the sine terms, as compared with Eq. (\ref{2plaqU1}), and where ${\cal N}$ is the normalization constant
$${\cal N}\,\equiv\,\int_{|A^u_b|<\pi/(ag)} \;\exp\left\{ -\,\dfrac{a^{d-4}}{g^2}\;\sum_{x,\mu<\nu}\,2\left(1\,-\,\cos (a^2gF_{\mu\nu}^{a}(x)) \right)\right\}\;\prod_b dA^u_b\,.$$ 

Using the scaled gauge field $ \chi_b\,=\, a^{(d-2)/2}\,A^u_b$ (see Eq. (\ref{chi})), we can rewrite the plaquette-plaquette correlation as
$$\lb{2plaque2}
\barr{lll}\langle {\cal F}^s_{\mu\nu}(x){\cal F}^s_{\rho\sigma}(y)\rangle&=&\dfrac1{\cal N^\prime}\,\dis\int_{| \chi_b|<\pi\left(a^{d-4}/g^2\right)^{1/2}}\,\dfrac{a^{d-4}}{g^2}\,\sin\left[\left(\dfrac{a^{d-4}}{g^2}\right)^{-1/2}\, \chi_{\mu\nu}(x)]\right]\;\,\sin\left[\left(\dfrac{a^{d-4}}{g^2}\right)^{-1/2}\, \chi_{\rho\sigma}(y)\right]\vspace{2mm}\\&&\times\;\exp\left\{-\,\dfrac{a^{d-4}}{g^2}\;\sum_{z,\mu<\nu}\,2\left(1\,-\,\cos \left(\dfrac{a^{d-4}}{g^2}\right)^{-1/2}\, \chi_{\mu\nu}(z) \right)\right\}\;\prod_b d \chi_b\,,
\earr$$
where $\cal N^\prime$ is the measure normalization.

Now, for the approximate model, we compute the plaquette-plaquette correlation exactly. We also show that its thermodynamic limit exists and that the correlation of Eq. (\ref{2plaque}) is bounded uniformly in $a\in(0,1]$. The continuum limit of $\langle {\cal F}^s_{\mu\nu}(x){\cal F}^s_{\rho\sigma}(y)\rangle$ also exists! [In the next subsection, we extend these results to the case of the nonabelian gauge group ${\rm U}(N)$, $N\geq 2$.]

More precisely,   for the approximate model, we will show that $\langle {\cal F}^s_{\mu\nu}(x){\cal F}^s_{\rho\sigma}(x)\rangle$ is bounded uniformly in $a\in(0,1]$ and $0<g^2\leq g_0^2<\infty$. The importance of this result is that it shows us that the coincident point ($x=y$) physical plaquette-plaquette correlation behaves as ${\rm const}/a^d$.

The $a^{-d}$ behavior is analogous to what occurs if we transform the physical massless scalar field $\phi^u(x)$, by a local scaling factor,  to a scaled field $\phi(x)\,=\,a^{(d-2)/2}\,(2d)^{1/2}\,\phi^u(x)$. The scaled field action is independent of the lattice spacing $a$ (see  Ref. \cite{M} and Appendix A for more details.) Moreover, the scaled   field correlations are bounded  at coincident points, uniformly in $a\in(0,1]$, for $d=3,4$, and the unscaled derivative field two-point correlation has the {\em exact} value $2/(da^d)$, for dimensions $d=2,3,4$.

In order to simplify the notation, like in Eq. (\ref{2plaque}), below ${\cal N}$ will mean the average of the constant which is identically 1, with the relevant measure appearing in the corresponding average integral, including the exponential density factor.

For the complete model   with gauge group $\mathcal G\,=\,\mathrm U(N)$, the integrals do {\em not} factorize, but for the approximate model they do factorize, which makes much easier the analysis of the plaquette-plaquette correlation. For this reason, from now on this subsection, we deal with only the approximate model. Note that we also take the two plaquettes, containing the external points, to be vertical (i.e. with at least one bond in the time direction).

To analyze the plaquette-plaquette correlation for the approximate model, we follow the same integration procedure employed before in our treatment of the partition function (see subsection \ref{sec3A}). The result is that {\em all} gauge integrals with densities given by the exponential of the actions factorize, provided they do not contain the external points $x$ and $y$. As before, the factorized terms correspond to single plaquette partition functions depending only on a single bond variable. They are present both in the numerator and the normalization integrals in the denominator in $\langle {\cal F}^s_{\mu\nu}(x){\cal F}^s_{\rho\sigma}(y)\rangle$. Thus, they do cancel out.

After this partial cancellation, we are left in the numerator with integrals whose coordinate supports contain the $x$ and $y$ external points. However, since the single plaquette field correlation is zero by the $A\rightarrow -A$ symmetry, the only nonzero contribution occurs when the points $x$ and $y$ coincide. 

For coincident points $x=y$, the contributions depend on a single bond variable $ \chi_b(x)$. Taking into account the partial cancellation between the numerator and denominator of the normalized plaquette-plaquette correlation, the infinite volume limit can then be taken. By translation invariance, the remaining integral does not depend on the lattice site point $x=y$ we fixed.   Thus, we can suppress $x$ and the bond lower index $b$ in $ \chi_b(x)$ and simply write $\chi$. Doing this, we obtain
\bequ\lb{coinMsquare}
\barr{llr}\!\langle \,[{\cal F}^s_{\mu\nu}(x)]^2\,\rangle&=&\dfrac1{\cal N}\,\dis\int_{| \chi|<(\pi/g)a^{(d-4)/2}}\,\left\{\left(\dfrac{a^{d-4}}{g^2} \right)\,\!\!\sin^2\left[\left(\dfrac{a^{d-4}}{g^2} \right)^{-1/2}\!\, \chi\right]\right\}\vspace{4mm}\\&&\times\;\,\exp\left\{ -\,\dfrac{a^{d-4}}{g^2}\,2\left[1\,-\,\cos\left[\left(\dfrac{a^{d-4}}{g^2} \right)^{-1/2} \chi\right] \right]\right\}\;d \chi\,,
\earr\eequ
where ${\cal N}$ denotes here the normalization with the integral over a single variable $\chi$, which is
$$
{\cal N}\,=\,\dis\int_{|\chi|<(\pi/g)a^{(d-4)/2}}\,\exp\left\{ -\,\dfrac{a^{d-4}}{g^2}\,2\left[1\,-\,\cos\left[\left(\dfrac{a^{d-4}}{g^2} \right)^{-1/2} \chi\right] \right]\right\}\;d \chi\,.
$$

Using the trigonometric inequalities of Eq. (\ref{trig}), for $a\in(0,1]$ and $0<g^2\leq g_0^2<\infty$, we have the bound
$$
\barr{lll}\!\langle \,[{\cal F}^s_{\mu\nu}(x)]^2\,\rangle&\leq&\dfrac1{{\cal N}_{1,0}}\,\dis\int_{| \chi|<\pi a^{(d-4)/2}/g}\, \chi^2\,\exp\left[-\,\dfrac4{\pi^2}\, \chi^2\right]\; d\chi\,,
\earr$$
where,   for ${\cal N}_{1}\,=\, \dis\int_{|\chi|<\pi a^{(d-4)/2}/g}\,\exp\left(- \chi^2\right)\,d \chi$, and the constant ${\cal N}_{1,0}$ is defined as ${\cal N}_1$ but with $g$ replaced by $g_0$ in the integral domain.

Similarly, we obtain the lower bound
$$
\langle \left[ {\cal F}^s_{\mu\nu}(0)\right]^2 \rangle\,\geq\,\dfrac{\dfrac{4}{\pi^2}\,\dis\int_{| \chi|<(\pi/2g)a^{(d-4)/2}}\, \chi^2\,e^{- \chi^2}\,d \chi}{\dis\int_\mathbb R\, e^{-(4/\pi^2) \chi^2}\,d \chi}\;,
$$
where the numerator is bounded below taking the integration domain to be $| \chi|\,\leq [(\pi\,a^{(d-4)/2})/(2g_0)]$. Thus, we see that the scaled plaquette-plaquette correlation at coincident points is uniformly bounded for $a\in(0,1]$ and $0<g^2\leq g_0^2$. Using these bounds and the relation given in Eq. (\ref{2plaque}), we see that the scaled plaquette-plaquette correlation at coincident points has the exact singular behavior $a^{-d}$ (rather than just an upper bound for the singular behavior!)

From these bounds, the continuum limit  $${\cal F}^2(x)\,\equiv\,\lim_{a\searrow 0}\langle \,[{\cal F}^s_{\mu\nu}(x)]^2\,\rangle\,,$$ exists and is given by
\bequ\lb{Msquare}
{\cal F}^2(x)\,=\,\left\{\barr{lll}\dfrac{\dis\int_{\mathbb R}\, \chi^2\,e^{- \chi^2}\,d \chi} {\dis\int_{\mathbb R}\,e^{- \chi^2}\,d\chi}\,=\,\dfrac12&,&\quad d=2,3\,;\vspace{4mm}\\\dfrac{\dis\int_{| \chi|\leq\pi/g}\,\left[\dfrac{\sin(g \chi)}{g}\right]^2\,e^{-2[1-\cos(g \chi)]/g^2}\,d \chi} {\dis\int_{|\chi|\leq\pi/g}\,e^{-2[1-\cos(g \chi)]/g^2}\,d \chi}    &,&\quad d=4\,. \earr\right.
\eequ
Furthermore, from Eq. (\ref{Msquare}), for $d=4$, the $g\searrow 0$ limit also exists and is $1/2$.

In the next subsection, considering the approximate model, we extend these exact and explicit results to scaled correlations with the nonabelian   gauge group $\mathrm U(N)$, $N\geq 2$. In the following sections, we obtain boundedness results for the YM model without approximation. The nonabelian case $N\geq 2$ is more difficult than the abelian $N=1$ case. One of the difficulties is that the gauge group Haar measure is much more complicated than the product Lebesgue measure of the abelian model \cite{Nelson,Holland,Marinov,Euler}. In our extension to the nonabelian case, rather than treat directly the correlations, we bound the two-point plaquette field scaled normalized generating function  (with the scaled partition function in the denominator). Bounds on correlations follow from this, using analyticity and Cauchy bounds for the source derivatives of the generating function at zero source field strengths.

To obtain bounds on the scaled generating function  which are independent of the number of lattice sites, we use the well known multiple reflection method (see Ref. \cite{GJ}). This method makes multiple use of the Cauchy-Schwarz inequality in the quantum mechanical physical Hilbert space of the associated quantum field theory.
%%%%%%%%%%%%%%%%%%%%%%%%%%%%%%%%%%%%%%%%%%%%%%%%%%%%%%%%%%%%%%%%%%%%
%%%%%%%%%%%%%%%%%%%%%%%%%%%%%%%%%%%%%%%%%%%%%%%%%%%%%%%%%%%%%%%%%
%%%%%%%%%%%%%%%%%%%%%%%%%%%%%%%%%%%%%%%%%%%%%%%%%%%%%%%%%%%%%%%%%
\subsection{Approximate Model: Plaquette Field Correlations for ${\mathrm U}(N\geq 2)$}
\lb{sec3C}
%%%%%%%%%%%%%%%%%%%%%%%%%%%%%%%%%%%%%%%%%%%%%%%%%%%%%%%%%%%%%%%%%
%%%%%%%%%%%%%%%%%%%%%%%%%%%%%%%%%%%%%%%%%%%%%%%%%%%%%%%%%%%%%%%%%
In this subsection, we analyze the plaquette field correlations for the  nonabelian case of the gauge group ${\cal G}={\mathrm U}(N)$.
The physical unscaled gauge-invariant plaquette field $\tr {\mathcal F}^u_{\mu\nu}$ for the plaquette $p\,=\,p_{\mu\nu}(x)$ is defined in Eq. (\ref{plaqfield}). Taking the physical parametrization, $U_b\,=\,\exp\{ iagA^u_b\}$, for small lattice spacing $a$, we have that $\tr\mathcal F^u_{\mu\nu}(x)\simeq \tr F^a_{\mu\nu}(x)$, where $F^a_{\mu\nu}$ is given by Eq. (\ref{fmunua}).

With $U_p\,=\,\exp(iX_p$), we also define the gauge-invariant scaled plaquette field by
$$
\tr {\cal F}^s_{\mu\nu}(x)\,=\, a^{d/2}\,\tr\mathcal F^u_{\mu\nu}\,=\,\left(\dfrac{a^{d-4}}{g^2}\right)^{1/2}\,\Im\tr (U_p-1)\,=\,\left(\dfrac{a^{d-4}}{g^2}\right)^{1/2}\,\tr (\sin X_p)\,.
$$
For small $a$, we have
$$\tr  {\cal F}^s_{\mu\nu}(x) \simeq a^{d/2} \,\tr F_{\mu\nu}^{a}(x)\,. $$

As explained in the previous subsection, when analyzing the plaquette-plaquette correlation in the $\mathcal G\,=\,{\mathrm U}(1)$, with external points $x$ and $y$, for the approximate model,whenever the external points $x$ and $y$ are not endpoints of the bonds, we have a factorization and cancellation of the single plaquette, single bond partition functions in the numerator and denominator of the scaled plaquette-plaquette correlations. By the left-right invariance of the Haar measure, the integrals associated with these factors are again over a single bond Haar measure and, by gauge integration properties, the only nonzero contributions are those with coincident points $x=y$. This property allows us to take the infinite volume limit $\Lambda\nearrow a \mathbb Z^d$. 

With this argument, we have that the usual truncated \cite{GJ} plaquette-plaquette correlation is then equal to the non-truncated one. The integrands are class functions, and we can apply the Weyl integration formula (see Refs. \cite{Weyl,Bump,Simon2,Far}) to pass from integrals over $N^2$, $N\times N$ matrix elements, to integrals over $N$ angular eigenvalues. Doing this [compare with Eq. (\ref{coinMsquare})], the coincident point scaled plaquette-plaquette correlation becomes, with $U$ being the single plaquette gauge variable,
$$\barr{lll}
\langle (\tr {\cal F}^s_{\mu\nu})^2\rangle&=&\dfrac1{{\cal N}_2}\,\dis\int_{{\mathrm U}(N)}\,\left[\left(\dfrac{a^{d-4}}{g^2}\right)^{1/2}\,\Im\tr (U-1)\right]^2\,\exp\left\{-\,2\,\left(\dfrac{a^{d-4}}{g^2}\right)\,\tr (1-U-U^\dagger)\right\}\,d\sigma(U)\vspace{2mm}\\&=& \dfrac1{{\cal N}_2}\,\dis\int_{(-\pi,\pi]^N}\,\left[\left(\dfrac{a^{d-4}}{g^2}\right)^{1/2}\,\sum_{j=1,\ldots,N}\,\sin\lambda_j\right]^2\,\exp\left\{-\,2\,\left(\dfrac{a^{d-4}}{g^2}\right)\,\,\sum_{j=1,\ldots,N}\,(1-\cos\lambda_j)\right\}\,\rho(\lambda)\,d\lambda\,.
\earr
$$
Note that the single plaquette correlation is obtained by replacing the squared bracket factor by the single bracket (power one!) in the above integrand. By the transformation of variables $\lambda_j \rightarrow (-\lambda_j)$, the single plaquette correlation $\langle \tr {\cal F}^u_{\mu\nu}\rangle\,=\,0$,  as asserted above.

In view of the recent result of Ref. \cite{AizDC}, on the triviality of the continuum limit of the $\phi^4_4$ model, we investigate whether or not the continuum limit of the approximate model is Gaussian. For this, we also want to consider the scaled four-plaquette correlation and, more generally, the $r$-th power of the scaled plaquette field at coincident points. Following the same gauge integration procedure as before, and after using the Weyl integration formula to pass to angular eigenvalues, the thermodynamic limit of the $r$-th power of the plaquette field at coincident points reduces to
\bequ\lb{rpower}
\langle (\tr {\cal F}^s_{\mu\nu})^r\rangle\,=\,\dfrac1{{\cal N}_r}\,\dis\int_{(-\pi,\pi]^N}\,\left[\left(\dfrac{a^{d-4}}{g^2}\right)^{1/2}\,\sum_{j=1,\ldots,N}\,\sin\lambda_j\right]^r\,\exp\left\{-\,2\,\dfrac{a^{d-4}}{g^2}\,\sum_{j=1,\ldots,N}\,(1-\cos\lambda_j)\right\}\,\rho(\lambda)\,d\lambda\,,
\eequ
where the ratio is taken over single plaquette single variable bond variable integrals. Here, ${\cal N}_r$ is a corresponding normalization constant and $\rho(\lambda)$ is given in Eq. (\ref{rho}). It is worth noticing that, for the abelian gauge group ${\mathrm U}(1)$, we have $\rho(\lambda)=1$.

From Eq. (\ref{rpower}), we easily see the the $r$-correlation is zero if $r$ is odd. For even $r$, making a change of variables, using elementary inequalities and the well-known Lebesgue integral convergence theorems, we obtain that the continuum limit of the above coincident point truncated correlations exists. With 
$${\hat \rho}(\lambda)\,=\,\prod_{1\leq j<k\leq N}\,|\lambda_j-\lambda_k|^2$$
and $T_r(g)\,\equiv\,\lim_{a\searrow 0}\,\langle [{\tr {\cal F}^s_{\mu\nu}]}^r\rangle$, for $d=2,3$, we obtain, letting $\lambda\,=\,\left(2a^{d-4}/g^2\right)^{-1/2}y$,
$$
T_\alpha(g)\,=\,\dfrac{2}{{\cal N}_2}\;\dis\int_{\mathbb R^N}\,\left(\sum_{j=1,\ldots,N}y_j\right)^r\;\hat\rho(y)\,\exp\left[ -\sum_{j=1\ldots,N} y_j^2\right]\;d^N y\,,
$$
with an associated measure normalization ${\cal N}_2$. For $d=4$, letting $\lambda\,=\,gy$, we obtain 
$$
T_r(g)\,=\, \dfrac 1{{\cal N}_4\;g^{N(N-1)}}\;\dis\int_{{(-\pi/g,\pi/g]}^N}
\,\left[\left(\sum_{j=1,\ldots,N}\dfrac{\sin(gy_j)}{g}\right)^r\,\right]
\;\rho(gy)\,\exp\left[-\sum_{j=1\ldots,N}\dfrac{2[1-\cos (gy_j)]}{g^2}\right]\,d^N y\,.
$$ 

For $d=4$, the $g\searrow 0$ limit $T_r$, of $T_r(g)$, is 
\bequ\lb{gue}
T_r\,=\,\dfrac 1{{\cal N}_4}\;\dis\int_{\mathbb R^N}
\,\left(\sum_{j=1,\ldots,N}y_j\right)^r\,\;\hat\rho(y)\,\exp\left[-\sum_{j=1\ldots,N} y_j^2\right]\,d^N y\,,
\eequ
with a normalization ${\cal N}_4$.

Note that, for $r=2,4$, the right-hand-side of Eq. (\ref{gue}) is, respectively, $\sum_{i,j,k,\ell=1,\ldots,N}\,\langle y_iy_j\rangle_G$ and $\sum_{i,j,k,\ell=1,\ldots,N}\,\langle y_iy_j,y_k,y_\ell\rangle_G$,  where $\langle \,\cdot\,\rangle_G$ is the expectation in the GUE (Gaussian Unitary Ensemble) (see e.g. Refs. \cite{Metha,Bump,Deift}).

Finally, for the case of an abelian gauge group ${\mathrm U}(1)$, we then see that the continuum limit is Gaussian for $d=2,3$. For $d=4$ the continuum limit followed by the $g\searrow 0$ limit is also Gaussian. 

From Eq. (\ref{gue}), for $d=4$ and taking the gauge group ${\mathcal G}\,=\,{\mathrm U}(2)$, we have
$$\barr{lll}
T_r&=&\dfrac 1{{\cal N}_4}\;\dis\int_{\mathbb R^2}
\,(y_1+y_2)^r\;(y_1-y_2)^2\;e^{-(y_1^2+y_2^2)}\;dy_1\,dy_2\vspace{2mm}\\
&=&\dfrac1{\xi}\;\int_{\mathbb R^2}\,\left(\sqrt{2}\eta\right)^r\; \left(\sqrt{2}\epsilon\right)^2\;e^{-(\eta^2+\epsilon^2)}\,d\eta\,d\epsilon\,,\earr
$$
with $\xi\,=\,\int_{\mathbb R^2}\;e^{-(\eta^2+\epsilon^2)}\,d\eta\,d\epsilon$, where we made the $(\pi/4)$ rotation change of variables $\sqrt{2}\epsilon\,=\,(y_1-y_2)$ and $\sqrt{2}\eta\,=\,(y_1+y_2)$. By performing the integrals in the denominator and the $\epsilon$ integral in the numerator, we obtain
$$
T_r\,=\,\dfrac{2^{r/2}}{\sqrt{\pi}}\;\int_{\mathbb R}\,\eta^\alpha\,e^{-\eta^2}\,d\eta\,,
$$
which shows a Gaussian, non-interacting behavior. Whether or not this is the behavior we have for any gauge group $\mathrm U(N>2)$ is still to be analyzed.
%%%%%%%%%%%%%%%%%%%%%%%%%%%%%%%%%%%%%%%%%%%%%%%%%%%%%%%%%%%%%%%%%
%%%%%%%%%%%%%%%%%%%%%%%%%%%%%%%%%%%%%%%%%%%%%%%%%%%%%%%%%%%%%%%%%
\section{Thermodynamic and Ultraviolet Stability Bounds: the General $\mathcal G={\mathrm U}(N\geq 1)$ Case}\lb{sec4}
%%%%%%%%%%%%%%%%%%%%%%%%%%%%%%%%%%%%%%%%%%%%%%%%%%%%%%%%%%%%%%%%%
%%%%%%%%%%%%%%%%%%%%%%%%%%%%%%%%%%%%%%%%%%%%%%%%%%%%%%%%%%%%%%%%%
We now obtain factorized stability bounds for the unscaled partition function $Z^{u,B}_{\Lambda}(a)$ of the complete model defined in Eq. (\ref{part}) with boundary condition $B$. In doing this, we are improving the proofs of Refs. \cite{YM,bQCD} and are extending the results to the periodic b.c. case. The bounds are factorized as a product. In the product, each factor is a single bond variable, single plaquette partition function. First, we give Lemma \ref{lema1}   which yields an exact representation for the Wilson plaquette action and is used to prove that the plaquette action upper bound is quadratic in each gluon field. This growth is in contrast to the classical Lagrangian action which has a quartic growth in the fields and which is used in Refs. \cite{MRS,Hairer}. The upper quadratic bound on $A_p$ is used to obtain the factorized lower bound on $Z^{u,B}_{\Lambda}(a)$. 

Again, as an example, it is worth recalling that, for the abelian gauge group ${\mathrm U}(1)$, the bound is obtained by elementary inequalities. Indeed, writing the unscaled plaquette gauge variable $U^u_p\,=\,\exp\left\{i(\theta_1+\theta_2-\theta_3-\theta_4) \right\}$, $|\theta_j|<\pi$, $j=1,2,3,4$, and using the first of Eq. (\ref{trig})], we obtain
$$
A_p\,=\,2[1-\cos(\theta_1+\theta_2-\theta_3-\theta_4)]\,\leq\,(\theta_1+\theta_2-\theta_3-\theta_4)^2\,\leq\,4(\theta_1^2+\theta_2^2+\theta_3^2+\theta_4^2)\,,
$$
where, we have expanded the square in the first inequality and used the bound $2uv\,\leq\,u^2\,+\,v^2$, $u,v\in\mathbb R$, to obtain the second inequality.

The following Lemma is a much improved version of Lemma 2 of Ref. \cite{bQCD}.
\begin{lemma}	\lb{lema1}
Let $U_j\,=\,e^{{\mathcal L}_j}$, where ${\mathcal L}_j\,=\,i\,\sum_{\alpha=1}^{N^2}\,x^j_\alpha\,\theta_\alpha$, so that $|\mathcal L_j|\,\leq\,\|\mathcal L_j\|_{H-S}\,=\,|x^j|$. Here, $x^j_\alpha$ is real and $x^j_\alpha\,=\,-i\,\tr \theta_\alpha\mathcal L _j$. Let also
$$U_p(\delta)\,=\,U_1(\delta)\,U_2(\delta)\,U_3^\dagger(\delta)\,U_4^\dagger(\delta)\,,$$
with $U_j(\delta)\,=\,e^{\delta\mathcal L_j}$, for $\delta\in[0,1]$. Then, by the Fundamental Theorem of Calculus, we have the representation
\bequ\lb{FTC}\barr{lll}
\!\!U_p-1\!\!&=&\!\!\dis\int_0^1d\delta\,\left[\mathcal L_1U_1(\delta)U_2(\delta)U_3^\dagger(\delta)U_4^\dagger(\delta)\,+\,U_1(\delta)\mathcal L_2U_2(\delta)U_3^\dagger(\delta) U_4^\dagger(\delta)\right. \vspace{2mm}\\\!\!&&\!\!\qquad\qquad\left.\!\!\! -\,U_1(\delta)U_2(\delta)\mathcal L_3U_3^\dagger(\delta)U_4^\dagger(\delta) \,-\,U_1(\delta)U_2(\delta)U_3^\dagger(\delta)\mathcal L_4U_4^\dagger(\delta) \right]\vspace{2mm}\\\!\!
&=&\!\!\,\dis\int_0^1\,\left\{ \mathcal L_1\,+\,\dis\int_0^\delta\,d\hat\delta\, \left[\mathcal L_1\mathcal L_1 U_1(\hat\delta)U_2(\hat\delta)U_3^\dagger(\hat\delta)U_4^\dagger(\hat\delta)+\ldots\right]\right\}\vspace{4mm}\\\!\!&&\!\!+\,
\dis\int_0^1\,\left\{ \mathcal L_2\,+\,\dis\int_0^\delta\,d\hat\delta\,\left[\mathcal L_1 U_1(\hat\delta)\mathcal L_2U_2(\hat\delta)U_3^\dagger(\hat\delta)U_4^\dagger(\hat\delta)+\ldots \right]\right\}\vspace{4mm}\\\!\!&&\!\!-\,
\dis\int_0^1\,\left\{ \mathcal L_3\,+\,\dis\int_0^\delta\,d\hat\delta\,\left[\mathcal L_1U_1(\hat\delta)U_2(\hat\delta)\mathcal L_3U_3^\dagger(\hat\delta)U_4^\dagger(\hat\delta)+\ldots \right]\right\}\vspace{4mm}\\\!\!&&\!\!-\,
\dis\int_0^1\,\left\{ \mathcal L_4\,+\,\dis\int_0^\delta\,d\hat\delta\,\left[\mathcal L_1U_1(\hat\delta)U_2(\hat\delta)U_3^\dagger(\hat\delta)\mathcal L_4U_4^\dagger(\hat\delta)+\ldots \right]\right\}
\earr\eequ

For a single retained unscaled plaquette action $A_p(U_p)\,=\,2\Re\,\tr(1\,-\,U_p)$, using the representation of Eq. (\ref{FTC}), the second equality holds without the isolated $\mathcal L_j$ terms, which give an imaginary trace, and we have the global quadratic upper bound
	\bequ\lb{lower1}
	\mathcal A^u_p\,=\,\|U^u_p-1\|^2_{H-S}\,\equiv\,|2\Re\,\tr (U^u_p-1)|\,\leq\, C^2 \,\sum_{1\leq j\leq 4}\, |x^j|^2\quad\:\:,\:\:\quad C=2\sqrt N\,,
	\eequ
	where $C^2=4N$.
	In particular, for the physical parametrization $U_b=\exp[igaA^u_b]$ and the scaled field parametrization $U_b=\exp[iga^{(4-d)/2}A_b]$, we have, respectively,
	\bequ\lb{actionbdlemma}\dfrac{a^{d-4}}{g^2}\,A_p\,\leq\,C^2\,
	a^{d-2}\,\sum_b\,|A^u_b|^2\quad{\mathrm and}\quad C^2\,\sum_b\,|A_b|^2\,,
	\eequ
	where $\sum_b$ runs over the bonds of the plaquettes and $C^2\,=\,4N$. Hence, the plaquette energy $\frac{a^{d-4}}{g^2}\,A_p$ is regular in $g^2$, for all $g^2\geq 0$, and has the quadratic growth bound in the fields.
	
	When there are only one, two or three retained bond variables in a plaquette, in the first equality, the sum over $j$ has, respectively, only one, two and three terms and the numerical factor $4$ in $C^2N$ is replaced by 1, 2 and 3, respectively. For the total unscaled action $A^{u,B}\,=\,\sum_p\,\mathcal A^u_p$, we have the global quadratic upper bound
	\bequ\lb{lower2}A^{u,B}\,\leq\,2(d-1)\,C^2\,\sum_b|x^b|^2\,=\,2(d-1)\,C^2\,\sum_b|\la_b|^2\,,
	\eequ
	where, whenever gauge fixing is applied, the sum runs over all $\La_r$ lattice bonds.
	
	Concerning the unscaled and scaled plaquette fields
	$$
	\tr \mathcal F^u_p\,=\, \dfrac1{a^2g}\, \Im\tr\,(U_p\,-\,1)\,,
	$$
	and
	$$
		\tr \mathcal F_p\,=\,	a^{d/2}\,\tr \mathcal F^u_p\,,
	$$
	we also have the representation of Eq. (\ref{FTC}), and the $\mathcal L_j$ terms are, respectively
	$$\barr{lll}
		\tr \mathcal F^u_p&=&a^{-d/2}\,\tr\left(A_1^u\,+\,A_2^u\,-\,A^u_3\,-\,A^u_4 \right)\quad,\quad\:\:\mathcal L_j\;{\mathrm terms\;only}\vspace{2mm}\\
		&=&a^{-d/2}\,\tr\left(\partial_\mu A^u_\nu\,-\,\partial_\nu A^u_\mu\right)\quad,\quad\:\:\mathcal L_j\;{\mathrm terms\;only}\,,\earr
	$$
	and
	$$
	\tr \mathcal F_p\,=\,\tr\left(\partial_\mu A_\nu\,-\,\partial_\nu A_\mu\right)\quad,\quad\:\:\mathcal L_j\;{\mathrm terms\;only}\,,
	$$
	The coefficients in the integral terms are proportional to $g$, so that the plaquette fields are also regular in $g$, for $g^2\,\geq\,0$.
\end{lemma}
\begin{rema}
From the second equality in Eq. (\ref{FTC}), and taking the real part of the trace, only the double integrals contribute. From this representation, the bound given in Lemma \ref{lema1} is obtained by inspection. We give a proof for the bound using only the first line. The iteration of the Fundamental Theorem of Calculus in Eq. (\ref{FTC}) produces the diverse terms in the Baker-Campbell-Haussdorff formula \cite{Far}. Here, instead, there is no question of convergence involved.
\end{rema}
\begin{rema}
	The apparent singularity at $g=0$, due to the action prefactor $(1/g^2)$ [see Eq. (\ref{partition})], which persists in the corresponding scaled expression, is removed  and the 'action' $A_p/g^2$ is regular at $g=0$, if we use the physical field parametrization $U_b=\exp[igaA^u_b]$, for the unscaled bond variable $U_b$, or the scaled bond field $U_b=\exp[iga^{(4-d)/2}A_b]$. In both cases, the action is bounded by a quadratic growth. This is in contrast to the classical Lagrangian, where cubic and quartic interactions are present, and the growth is quartic. This growth behavior is also present in the analysis of the existence of YM models in Refs. \cite{MRS,Hairer}. Besides, in these references, an explicit infrared regulator is introduced contrary to the Wilson YM case where an infrared cutoff is {\sl not} needed for periodic and free b.c.
\end{rema}

For completeness of the present paper, we give the proof of Lemma \ref{lema1} in section \ref{sec6}. A preliminary version of the stability bounds was given in Ref. \cite{bQCD}. The following four theorems are also proved in section \ref{sec6}. Our results on stability and boundedness of the generating function and correlations are for the Wilson YM model. We sometimes use gauge fixing but there are no additional infrared regulator terms  added to the action like in Refs. \cite{MRS,Hairer}. 

Our stability bounds on the unscaled partition function $Z^{u,B}_{\Lambda}(a)$, leading to TUV stability bounds for the scaled partition function $Z^{s,B}_{\Lambda}(a)$ are given by
\begin{thm} \lb{thm1}
	The unscaled partition function $Z^{u,B}_{\Lambda}(a)$ verifies the following stability bounds:  \vspace{2mm}\\
	\noindent\underline{Free b.c.}: For $B$= blank, we have
	\bequ\lb{sbfree}z_\ell^{\Lambda_r}\,\leq\,Z^u_{\Lambda}(a)\,\leq z_u^{\Lambda_r}\,,\eequ
	\noindent\underline{Periodic b.c.}: For $B=P$, we obtain 
	\bequ\lb{sbperiodic}z_\ell^{\Lambda_r+\Lambda_e}\,\leq\,Z^{u,P}_{\Lambda}\,\leq\,Z^u_{\Lambda}(a)\,\leq \,z_u^{\Lambda_r}\,,\eequ        
	where
	\bequ\lb{zu}
	z_u\,=\,\int\, \exp\left[-2(a^{d-4}/g^2)\,\Re\tr (1-U)]\right]\; d\sigma(U)\,.
	\eequ
    Also,   we have $U= e^{iX}$, $C^2=4N$, $X\,=\,\sum_{\alpha=1,\ldots,N^2}\, x_\alpha\theta_\alpha$   and then
	$$
	\tr X^2\,=\,\sum_{\alpha=1,\ldots,N^2}\,x_\alpha^2\,=\,\sum_{k=1,\ldots,N}\,\,\lambda_k^2\,,
	$$
	where $\lambda_1,\ldots,\lambda_N$ are the angular eigenvalues of $U$.   Finally,
	\bequ\lb{zl}
	z_\ell\,=\,\int\, \exp\left[-2C^2(a^{d-4}/g^2)\,(d-1)\,\tr X^2\right]\; d\sigma(U)\,.
	\eequ
\end{thm}
\begin{rema}\lb{rema3}
Using Jensen's inequality, we obtain the factorized lower bound $Z^u_{\Lambda}(a)\geq \xi^{\Lambda_p}$, where 
$$
\xi\,=\,\exp\left\{- \dfrac{a^{d-4}}{g^2}\,\dis\int\,\|U-1\|_{H-S}^2\,d\sigma(U) \right\}\,\geq\,\exp\left[- 2N\,\dfrac{a^{d-4}}{g^2} \right]\,,
$$
where we recall $\Lambda_p$ is the number of plaquettes in $\Lambda$. We have $\Lambda_p=\Lambda_r$, for $d=2$; $\Lambda_p\simeq 3L^3,6L^4$, respectively, for $d=3,4$. In Theorem \ref{thm2} below, we obtain factorized lower and upper bounds with $\Lambda_r=(d-1)L^d$ factors. In both the upper and lower bound a factor of $[(a^{d-4}/g^2)^{-N^2/2}]$ is extracted. This factor dominates the $a$, $g^2$ dependence.
\end{rema}

We continue by giving more detailed bounds for $z_u$ and $z_\ell$.  In these bounds, we extract a factor of $[(a^{d-4}/g^2)^{-N^2/2}]$ from both $z_u$ and $z_\ell$.  Note that the integrands of both $z_u$ and $z_\ell$ only depend on the angular eigenvalues of the gauge variable $U$; they are class functions on $\mathcal G$.  The $N^2$-dimensional integration over the group can be reduced to an $N$-dimensional integration over the angular eigenvalues of $U$ by the Weyl integration formula of Eq. (\ref{weyl1})    (see Refs. \cite{Weyl,Bump,Simon2,Far}). For the group ${\mathrm U}(N)$, we explicitly  have
\bequ\lb{weyl2} \int_{\mathrm U(N)}\,f(U)\;d\sigma(U)\,=\,\dfrac1{\mathcal N_C(N)}\,\int_{(-\pi,\pi]^N}\,f(\lambda)\,\rho(\lambda)\,d^N\lambda\,,
\eequ
where $\mathcal N_C(N)=[(2\pi)^N\,N!]$, $\lambda=(\lambda_1,\ldots,\lambda_N)$, $d^N\lambda=d\lambda_1\ldots d\lambda_N$ and $\rho(\lambda)=\prod_{1\leq j<k\leq N}\,|e^{i\lambda_j}-e^{i\lambda_k}|^2$.

In our stability and generating function bounds in the $\mathcal G\,=\,\mathrm U(N)$ case, the following integrals of the Gaussian unitary ensemble (GUE) and Gaussian symplectic ensemble (GSE) probability distributions (see \cite{Metha,Deift}), of random matrix theory, arise naturally.  Let, for $\beta=2,4$ and $u>0$,
\bequ\lb{Ibeta}I_\beta(u)\,=\,\dis\int_{(-u,u)^N}\,\exp\left[-(1/2)\,\beta\,\sum_{1\leq  j\leq N}\,y_j^2\right]\;\hat\rho^{\beta/2}(y)\, d^Ny\,,\eequ
where $\hat\rho(y)=\prod_{1\leq j<k\leq N}\,(y_j-y_k)^2$, $I_\beta(u)<I_\beta(\infty)=\mathcal N_\beta$, is the normalization constant for the GUE and the GSE probability distributions for $\beta=2,4$, respectively. Explicitly, we have
$\mathcal N_G\,= [(2\pi)^{N/2}\,2^{-N^2/2}\,\prod_{1\leq j\leq N}\,j!\,]$ and  $\mathcal N_S\,= [(2\pi)^{N/2}\,4^{-N^2}\,\prod_{1\leq j\leq N}\,(2j)!\,]$.

For the upper bound on $z_u$ and lower bound on $z_\ell$, we have the following result. 
\begin{thm} \lb{thm2}
For $C^2=4N$, we have the bounds on $z_u$ and $z_\ell$ appearing in Theorem \ref{thm1}   
	\bequ\lb{uzu}\begin{array}{lll}
		z_u&=& \mathcal N_C^{-1}\,\int_{(-\pi,\pi]^N}\,\exp[-2(a^{d-4}/g^2)\,\sum_{1\leq j\leq N}\,(1-\cos\lambda_j)]\;\rho(\lambda)\;d^N\lambda\vspace{1mm}\\
		&\leq&(a^{d-4}/g^2)^{-N^2/2}\,(\pi/2)^{N^2}\,\mathcal N_G(N)\mathcal N^{-1}_C(N)\vspace{1mm}\\
		&\equiv&(a^{d-4}/g^2)^{-N^2/2}\,e^{c_u}\,,\end{array}\eequ
	and
	\bequ\lb{lzl}\begin{array}{lll}
		z_\ell&=& \mathcal N_C^{-1}\,\int_{(-\pi,\pi]^N}\,\exp[-2C^2(d-1)(a^{d-4}/g^2)\,\sum_{1\leq j\leq N}\,\lambda_j^2]\;\rho(\lambda)\;d^N\lambda\vspace{1mm}\\
		&\geq&(a^{d-4}/g^2)^{-N^2/2}\,\mathcal N^{-1}_C(N)\,(4/\pi^2)^{N(N-1)/2}\,[2(d-1)C^2]^{-N^2/2}\,I_\ell\,,\vspace{1mm}\\
		&\equiv&(a^{d-4}/g^2)^{-N^2/2}\,e^{c_\ell}\,,\end{array}\eequ
	where, recalling Eq. (\ref{Ibeta}),  $I_\ell\equiv I_2(\pi[2(d-1)C^2]^{1/2}/(2g_0))$. The constants $c_u$ and $c_\ell$ are real and finite, and independent of $a$, $a\in(0,1]$ and $g^2\in(0,g_0^2]$, $0<g_0<\infty$.
\end{thm}

Concerning the existence of the thermodynamic and continuum limits of the scaled free energy we define the scaled partition function by
\bequ\lb{nZ}Z^{s,B}_{\Lambda}(a)\,=\,(a^{d-4}/g^2)^{(N^2/2)\Lambda_r}\; Z^{u,B}_{\Lambda}(a)\,,\eequ
and a finite lattice scaled free energy by
\bequ\lb{nf}f^{s,B}_{\Lambda}(a)\,=\, \dfrac1{\Lambda_r}\, \ln Z^{s,B}_{\Lambda}(a)\,.\eequ

Using Theorems \ref{thm1} and  \ref{thm2}, together with the Bolzano-Weierstrass theorem \cite{Rudin}, we prove the following Theorem. 
\begin{thm}	\lb{thm3}
	The scaled free energy $f^{s,B}_{\Lambda}(a)$ converges subsequentially, at least, to a thermodynamic limit
	$$f^{s,B}(a)\,=\,\lim_{\Lambda\nearrow a\mathbb Z^d}\; f^{s,B}_{\Lambda}(a)\,,$$ and, subsequently, again, at least subsequentially, to a continuum limit $$f^{s,B}\,=\,\lim_{a\searrow 0}\,f^{s,B}(a)\,.$$ Besides, $f^{s,B}(a)$ satisfies the bounds
	\bequ\lb{bdsf}
	-\infty \,<\, c_\ell\,\leq\, f^{s,B}(a)\,\leq c_u<\infty\,.
	\eequ
	and so does its subsequential continuum limit $f^{s,B}$. The constants $c_\ell$ and $c_u$ are finite real constants independent of $a\in(0,1]$ and $g^2\in(0,g^2_0]$, $0 < g_0< \infty$.
\end{thm}
\begin{rema}
For the case of the free scaled scalar field, the TUV stability bound follows by bounding  spectral representations [check Eq. (\ref{2pfscalar})] and their finite lattice counterparts. For the case of bond independent couplings, the TUV stability bound is proven in Appendix C. The proof uses a multiple reflection bound in addition to the H\"older inequality to decouple the $d$ coordinate directions. In this way, the proof of TUV is reduced to a TUV bound for the partition function $Z_c$ of a one-dimensional chain. In turn, this bound is obtained by bounding a single bond `transfer matrix´. The bound can also be obtained by successive integration for free boundary conditions. The free boundary conditions serve as an infrared regulator, i.e. no mass term is needed in the action to exclude the zero mode.
\end{rema}

%==========================================================================================
%==========================================================================================
%==========================================================================================
\section{Generating Function for Plaquette Field Correlations}\lb{sec5}
%============================================================================================%%%%%%%%%%%%%%%%%%%%%%%%%%%%%%%%%%%%%%%%%%%%%%%%%%%%%%%%%%%%%%%%%%%%
Here, we obtain bounds for the generating function  of gauge-invariant plaquette field correlations. Bounds for the field correlations follow from analyticity in the source field strengths, using Cauchy estimates on the generating function. The same hypercubic lattice $\Lambda$ is maintained, with periodic b.c., and we use the multiple reflection method \cite{GJ}. Our choice of correlations is guided by the E-M spectral results for lattice YM with strong gauge coupling $g^2\,\gg 1$ (see Ref. \cite{Schor}). We fix the lattice spacing $a=1$ and denote the plaquette coupling constant by $\gamma=a^{d-4}/g^2$.  For $0<\gamma\ll 1$, a lattice quantum field theory is constructed via a Feynman-Kac formula. By polymer expansion methods, infinite lattice correlations exist and are analytic in $\gamma\in\mathbb C$, $|\gamma|\ll 1$ (see Ref. \cite{Sei}).  In Ref. \cite{Schor}, for $0<\gamma\ll 1$, it is shown that, associated with the truncated plaquette-plaquette correlation, there is an isolated particle (glueball) state in the low-lying E-M spectrum, with mass of order $(–8\ln \gamma)$. Furthermore, it is proved that the low-lying spectrum is generated by limits of local polynomials in the plaquette field. No more general loop variables are needed. The isolated dispersion curve of the glueball is the {\em only} low-lying spectrum that is present.

Returning to our model, we consider the generating function for the correlation of $r\in\mathbb N$ gauge-invariant real plaquette fields. For $p=p_{\mu\nu}(x)$, as defined in Eq. (\ref{plaqfield}), with $U_p=e^{iX_p}\in\mathrm U(N)$, 
the physical, unscaled plaquette field is given by
$$\barr{lll}
\tr \mathcal F^u_{\mu\nu}(x)&=&\dfrac1{a^2g}\,\Im\tr(U_p-1)\,=\,-\dfrac{i}{2a^2g}\,\tr(U_p\,-\,(U_p)^\dagger)\,=\,\dfrac1{a^2g}\,\tr \sin X_p\,.
\earr$$

Recalling Eq. (\ref{scaledplaq}), parametrizing $U_p$ by the physical, unscaled gauge field $U_b\,=\,\exp\{iagA^u_b \}$, we have 
$$
\tr \mathcal F^u_{\mu\nu}(x)\,\simeq\, \tr F^a_{\mu\nu}(x)\,=\, \tr\left[\partial^a_\mu A^u_\nu(x)\,-\,\partial^a_\nu A^u_\mu(x) \right]\,,
$$
where $F^a_{\mu\nu}(x)\,=\,\partial^a_\mu A^u_\nu(x)\,-\,\partial^a_\nu A^u_\mu(x) \,+\,ig\left[ A^u_\mu(x),A^u_\nu(x) \right]$, with a commutator in the Lie algebra of $\mathrm U(N)$.

Next, define the gauge-invariant scaled plaquette field by
\bequ\lb{Mmunu}
\tr {\cal F}^s_{\mu\nu}(x)\,=\,a^{d/2}\,\tr \mathcal F^u_{\mu\nu}\,=\,\left(\dfrac{a^{d-4}}{g^2}\right)^{1/2}\,\Im\tr (U_p-1)\,.
\eequ
Using once more the physical, unscaled field parametrization, we have that
$$
\tr {\cal F}^s_{\mu\nu}(x)\,\simeq\,\,a^{d/2}\,\tr\left[\partial^a_\mu A^u_\nu(x)\,-\,\partial^a_\nu A^u_\mu(x) \right]\,.
$$
With our choice of the scaling factor $[(a^{d-4}/g^2)^{1/2}]$, the generating function  for scaled plaquette field correlations is finite, uniformly in $a\in(0,1]$. It may seem surprising that the generating function is pointwise bounded. However, it is known that a similar phenomenon occurs in the case of a free massless or massive scalar field in $d=3,4$. Namely, as analyzed in Ref. \cite{MP2}, if instead of the given physical field $\phi^u(x)$, we use a locally scaled field $\phi(x)\simeq a^{(d-2)/2}\phi^u(x)$, then the $r$–point correlation for the scaled $\phi$ fields is bounded pointwise, uniformly in $a\in(0,1]$. No smearing by a smooth test function is needed to achieve boundedness! We give more details regarding the properties of scalar fields in  Appendix A.
\begin{rema}\lb{remanew1}
	We can also define other plaquette fields and their associated scaled fields. For instance, we can also work with the field $$\tr {\mathcal H}_{\mu\nu}(x)\,=\,\dfrac{1}{a^4g}\,A_p\,\simeq\,\tr \left[F_{\mu\nu}^a(x)\right]^2\,,$$ and the associated scaled fields given by $\tr S_{\mu\nu}(x)\,\equiv\,a^d\,\tr{\mathcal H}_{\mu\nu}(x)$. The results and proofs obtained below for the generating function of correlations of the scaled field $\tr M_{\mu\nu}(x)$ carry over to $\tr S_{\mu\nu}(x)$. 
\end{rema}

The $r$-plaquette scaled field generating function, associated with the field of Eq. (\ref{Mmunu}), is defined by
$$G_{r,\Lambda,a}(J^{(r)}) = \dfrac1{Z^{s,P}_{\Lambda}(a)}\;Z^{s,P}_{r,\Lambda}(a,J^{(r)})\,,$$
where, for the source strengths $J_j$,  $j = 1,\ldots,r$, we have  $J^{(r)}\,=\,(J_1,\dots,J_r)$ and $Z^{s,P}_{r,\Lambda}(a,J^{(r)})$ is   defined similarly to $Z_{\Lambda}^{s,P}(a)$ (see Eq. \ref{part}), but with the inclusion of $r$ local source factors in the integrand given by 
$$\exp\left[\sum_{x\in\Lambda}\sum_{1\leq j\leq r}\,J_j(x_j)\,\tr {\cal F}^s_{p_j}(U_{p_j})\right]\,,$$ 
Here, we adopt the convention that the plaquette $p_j$ originates at the lattice point $x_j$ and $p_j$ is a shorthand for $p_j\,=\,p_{\mu_j,\nu_j}(x_j)$. The $r$-plaquette correlation, with a set $y_E\,=\,(y_1,\dots,y_r)$ of $r$ lattice external points in $\Lambda$ is given by  
$$\left.\dfrac{\partial^r}{\partial J_1(y_1)\ldots\partial J_r(y_r)}\,G_{r,\Lambda,a}(J^{(r)})\right|_{J_j=0}\,.$$

Our factorized bound is given in the next Theorem. For simplicity of notation, from now on, we set $J_i\,\equiv\,J_i(y_i)$.
\begin{thm} \label{thm4}
Considering the model with periodic b.c., we have:
\begin{enumerate}
\item The $r-$plaquette scaled field generating function  is bounded by
\bequ\lb{Gb}|G_{r,\Lambda,a}(J^{{(r)}})|\,\leq\,\prod_{1\leq j\leq r}\,\dfrac{\left|z_u(rJ_j)\right|^{2^d\Lambda_r/(r\Lambda_s)}}{z_\ell^{2^d(\Lambda_r+\Lambda_e)/(r\Lambda_s)}}\,.\eequ  
\item From this, if $G_{r,a}(J^{(r)})$ denotes a sequential or subsequential thermodynamic limit $\Lambda\nearrow a\mathbb Z^d$, then 
$$\left|G_{r,a}(J^{(r)})\right|\,\leq\, \prod_{1\leq j\leq r}\,\left|z_u(rJ_j)/z_\ell\right|^{2^d(d-1)/r}\,,$$
with
\bequ\lb{Gb2}\begin{array}{lll}
\left|z_u(J)\right|&=&\dis\int\,\exp\left[\,|J|\,(a^{d-4}/g^2)^{1/2}\,|\Im Tr(U-1)|\,-\,(a^{d-4}/g^2)\,A^u_p(U)\,\right]\, d\sigma(U)\,\vspace{1mm}\\
&=& (\mathcal N_c)^{-1}\,\dis\int\,exp\left[|J|\,(a^{d-4}/g^2)^{1/2}\,\sum_{1\leq j\leq N}\, |\sin \lambda_j|\,-\,2a^{d-4}/g^2 \sum_{1\leq j\leq N}\,(1 – \cos \lambda_j )\right]\,\rho(\lambda)\,d^N\lambda\vspace{2mm}\\
&\leq&\dfrac{(a^{d-4}/g^2)^{-N^2/2}\,\pi^{N^2+N/4}\,\mathcal N_S^{1/2}}{\mathcal N_C}\:\exp[(\pi^2/8)N|J|^2]\vspace{2mm}\\
&\equiv&(a^{d-4}/g^2)^{-N^2/2}\,\exp(c_u^\prime+\pi^2/8N|J|^2)\,.\end{array}\,,\eequ
where $\exp c_u^\prime\,=\,\pi^{N^2+N/4}\,+\,\dfrac{\sqrt{\mathcal N_s}}{\mathcal N_C}$. 
Recalling $C^2\,=\,4N$ and using Eq. (\ref{lzl}) of Theorem \ref{thm2}, we obtain
$$
\barr{lll}
z_\ell&=&{\cal N}_C^{-1}\,\dis\int_{(-\pi,\pi]^N}\,\exp\left\{-\left[2C^2(d-1) a^{d-4}/g^2\right]\,\sum_{j=1,\ldots,N}\,\lambda_j^2\right\}\;\rho(\lambda)\,d^N\lambda\vspace{2mm}\\
&\geq&{\cal N}_C^{-1}\,\left(\dfrac{2(d-1)C^2a^{d-4}}{g^2} \right)^{-N^2/2}\,\left( \dfrac4{\pi^2}  \right)^{N(N-1)/2}\,I_\ell\vspace{2mm}\\&\equiv&\left( a^{d-4}/g^2 \right)^{-N^2/2}\,e^{c_\ell}\,, 
\earr
$$
where $c_\ell$ is defined in Theorem \ref{thm2} and $I_\ell\,\equiv\,
I_2(\pi C\sqrt{2(d-1)}/(2g_0))$ and $I_2$ is the function defined in Eq. (\ref{Ibeta}).

Hence, from the bounds of Eqs.(\ref{Gb}) and (\ref{Gb2}), it follows that $G_{r,\Lambda,a}(J^{{(r)}})$ is a jointly analytic, entire complex function of the source field strengths $J_j\in\mathbb C$.  
\item Letting $G_{r}(J^{(r)})$ denote a sequential or subsequential continuum limit $a\searrow 0$ of $G_{r,a}(J^{(r)})$, then 
$$\left|\,G_{r}(J^{(r)})\,\right|\,\leq\, \exp\left[\dfrac{2^d}r\,(d-1)\,(c^\prime_u-c_\ell)\,+\,(\pi^2/8)\,Nr\,\sum_{1\leq j\leq r}\,    |J_j|^2\right]\,.$$
\end{enumerate}
This bound is independent of the location and orientation of the $r$ plaquettes, and independent of the value of $a\in(0,1]$ and $g^2$.
\end{thm}
\begin{rema}
Our method to prove the generating function bound uses multiple reflection	bounds \cite{GJ}. Remember the bound has to be uniform in the volume $\Lambda_s$ (number of lattice sites) and in the lattice spacing $a\in(0,1]$. For the case of the free scaled scalar field, the multiple reflection method does not work, as we now explain. Using multiple reflection with a uniform source field, we have to bound $\langle \exp[J\sum_x \phi(x)]\rangle$. The bound should be in the form $\exp[_u(J)\,\Lambda_s]$, where $c_u(J)$ is bounded uniformly in $\Lambda_s$, $a\in(0,1]$ and,for finite $J$ or, at least, for small $|J|$. However, as the free field is Gaussian, 
$$
\langle\,exp\left[J\,\sum_x\phi(x) \right]\,\rangle\,=\,\exp\left[ \dfrac12\,J^2\,\sum_{x,y}\,C(x,y)\right]\,,$$
where $C(x,y)$ is the 2-point correlation for the scaled scalar field and is given in Eq. (\ref{2pfscalar}), in the thermodynamic limit. Using translation invariance, $\sum_{x,y}\,C(x,y)\,=\,\Lambda_s\,\sum_x\,C(x,0)$. The sum $\sum_x\,C(x,0)$ gives the zero momentum value of the Fourier transform. This quantity in {\sl not} bounded uniformly in $a\in(0,1]$, and blows up as $1/a^2$ as seen from Eq. (\ref{2pfscalar}). Of course,  $\langle\,exp\left[J\,\phi(x) \right]\,\rangle\,=\,\exp\left[ \frac12\,J^2\,C(x,x)\right]$  and $C(x,x)$ are bounded uniformly in $a\in(0,1]$ which can again be seen using Eq. (\ref{2pfscalar}).
\end{rema}
\begin{rema}\lb{remm}
For comparison, the $r-$point  correlation for the free scaled scalar field, in the thermodynamic limit, is given by
$$\barr{lll}
\left\langle\,\exp\,\left[\,\sum_{j=1}^r\,J_j\,\phi(x_j)\right]\,\right\rangle&
=&\exp\left[ \dfrac12\,\sum_{j,k=1}^r\,J_j\,C(x_j,x_k)\,J_k\right]\, \leq\,\exp\left\{ \dfrac{2d\,+\,(m_u/\kappa_u)^2}{4d}\,\Delta_1^{-1}(0,0)\,\sum_{j,k=1}^r\, |J_j|\,|J_k|\right\}\vspace{2.5mm}\\&\leq&\exp\left[\ r\,\,\dfrac{2d\,+\,(m_u/\kappa_u)^2}{4d}\,\Delta_1^{-1}(0,0)\,\sum_{j=1}^r\, J_j^2\right]\,,\earr $$
where $\Delta^{-1}_1(0,0)$ is given in Eq. (\ref{DDelta}) and is finite for $d\geq 3$. By Griffths I \cite{GJ}, this bound extends to complex $J$.
\end{rema}
\begin{rema}\lb{rema4}
  The generating function extends to an entire jointly analytic function of the source strengths $J_i$, $i=1,\ldots,r$. By Cauchy estimates, it can be used to bound the $r$-plaquette scaled field correlations. We use the $\mathbb C^r$ version of the Cauchy bounds. Recall that, for $\mathbb C$, if $f(z)$ is analytic in the disk $|z|\,<\, R$, $R>0$, then $|(d^n f/dz^n)(z=0)|\,\leq\,n!\;[\sup_{z;\,|z|=R_0}\,|f(z)|\,]/R_0^n$, for any $0<R_0<R$ (see e.g. Ref. \cite{Hille}). In particular, the coincident point plaquette-plaquette physical field correlation is bounded by $\mathrm{const}\,a^{-d}$. The $a^{-d}$ factor is the same small $a$ behavior of the coincident point, two-point correlation of the derivative of the real scalar physical free field (see Appendix A). Using the free scalar field as a reference, this singular behavior is a measure of the ultraviolet asymptotic freedom.
\end{rema}
\begin{rema}\lb{rema5}
In obtaining the bounds on the scaled plaquette field generating function and correlations, we have used the group   bond variable parametrization $U_b\,=\,\exp\left\{iga^{-(d-4)/2}\, \chi_b \right\}$. In the physically relevant $d=4$ case, $U_b=e^{ig \chi_b}$ and $\langle[\tr {\cal F}^s]^r\rangle_{\Lambda,a,g}$ is independent of the lattice spacing $a$, so that
$$
\langle[\tr {\cal F}^s]^r\rangle_{\Lambda,g}\,\equiv\,\langle[\tr {\cal F}^s]^r\rangle_{\Lambda,a,g}\,=\,a^{dr/2}\,\langle[\tr{\cal F}^u]^r\rangle_{\Lambda,a,g}\,.
$$
For the thermodynamic limit or subsequential limit, we drop the subscript $\Lambda$, so that we have
$$
\langle[\tr {\cal F}^s]^r\rangle_{g}\,=\, a^{dr/2}\,\langle[\tr{\cal F}^u]^r\rangle_{a,g}\,.
$$
Of course, the continuum limit of the left-hand-side is $\langle[\tr {\mathcal F}^s]^r\rangle_{g}$ and
$$
\langle[\tr {\cal F}^u]^r\rangle_{g}\,=\,a^{-dr/2}\,\,\langle[\tr {\cal F}^s]^r\rangle_{g}\,,
$$
which displays the {\em exact} dependence on the lattice spacing $a$ as a multiplicative factor.
\end{rema}

Lemma \ref{lema1} and Theorems $1-4$ are proved in the next section.
%=====================================================================================
%=====================================================================================
%=====================================================================================
\section{Proofs of the Lemma and Theorems}\lb{sec6}
%=====================================================================================
%=====================================================================================
%=====================================================================================
Here, following Ref. \cite{bQCD}, we give a proof of Lemma \ref{lema1}. We also prove Theorems $1-4$. The enhanced temporal gauge is sometimes used for proving these theorems. The proof of the upper stability bound on the partition function actually does not depend on this choice.
%=====================================================================================
%=====================================================================================
\subsection{Proof of Lemma \ref{lema1}}\lb{sec6A}
%=====================================================================================
%=====================================================================================
For simplicity, we consider the case where we have four bonds in a plaquette. The other cases, when only one, two, or three bonds are retained, are similar. Recalling that, if $A$ and $B$ are self-adjoint, then $\tr (AB)\,=\,\tr (BA)\,=\,\tr(AB)^\dagger$ is real, we define, for $1\leq j\leq 4$, $\mathcal L_j=i\sum_{1\leq \alpha\leq N^2} x^j_\alpha\theta_\alpha$, so that  $U_j=e^{\mathcal L_j}$ and $U_p=U_1U_2U_3^\dagger U^\dagger_4$.

Since $\|\mathcal L_j\|\,\leq\,\|\mathcal L_j\|_{H-S}\,=\,|x^j|$ and letting $U_p(\delta)\,=\,U_1(\delta)U_2(\delta)U^\dagger_3(\delta)U^\dagger_4(\delta)$, $U_j(\delta)\,=\,e^{\delta\mathcal L_j}$, for $\delta\in[0,1]$, by the fundamental theorem of calculus, suppressing $\delta$,
$$
U_p\,-\,1\,=\,\dis\int_0^1\,d\delta\,\left[\mathcal L_1U_1U_2U_3^\dagger U_4^\dagger\,+\,U_1\mathcal L_2U_2U_3^\dagger U_4^\dagger\,-\,U_1U_2\mathcal L_3U_3^\dagger U_4^\dagger\,-\,U_1U_2U_3^\dagger\mathcal L_4 U_4^\dagger\right]\,.
$$
Using the triangle and Cauchy-Schwarz inequalities, we obtain
$$
\|U_p\,-\,1\|\,\leq\,\sum_{j=1}^4\,\|\mathcal L_j\|\,\leq\,\sum_{j=1}^4\,\|\mathcal L_j\|_{H-S}\,=\,\sum_{j=1}^4\,|x^j|\,\leq\,2\,\left[\sum_{j=1}^4\,|x^j|^2\right]^{1/2}\,.
$$
But, $\|U_p\,-\,1\|\,\geq\,N^{-1/2}\,\|U_p-1\|_{H-S}$. Hence,
$$
\mathcal A_p\,=\,\|U_p-1\|^2_{H-S}\,\leq\,4N\,\sum_{j=1}^4\,|x^j|^2\,.
$$
By considering the number   of terms in the sum over $j$, the factor $4$ in $C^2$ is replaced by $1$, $2$ and $3$, respectively, when only one, two or three retained bond variables appear in a retained plaquette.

Alternatively, we can bound the terms of the second equality of Eq. (\ref{FTC}), for $(U_p\,-\,1)$ directly.

Using this upper bound on the single plaquette action, we sum over the plaquettes. Noting that, fixing a given lattice bond, there are at most $[2(d-1)]$ plaquettes that have this bound in common, the result for the total action follows.\qed
%%%%%%%%%%%%%%%%%%%%%%%%%%%%%%%%%%%%%%%%%%%%%%%%%%%%%%%
%%%%%%%%%%%%%%%%%%%%%%%%%%%%%%%%%%%%%%%%%%%%%%%%%%%%%%%
\subsection{Proof of Theorem \ref{thm1}}\lb{sec6B}
%%%%%%%%%%%%%%%%%%%%%%%%%%%%%%%%%%%%%%%%%%%%%%%%%%%%%%%
%%%%%%%%%%%%%%%%%%%%%%%%%%%%%%%%%%%%%%%%%%%%%%%%%%%%%%%
\noindent{\em The Case of Free b.c.}:\vspace{1mm}\\
\noindent\underline{Upper Bound}:  For ease of visualization we carry it out explicitly for $d=3$. An upper bound is obtained by discarding all horizontal plaquettes from the action, except those with temporal coordinates $x^0=1$.  We now perform the horizontal bond integration.  Integrate over successive planes of horizontal bonds starting at $x^0=L$ and ending at $x^0=2$. For the $x^0=1$ horizontal plane, integrate over successive lines in the $\mu=2$ direction, starting at $x^1=L$ and ending at $x^1=2$. For each horizontal bond variable, the bond variable appears in only one plaquette in the action. After the integration, in principle, the integral still depends on the other bond variables of the plaquette.   However, using the left or right invariance of the Haar measure, the integral is independent of the other variables. In this way, we extract a factor $z_u$. In the total procedure, we integrate over the $\Lambda_r$   horizontal bonds (see Eq. (\ref{lambdar})), so that we extract a factor $z_u^{\Lambda_r}$.\vspace{2mm}\\
\noindent\underline{Lower Bound}: Using Lemma 1 gives the factorization and $z_\ell$.\vspace{4mm}

\noindent{\em The Case of Periodic b.c.}:\vspace{1mm}\\
\noindent\underline{Upper Bound}: Considering the positivity of each term in the model action of Eq. (\ref{action}), since $A^P\geq A$, we have
$$Z^P_{\Lambda,a}\,\leq\,\int\,e^{-A}\,dg^B\,=\,\int\, e^{-A}\,dg\,=\,Z_{\Lambda,a}\,\leq\,z_u^{\Lambda_r}\,.$$ 
\vspace{2mm}\\
\underline{Lower Bound}: Use the global quadratic upper bound of Lemma \ref{lema1} on all $\Lambda_r\cup\Lambda_e$ bond variables. Thus, we have
$$Z^P_{\Lambda,a}\,\geq\,z_\ell^{\Lambda_r+\Lambda_e}\,,$$
where $U = exp(iX)$, $X =\sum_\alpha x_\alpha\theta_\alpha$.\qed
%%%%%%%%%%%%%%%%%%%%%%%%%%%%%%%%%%%%%%%%%%%%%%%%%%%%%%%%%%%%%%%%%%%%%%%%%%%%%
%%%%%%%%%%%%%%%%%%%%%%%%%%%%%%%%%%%%%%%%%%%%%%%%%%%%%%%%%%%%%%%%%%%%%%%%%%%%%
\subsection{Proof of Theorem \ref{thm2}}\lb{sec6C}
%%%%%%%%%%%%%%%%%%%%%%%%%%%%%%%%%%%%%%%%%%%%%%%%%%%%%%%%%%%%%%%%%%%%%%%%%%%%%
%%%%%%%%%%%%%%%%%%%%%%%%%%%%%%%%%%%%%%%%%%%%%%%%%%%%%%%%%%%%%%%%%%%%%%%%%%%%%
  In Theorem \ref{thm2}, the first line for $z_u$ [see Eq. (\ref{uzu})] is the application of the Weyl integration formula of Eq. (\ref{weyl2}) (see Refs. \cite{Weyl,Bump,Simon2,Far}). Use the inequality (see \cite{Simon3}) $(1-\cos x) \geq 2x^2/\pi^2$, $x \in [-\pi,\pi]$, in the action, and the inequality $(1-\cos x)\leq x^2/2$ in each factor of $\rho(\lambda)$. After making the change of variables $y \,=\,2[a^{(d-4)/2}/(\pi g)]\,\lambda$ and using the monotonicity of the integral, the result follows. 

  To obtain Eq. (\ref{lzl}) for $z_\ell$, apply the Weyl integration formula and use the inequality $2[1-\cos(\lambda_j-\lambda_k)]\,\geq\,(4/\pi^2)\, (\lambda_j-\lambda_k)^2$, $|\lambda_\ell|<\pi/2$ in each factor of the density $\rho(\lambda)$. Then, use the positivity of the integrand and restrict the domain of integration to $(-\pi/2,\pi/2]^N$. In making the change of variables $y\,=\, [a^{(d-4)/2}/g]\,C\,\sqrt{2(d-1)}\,\lambda$, the integral $I_2([a^{(d-4)/2}/g]\,C\,\sqrt{2(d-1)})\pi/2)$ appears (see Eq. (\ref{Ibeta})). Since $I_2(u)$ is monotone increasing, the integral assumes its smallest value for $a = 1$ and $g^2=g_0^2$.  
\qed
%%%%%%%%%%%%%%%%%%%%%%%%%%%%%%%%%%%%%%%%%%%%%%%%%%%%%%%%%%%%%%%%%%%%%%%%%%%%%
%%%%%%%%%%%%%%%%%%%%%%%%%%%%%%%%%%%%%%%%%%%%%%%%%%%%%%%%%%%%%%%%%%%%%%%%%%%%%
\subsection{Proof of Theorem \ref{thm3}}\lb{sec6D}
%%%%%%%%%%%%%%%%%%%%%%%%%%%%%%%%%%%%%%%%%%%%%%%%%%%%%%%%%%%%%%%%%%%%%%%%%%%%%
%%%%%%%%%%%%%%%%%%%%%%%%%%%%%%%%%%%%%%%%%%%%%%%%%%%%%%%%%%%%%%%%%%%%%%%%%%%%%
For periodic b.c. and the lower bound, using Theorem \ref{thm1}, we have the finite volume lattice normalized free energy
$$\barr{lll}
f^{P,n}_{\Lambda,a}&=&\dfrac1{\Lambda_r}\,\ln Z^{P,n}_{\Lambda,a}\,=\,\dfrac1{\Lambda_r}\,\ln\left[ \dfrac{a^{d-4}}{g^2}\right]^{N^2\Lambda_r/2}\,+\,\dfrac1{\Lambda_r}\,\ln Z^{P}_{\Lambda,a}\vspace{2mm}\\
&\geq&\dfrac1{\Lambda_r}\,\ln\left[ \dfrac{a^{d-4}}{g^2}\right]^{N^2\Lambda_r/2}\,\,+\,\dfrac1{\Lambda_r}\,\ln z_\ell^{\Lambda_r+\Lambda_\ell}\,.\earr
$$

Continuing the inequality and using Theorem \ref{thm2}, we have
$$\barr{lll}
f^{P,n}_{\Lambda,a}&\geq&\dfrac1{\Lambda_r}\,\ln\left[ \dfrac{a^{d-4}}{g^2}\right]^{N^2\Lambda_r/2}\,+\,\dfrac{\Lambda_e+\Lambda_r}{\Lambda_r}\,\ln \left[\left(\dfrac{a^{d-4}}{g^2}\right)^{-N^2/2}\,e^{c_\ell}\right]\vspace{2mm}\\
&\geq&\ln\left[ \dfrac{a^{d-4}}{g^2}\right]^{N^2/2}\,\,+\,\dfrac{\Lambda_e+\Lambda_r}{\Lambda_r}\,\left[\ln \left(\dfrac{a^{d-4}}{g^2}\right)^{-N^2/2}\,+\,c_\ell \right]\,\earr
$$
which gives, when $\Lambda\rightarrow a\mathbb Z^d$,
$$
f^{P,n}_{a}\,\geq\,c_\ell\,.
$$

  A similar calculation for the upper bound, setting to zero the number of extra bonds in the lattice with periodic b.c., $\Lambda_e=0$, proves the theorem for the upper bound. Of course, for free b.c., set $\Lambda_e=0$ in the above calculations.\qed
%%%%%%%%%%%%%%%%%%%%%%%%%%%%%%%%%%%%%%%%%%%%%%%%%%%%%%%%%%%%%%%%%%%%%%%%%%%%%
%%%%%%%%%%%%%%%%%%%%%%%%%%%%%%%%%%%%%%%%%%%%%%%%%%%%%%%%%%%%%%%%%%%%%%%%%%%%%
\subsection{Proof of Theorem \ref{thm4}}\lb{sec6E}
%%%%%%%%%%%%%%%%%%%%%%%%%%%%%%%%%%%%%%%%%%%%%%%%%%%%%%%%%%%%%%%%%%%%%%%%%%%%%
%%%%%%%%%%%%%%%%%%%%%%%%%%%%%%%%%%%%%%%%%%%%%%%%%%%%%%%%%%%%%%%%%%%%%%%%%%%%%
To prove Theorem \ref{thm4}, first use the generalized H\"older's inequality to bound $G_{r,\Lambda, a}(J^{(r)})$ by a product of single plaquette generating functions, i.e.
$$|G_{r,\Lambda,a}(J^{(r)})|\,\leq\,\prod_{1\leq j\leq r}\,|G_{1,\Lambda, a}(rJ_j)|^{1/r}\,.
$$
Now, since we are adopting periodic b.c., we can apply the multireflection method (see \cite{GJ}) to bound each factor in the product. To this end, we make a shift in the lattice by $(1/2a)$ in each coordinate direction. Also, we use the $\pi/2$ lattice rotational symmetry and translational symmetry to put the single plaquette in the $\mu\nu=01$   coordinate plane in the first quadrant, with lower left vertex at $(a/2,a/2,\ldots,a/2)$. Then, we apply the multireflection method to obtain the bound 
$$|G_{1,\Lambda, a}(rJ_j)|\,\leq\,|G_{\Lambda,a}(rJ_j)|^{2^d/\Lambda_s}\,,$$
where $G_{\Lambda,a}(J)\,=\,\left[Z^P_{\Lambda,a}\right]^{-1}\,Z^P_{\Lambda,a}(J)$, with $Z^P_{\Lambda,a}(J)$ denoting $Z^P_{\Lambda,a}$ with a source of uniform source strength $J$. The source factor is given by $exp[J\sum^\prime_p\tr {\cal F}_p(U_p)]$, where the sum is over an array of plaquettes. The array consists of planes of plaquettes that are parallel to the $01$   coordinate plane. In each plane, they are only alternating, i.e. like considering only squares of a same color on a chessboard.  We obtain a greater upper bound by noting that
$$\barr{lll}
|J \;\tr {\cal F}
_p(U_p) |&\leq& |J|\,[a^{(d-4)/2}/g]\,|\Im \tr(U_p-1)|\\
&\leq&|J|\,[a^{(d-4)/2}/g]\,|\tr(U_p-1)|\\
&\leq&|J|\,[a^{(d-4)/2}/g]\,N^{1/2}\,\|U_p-1\|_{H-S}\,,\earr
$$
where we have used the Cauchy-Schwarz inequality in the Hilbert-Schmidt inner product.

We also increase the bound by summing over all plaquettes in the lattice $\Lambda$ that are parallel to the $01$   coordinate plane. We denote this sum by $\sum_p^{\prime\prime}$. In this way, we obtain the upper bound
$$
\left|Z^P_{\Lambda,a}(J)\right|\,\leq\,\int\, exp\left[|J| a^{(d-4)/2}g^{-1}N^{1/2}\sum^{\prime\prime}_p\,\|U_p-1\|_{H-S}\,-\,a^{d-4}A^P/g^2\right]\,dg^P\,.
$$

As in the proof of the upper stability bound, for the periodic model, given above, we discard plaquette actions in $A^P$, for plaquettes that are not in $\Lambda$ so that 
$$
\left|Z^P_{\Lambda,a}(J)\right|\,\leq\,\int\, exp\left[|J| a^{(d-4)/2}g^{-1}N^{1/2}\sum^{\prime\prime}_p\,\|U_p-1\|_{H-S}\,-\,a^{d-4}A/g^2\right]\,dg\,.
$$

We bound the integral as we did for the upper stability bound for the free b.c. case.  In this manner, we obtain the factorized bound
$$
\left|Z^P_{\Lambda,a}(J)\right|\,\leq\,[z_u(J)]^{\Lambda_r}\,,
$$
and the factorized bound of   Theorem \ref{thm4}for $G_{r\Lambda a}(J^{(r)})$ is proved. Here, we have used the factorized lower bound of Theorems \ref{thm1} and \ref{thm2} for $Z^P_{\Lambda,a}$.

Now, recalling that $\Lambda_s=L^d$, $\Lambda_r\simeq (d–1) L^d$ and $\Lambda_e=dL^{d-1}$, the factorized bound for $G_{ra}(J^{(r)})$ follows.

Application of the Weyl integration formula of Eq. (\ref{weyl2}) \cite{Weyl,Bump,Simon2,Far} gives the  $\lambda$ integral for $z_u(J)$. Using the bounds $|\sin \lambda_j|\leq |\lambda_j|$, for all $j$, and  $|\exp(i\lambda_j)-\exp(i\lambda_k)|^2= 2[1 –\cos(\lambda_j-\lambda_k)]\leq (\lambda_j-\lambda_k)^2$, for each factor of $\rho(\lambda)$ gives the inequality
$$
z_u(J)\,\leq\,(1/\mathcal N_c)\,\displaystyle\int_{(-\pi,\pi]^N)}\,\exp\left[|J| (a^{(d-4)/2}/g)\,\sum_{1\leq j\leq N}|\lambda_j|\,-\,4a^{d-4}/(g^2\pi^2)\,\sum_{1\leq j\leq N}\,\lambda_j^2\right]\,\hat\rho(\lambda)\,d^N\lambda\,.
$$

Making the change of variables $y_k=[2a^{(d-4)/2}/(g\pi)]\,\lambda_k$, a factor of $[a^{(d-4)/2}/g]^{-N^2}$ is extracted and the remaining integral is bounded by, with $y^2=\sum_j\,y_j^2$,
$$
\dis\int_{\mathbb R^N}\,exp\left[\pi |J|\sum_{1\leq j\leq N} |y_j|/2\,-\,\sum_{1\leq j\leq N}y_j^2\right]\,\hat\rho(y)\,d^Ny\,.
$$

Writing $\exp(-y^2)=\exp(-y^2/2) \exp(-y^2/2)$ and using the Cauchy-Schwarz inequality, the integral is bounded by
$$
\left[\dis\int_{\mathbb R^N}\,exp\left(\pi\,|J|\sum_{1\leq j\leq N} |y_j|\,-\,\sum_{1\leq j\leq N}y_j^2\right)\,d^Ny\right]^{1/2}\;\left[ \dis\int_{\mathbb R^N}\,exp\left(-\,\sum_{1\leq j\leq N}y_j^2\right)\,\hat\rho^2(y)\,d^Ny\right]^{1/2}\,.
$$
Using the inequality $e^{s|y_k|}\leq e^{sy_k} + e^{-sy_k}$, $s>0$, the Gaussian integral of the bound of the integral of the first factor is carried out explicitly.  For the integral of the second factor, after making the change of variables $w_k=(y_k/\sqrt{2})$ and, up to a numerical factor, the resulting integral is the normalization constant  $\mathcal N_S$ for the GSE ensemble (see \cite{Metha,Deift}). Keeping track of the numerical factors gives the final inequality for $z_u(J)$ and the proof of Theorem \ref{thm4} is complete.\qed      
%%%%%%%%%%%%%%%%%%%%%%%%%%%%%%%%%%%%%%%%%%%%%%%%%%%%%%%%%%%%%%%%%%%%%%%%%%%%%%
%%%%%%%%%%%%%%%%%%%%%%%%%%%%%%%%%%%%%%%%%%%%%%%%%%%%%%%%%%%%%%%%%%%%%%%%%%%%%%
%%%%%%%%%%%%%%%%%%%%%%%%%%%%%%%%%%%%%%%%%%%%%%%%%%%%%%%%%%%%%%%%%%%%%%%%%%%%%%
\section{Concluding Remarks}\lb{sec7}
%===================================================================================
%%%%%%%%%%%%%%%%%%%%%%%%%%%%%%%%%%%%%%%%%%%%%%%%%%%%%%%%%%%%%%%%%%%%
We consider the Yang-Mills quantum field theory in an imaginary-time functional integral formulation and on an hypercubic lattice $\Lambda\subset a\mathbb Z^d$, $d=2,3,4$, $a\in (0,1]$, with the Wilson partition function defined with gauge coupling $0\,<\,g^2\,<\,g_0^2$, for $g_0$ positive and finite. This means our results are {\em not} restricted to small couplings $g_0$. The lattice $\Lambda$ has $L$ (even) sites on a side, $\Lambda_s=L^d$ sites, and we use both free and periodic boundary conditions. Letting $0$ label the temporal direction, $x=(x^0,\ldots,x^{d-1})$ denotes a lattice site and  $e^\mu$, $\mu=0,\ldots,(d-1)$ is a unit vector in the positive $\mu$ direction. We are concerned with finiteness properties of physically relevant quantities and our main goal is to derive stability bounds to control the thermodynamic limit ($\Lambda\nearrow a\mathbb Z^d$) and the continuum limit ($a\searrow 0$) of the free energy and correlations. 

In the Wilson formulation, there is a matrix, gauge variable $U_b$ for each positively oriented lattice bond $b$. A positively oriented bond $b_\mu(x)$ is a segment $[x,x^+_\mu\equiv x+ae^\mu]$ connecting the $\Lambda$ site $x$ to $x^+_\mu$ in the $\mu$th positive axis direction, and we take $U_b$ to be an element of the compact Lie gauge groups $\mathcal G={\mathrm U}(N),\;\mathrm {SU}(N)$. With this, we write the unitary gauge variable $U_b\,=,e^{iX_b}$, with a self-adjoint element $X_b$ of the $\mathcal G$ Lie algebra. We also use what we call the physical parametrization $U_b\,=\,e^{iagA_b}$, and if $b\,=\,b_\mu(x)\,=\,[x,x_\mu^+\equiv x+ae^\mu]$, we set $A_b\,=\,A_\mu(x)$. The physical gauge potentials (gluon fields) $A_\mu(x)$ then lie in the Lie algebra of $\mathcal G$.
 
The partition function for type $B$ (free$\,=\,$no superscript or periodic$\,=P\,$) b.c. is given by  
$$Z^B_{\Lambda,a}\,=\, \int\; \exp[(-a^{d-4}/g^2)\,A^B]\, d\tilde g^B\,.$$
A lattice plaquette (minimal square)  $p\,=\,p_{\mu\nu}(x)$, with positively oriented bonds $b_1\,=\,[x\,,\,x^\mu_+]$, $b_2\,=\,[x^\mu_+\,,\,x^\mu_+\,+\,ae^\nu]$, $b_3\,=\,[x_+^\nu\,,\,x^\nu_+\,+\,ae^\mu]$ and $b_4\,=\,[x, x^\nu_+]$, in the $\mu\nu$ coordinate plane ($\mu<\nu$), is associated with the plaquette variable $U_p\,=\,U_{b_1}U_{b_2}U^\dagger_{b_3}U^\dagger_{b_4}\,=\,e^{iX_p}$, since $U_b$ is unitary. The total Wilson action $A^B$ is a sum over all distinct lattice $\Lambda$ single plaquette actions $A_p$ given by $$A_p\,=\,\|U_p-1\|^2_{H-S}\,=\,2\,\Re\tr(1-U_p)\,= \,2\tr(1–\cos X_p)\,.$$ Here, $\|\,\cdot\,\|_{H-S}$ is the Hilbert-Schmidt norm. Note that $A_p$ is pointwise nonnegative and so is $A^B=\sum_p\,A_p$. Last, a copy of the gauge group, denoted by $\mathcal G_x$ is attached to each lattice point $x\in\Lambda$ and the gauge group measure $d\tilde g^B$ is a product measure over bonds of a gauge group $\mathcal G$ Haar measure. Whenever periodic b.c. is employed, as usual, we add extra bonds to $\Lambda$ connecting the $\Lambda$ {\em endpoints} of the boundary $\partial \Lambda$ of $\Lambda$ to the {\em initial} points of the boundary $\partial\Lambda$ in each spacetime direction $\mu=0,1,\ldots,(d-1)$. The periodic plaquettes are those that can be formed from the totality of periodic bonds.

Formally, for small $a\in(0,1]$, $A_p\,\simeq\,a^4g^2\; \tr [F^a_{\mu\nu}(x)]^2$, where, with finite difference derivatives given in Eq. (\ref{partiala}), we have $F^a_{\mu\nu}(x) = \partial^a_\mu A_\nu(x)\, –\,\partial^a_\nu A_\mu(x)\,+\,ig \,[A_\mu(x),A_\nu(x)]$, and the commutator is taken over the Lie algebra of $\mathcal G=\mathrm U(N)$. 
Thus, for $\mu\,<\,\nu$, $(a^{d-4}/g^2)\,\sum_p A_p\,\simeq\,\sum_{x\in\Lambda}\;a^d\, \sum_{\mu,\nu=0}^{d-1}\,\tr [F^a_{\mu,\nu}(x)]^2$ is the Riemann sum approximation to the smooth field continuum Yang-Mills action $\sum_{\mu<\nu}\,\int_{[-La,La]^d}\,\tr [F_{\mu\nu}(x)]^2\,d^dx$, where $F_{\mu\nu}(x)$ is defined as above, but with usual partial derivatives.

The Wilson action is invariant under local gauge transformations from the group $\prod_{x\in\Lambda}\,\mathcal G_x$ [see Eq. (\ref{gt})]. By this invariance, initially, there is an excess of degrees of freedom. Then, by a gauge fixing procedure, certain gauge variables are gauged away, i.e. they are fixed to the identity matrix, and have trivial gauge integration. In this procedure, there are only $\Lambda_r$ retained bond variables and the value of the partition function is not changed if the gauged away bonds do {\em not} form a loop on $\Lambda$ \cite{GJ}. We sometimes adopt the {enhanced temporal} gauge, in which case all temporal bonds in $\Lambda$ are gauged away (i.e. set to the identity with a trivial corresponding gauge Haar integral) as well as some specified spatial bonds on the boundary of $\Lambda$. Here, $\Lambda_r$ is approximately the number of non temporal (or spatial) bonds, namely, $\Lambda_r\simeq (d-1)L^d$.

Since there are $\delta_N$ (the group Lie algebra dimension) components of the gauge potential, the $\Lambda$ lattice total number of degrees of freedom in the Yang-Mills model is $\delta_N\Lambda_r$. Instead of working with the unscaled, physical fields $A_\mu(x)\,\equiv\,A^u_\mu(x)$, if we work with {\em scaled fields} $A^s_\mu(x)\,=\,a^{(d-2)/2}\,A^u(x)$, we find that the Wilson action becomes more regular in $a\in(0,1]$ and $0\,<\,g^2\,<\,g_0^2\,<\,\infty$. If we define a scaled field partition function $Z^s_{\Lambda}(a)$ by $$Z^s_{\Lambda}(a)\,=\,\left(a^{(d-4)/2}/g\right)^{\delta_N\Lambda_r}\,Z^u_{\Lambda}(a)\,,$$ then we find that $Z^s_{\Lambda}(a)$ obeys the thermodynamic and ultraviolet stability bounds (TUV)
$$
e^{c_\ell\Lambda_r}\,\leq\,Z^s_{\Lambda}(a)\,\leq\,e^{c_u\Lambda_r}\,,
$$
with finite constants $c_\ell$ and $c_u$, independent of $a$ and $g^2$.
By the Bolzano-Weierstrass theorem, this bound leads to the existence of the thermodynamic followed by the continuum limit of the scaled free energy $f^s_{\Lambda}(a)$, at least in the subsequential sense. Note that the scaling we use is non-canonical and that it does not affect the underlying quantum mechanical energy-momentum spectrum and particle content of the model, since it does not alter the decay rate of correlations.

What about correlations? For the plaquette $p=p_{\mu\nu}$, as given above, we define a gauge-invariant unscaled plaquette field, with $U_p\,=\,e^{iX_p}$,
$$
\tr\,\mathcal F^u_{\mu\nu}(x)\,=\, \dfrac 1 {a^2g}\,\Im\tr(U_p-1)\,=\, -\dfrac i{2a^2g}\,\tr\left(U_p\,-\,U_p^\dagger\right)\,=\,\dfrac1{a^2g}\,\tr\,\sin X_p\,.
$$
The gauge invariance of the plaquette field results from Eq. (\ref{gt}) and, more directly, from the gauge invariance of $\tr U_p$.

With the physical parametrization $U_b\,=\,e^{igaA_b}$, for small $a$, we have $\mathcal \tr\,\mathcal F^u_{\mu\nu}(x)\,\simeq\, \tr\,F^a_{\mu\nu}(x)$. We also define a scaled plaquette field
$$
\tr\,\mathcal F^s_{\mu\nu}(x)\,=\,a^d\, \tr\,\mathcal F^u_{\mu\nu}(x)\,=\,\dfrac {a^{(d-4)/2}}{g}\:
\Im\tr(U_p-1)\,.
$$

Our results on stability bounds on the partition function and boundedness of the generating function and correlations are for the Wilson Yang-Mills model with a priori scaled fields. Sometimes we apply gauge fixing, imposing what we call the enhanced temporal gauge, but there are no additional infrared regulator terms  added to the action like in Refs. \cite{MRS,Hairer}. An infrared cutoff is not needed in the Wilson model with free and periodic b.c..

We have applied our scaled field method to many Bose and Fermi models. Scaling improves regularity of the physical action. The corresponding partition function and model correlations are bounded. Correlations are bounded even at coincident points. In models which are perturbations of the free field, the usual subtraction terms have finite coefficients.

In particular, for the real scalar field, these properties are proved in Appendix A, for the free field, and in Appendix B, for a perturbatively non-Gaussian scalar field model. Denote by $\phi^u(x)$ and $\phi^s(x)$ the unscaled, physical and the scaled scalar field. They satisfy the relation $\phi^s(x)\,=\,s(a)\,\phi^u(x)$, where $s(a)\,=\,a^{(d-2)/2}\,\left(2d\kappa_u^2\,+\,m_u^2a^2\right)^{1/2}$ is a noncanonical $x$-independent scaling factor. With this, the scaled and unscaled partition functions verify $Z^s_{\Lambda}(a)\,=\,[s(a)]^{\Lambda_s}\,Z^u_{\Lambda}(a)$. The scaled partition function $Z^s_\Lambda(a)$ obeys TUV stability bounds with the exponent $\Lambda_s$ (number of sites) and {\em not} the $\mathbb R^d$ volume $(La)^d$. The scaled free energy 
$f^s_\Lambda(a)\,=\, [\ln Z^s_\Lambda(a)]/\Lambda_s$ is bounded uniformly in $L$ and $a\in (0,1]$. $f^s_\Lambda(a)$ then converges to a thermodynamic limit $f^s(a)$ and then to a continuum limit $f^s$, for dimensions $d=3,4$. 

Moreover, the unscaled, physical two-point normalized correlation is related to the scaled one by $\langle \phi^s(x)\,\phi^s(y) \rangle^s\,=\,s^2(a)\,\langle \phi^u(x)\,\phi^u(y) \rangle^u$ and the scaled correlation is bounded uniformly in $\Lambda$ and $a$, even at coincident points. This property leads to the existence of its thermodynamic and, subsequently, continuum limits. The unscaled, physical two-point correlation at coincident points presents the singular behavior  $a^{(d-2)/2}$, which can be taken as a lattice characterization of UV asymptotic freedom. The singular behavior of derivative fields is different and we have, with finite difference derivatives and $\langle \partial^a_\rho\phi^u(x)\,\partial^a_\sigma\phi^u(y) \rangle^u\,=\,\simeq\, [1/(s^2(a)\,a^2)]$. For $x=y$, we obtain the behavior $a^{-d}$, $d=1,2,3,4$. Again, this behavior can be taken as a characterization of UV asymptotic freedom. The same type of analysis can be used to treat the complete $\phi^4_3$-model and we also notice that singularities present in the construction of $\phi^4_3$ in Ref. \cite{BFSo}, and the stochastic PDE construction of Ref. \cite{GH} are greatly reduced using our scaled field method.

The scaled field method also applies to treat interacting Fermi fields, and may be a useful tool to extend our results for Yang-Mills models to QCD. For fermionic models TUV bounds, etc, also hold for scaled free field fermionic models \cite{MP,MP2}. The key fact is that the fermionic Gaussian integral, written here symbolically as ($\tilde\psi$ meaning a $\psi$ or a $\bar\psi$ field, where we are suppressing lattice and internal indices (such as spin, flavor/isospin, ...) 
$$
\int\,M(\tilde\psi)\,e^{\bar\psi\psi}\,d\bar\psi\,d\psi\,=\,-1,\,0,\,1\,,
$$
where $M(\tilde\psi)$ is a coefficient one monomial in the Fermi (Grassmann) fields $\bar\psi$ and $\psi$. The integrand and the measures factorizes over lattice sites and internal indices. Thus, for example, considering a model with partition function \bequ\lb{Zf}\barr{lll}Z_\Lambda&=&\dis\int\,e^{h(\tilde\psi)}\,e^{V(\tilde\psi)}\,e^{-\bar\psi\psi}\, d\bar\psi\,d\psi\vspace{2mm}\\
&=&\dis\int\,\left[\prod_{b\in\Lambda}\,\left( 1\,+\,\sum_{n(b)}\,c^b_{n(b)}\,M_{n(b)}(\tilde\psi_b)\right) \right]\:\left[\prod_{s\in\Lambda}\,\left( 1\,+\,\sum_{m(s)}\,c^s_{m(s)}\,M_{m(s)}(\tilde\psi_s)\right) \right]\;e^{-\bar\psi\psi}\, d\bar\psi\,d\psi\,,\earr
\eequ
where $b$ denotes a lattice bond and $s$ a lattice site. To arrive at the last equality, we expand each free field nearest neighbor hopping factor $e^{h(\tilde\psi)}$ and each factor in $e^{V(\tilde\psi)}$, of the local interacting potential $V(\tilde\psi)$. Due to Pauli repulsion, the expansions are finite, in the finite lattice. For the terms in the bond $b$ expansion, we make a correspondence with the integers $n(b)\,=\,1\,,2\,,\ldots$ For the terms in the site $s$ expansion,  a correspondence with the integers $m(s)\,=\,1\,,\,2\, ,\ldots$ is also made. The scaled field Fermi bond coupling is (see Ref. \cite{PMJMP})
\bequ\lb{kf}\kappa\,=\,\dfrac1{1\,+\,(m_ua/\kappa_u)}\,.\eequ
and each $c_{n(b)}$ has at least one factor of $\kappa$.

In the two above square brackets, the $c$'s are coefficients (with a possible dependence on $a$, $\kappa$) and the $M$'s are {\sl coefficient one} monomials in the Fermi fields. Expand the products and use the basic fact to perform the fermionic integrals. In this way, we obtain a numerical partition function. Bounding each of the $c$' s by its absolute value, we obtain the upper stability bound on the fermionic $Z_\Lambda$ 
\bequ\lb{ubZf}\barr{lll}
|Z_\Lambda|&\leq & \left[\prod_{b\in\Lambda}\,\left( 1\,+\,\sum_{n(b)}\,|c^b_{n(b)}|\right) \right]\:\left[\prod_{s\in\Lambda}\,\left( 1\,+\,\sum_{m(s)}\,|c^s_{m(s)}|\right) \right]\vspace{3mm}\\
&\leq &\exp\left[\Lambda_b\,\ln \left(1\,+\,C_b \right) \right]\:\exp\left[\Lambda_s\,\ln \left(1\,+\,C_s \right) \right]\,,\earr
\eequ
with $C_b\,\equiv\, \max_{b\in\Lambda}\,\sum_{n(b)}|c^b_{n(b)}|$ and $C_s\,\equiv\, \max_{s\in\Lambda}\,\sum_{m(s)}\,|c^s_{(s)}|$.

For the bound to be independent of the lattice spacing $a\in(0,1]$, renormalization of the coefficients may be required. For large enough $(m_ua/\kappa_u)$ (hence, a small scaled hopping parameter $\kappa$, as in Eq. (\ref{kf}), after dividing by the integral of $e^{V(\tilde\psi)}$, which factorizes over sites, the free energy and correlations admit convergent polymer expansion and the thermodynamic limit. 

Returning to the Yang-Mills model, we take periodic conditions and apply the multireflection method. We analyzed the generating function of $r\in\mathbb N$ of the above scaled plaquette field correlations. It is given by (here, $Z^P_\Lambda$ denotes the measure normalization and the $J$' s are source strengths)
$$
\barr{lll}
\left\langle \exp\left\{ \sum_{j=1}^r\,J_j\,\tr\,{\mathcal F}^s_{p_j}(x_j)\right\}  \right\rangle&=&\dfrac 1{Z^P_\Lambda}\,\dis\int\,\exp\left\{ \sum_{j=1}^r\,J_j\,\tr\,{\mathcal F}^s_{p_j}(x_j)\,-\,\dfrac{a^{d-4}}{g^2}\,\dis\sum_{p\in\Lambda}\,A_p(U_p)\right\}\,d\mu(U)\vspace{2mm}\\&
\equiv& G_{r,\Lambda}(J^{(r),x})\,\equiv G_{r}(\Lambda,a,J^{(r),x})\,.
\earr
$$
Here, $p_j$ is a shorthand for $p_{\mu_j\nu_j}(x_j)$. We proved that $G_{r,\Lambda}(J^{(r),x})$ is uniformly bounded in the $x$'s and $J$'s and possess a finite thermodynamic and continuum limits, at least in the subsequential sense. Moreover, the generating function extends to an entire function of the $J$'s and Cauchy bounds are applied to derive bounds on $r\in\mathbb N$ scaled plaquette field correlations. In particular, our analysis shows that the two unscaled plaquette field correlation presents a singular behavior which is bounded by $a^{-d}$, which is less than or equal to the singular behavior of the UV asymptotic freedom behavior. In particular, in $d\,=\,4$, the unscaled coincident point two-plaquette correlation is equal to $a^{-d}\,M(g)$, with $M(g)$ uniformly bounded for $g^2\in(0,\,g_0^2<\infty]$.

In deriving the above results, we have obtained a new global upper bound on the Wilson plaquette action, which is quadratic in the gluon fields. This is {\sl not} what happens in the classical Lagrangian version of the model, which was used in the analysis of Yang-Mills models in Refs. \cite{MRS,Hairer}, and is also a surprise here,  since the small $a$ naive approximation has positive quartic terms. This quadratic bound on the Wilson plaquette action is used to obtain a lower bound on the partition function.

Furthermore, the bound on the  partition function factorizes and each factor is the partition function of a single plaquette action with a single bond
partition function. The factorization occurs by setting to zero the actions of the spatial plaquettes. Such a local factorization does not occur in typical bosonic and fermionic models. The factorization suggests an avenue for a cluster or polymer expansion. In this case, the unperturbed partition function has only temporal plaquettes (plaquettes with at least one bond in the temporal direction) and the perturbation only has spatial plaquettes.

Concerning the integral over the gauge group Haar measure of the single bond partition function, the integrand is a class function. In this case, with the help of the Weyl' s integration formula, the $N^2$ dimensional group integral can be reduced to an $N$-dimensional integral over the angular eigenvalues.During this process, the random matrix probability distributions of the the circular unitary ensemble (CUE) and the Gaussian unitary ensemble (GUE)  
appear.

An open and important question is whether or not the class of Yang-Mills models we analyzed here has a mass gap in the spectrum.

Lastly, we hope that our methods and techniques, combined with other methods, will be useful to accomplish a complete construction of the $d=4$ Yang-Mills and QCD models, including the verification of the Osterwalder-Schrader axioms.
%%%%%%%%%%%%%%%%%%%%%%%%%%%%%%%%%%%%%%%%%%%%%%%%%%%%%%%%%%%%%%%%%%%%%%%%%%%%%
%%%%%%%%%%%%%%%%%%%%%%%%%%%%%%%%%%%%%%%%%%%%%%%%%%%%%%%%%%%%%%%%%%%%%%%%%%%%%
\appendix{\begin{center}{\bf APPENDIX A: Unscaled (Physical) and Scaled Real Scalar Free Field Model}\end{center}}
%%%%%%%%%%%%%%%%%%%%%%%%%%%%%%%%%%%%%%%%%%%%%%%%%%%%%%%%%%%%%%%%%%%%%%%%%%%%%
%%%%%%%%%%%%%%%%%%%%%%%%%%%%%%%%%%%%%%%%%%%%%%%%%%%%%%%%%%%%%%%%%%%%%%%%%%%%%
%%%%%%%%%%%%%%%%%%%%%%%%%%%%%%%%%%%%%%%%%%%%%%%%%%%%%%%%%%%%%%%%%%%%%%%%%%%%%%
\lb{appA}
\setcounter{equation}{0}
\setcounter{lemma}{0}
\setcounter{thm}{0}
\setcounter{rema}{0}
\renewcommand{\theequation}{A\arabic{equation}}
\renewcommand{\thethm}{A\arabic{thm}}
\renewcommand{\thelemma}{A\arabic{lemma}$\:$}
\renewcommand{\therema}{{\em{A\arabic{rema}$\:$}}}\vspace{.5cm}
%%%%%%%%%%%%%%%%%%%%%%%%%%%%%%%%%%%%%%%%%%%%%%%%%%%%%%%%%%%%%%%%%%%%%%%%%%%%%
%*************************************************************************
%%%%%%%%%%%%%%%%%%%%%%%%%%%%%%%%
Here, we consider the case of an one-component ($N=1$) real free scalar field and obtain the relation between the physical, unscaled partition function and the scaled partition function. Correlations are also analyzed. Considering the scaled partition function and correlations, the thermodynamic and continuum limits are obtained. These quantities also admit a convergent expansion in the scaled hopping parameter $\kappa^2$. The convergence is absolute, up to and including the critical point $\kappa^2\,=\,\kappa_c^2\,=\,(1/2d)$. The use of scaled fields removes the UV divergences from the scaled free energy and correlations, even at coincident points. The use of free b.c. is proved to provide infrared (IR) regularization and no global local mass term is needed. The relation between the original physical, unscaled fields and scaled field correlations at coincident points provides a characterization of UV asymptotic freedom. 

Our analysis here is given in more detail in Appendix A of Ref. \cite{bQCD}, where we considered $N$ component fields and to which we refer the reader. Concerning the lattice $\Lambda$, we use the same as in the text and recall that $\Lambda_s\,=\,L^d$, $L$ even, is the total number of sites.\vspace{2mm}

\subsection{Partition Function:}\vspace{2mm}
We consider the hypercubic lattice $\Lambda\,\subset\,a\mathbb Z^d\subset\mathbb R^d$, $a\in(0,1]$, with $L\in\mathbb N$, $L$ even sites on a side and  periodic boundary conditions. The total number of sites in $\Lambda$ is denoted by $\Lambda_s\,=\,L^d$. For the real scalar field model, the physical or unnormalized finite lattice partition function is (the lowercase/uppercase index $u$ denotes unscaled)
\bequ\lb{scpart}
Z^u_\Lambda\,=\,\dis\int\,e^{-A^u(\phi^u)}\,D\phi^u\,,
\eequ
where $D\phi^u\,=\,\prod_{x\in\Lambda}d\phi^u(x)/\sqrt{2\pi}$, with a Lebesgue measure $d\phi^u(x)$ for the unscaled field at each lattice site. Also, up to boundary terms, for $\mu=0,\ldots,(d-1)$ denoting a lattice spacetime direction and for any site $x\in\Lambda$, the model action is given by (as before $x^+_\mu\,=\,x\,+\,ae^\mu$, $e^\mu$ being the unit vector of the $\mu$ spacetime direction)
\bequ\lb{scaction}
\barr{lll}
A^u(\phi^u)&=&\dfrac12\, \kappa_u^2 a^{d-2}\,\sum_{x,\mu}\,\left( \phi^u(x_\mu^+)\,-\,\phi^u(x)\right)^2\,+\,\dfrac12\,m_u^2a^d\,\sum_{x}\,[\phi^u(x)]^2\vspace{2mm}\\
&=&-\kappa_u^2a^{d-2}\,\sum_{x,\mu}\,\left( \phi^u(x)\phi^u(x_\mu^+)\right)\,+\, \dfrac12\,\left(m_u^2a^d\,+\,2d\kappa_u^2a^{d-2}\right)\,\sum_{x}\,[\phi^u(x)]^2\,,
\earr
\eequ
where $\kappa_u,\,m_u\,>\,0$. The mass associated with this action and corresponding to the partition function is defined as the infinite time limit of the exponential decay rate of the two-point correlation. Equivalently, it is the first isolated point in the E-M spectrum of the associated lattice QFT, lying above the vacuum and at zero spatial momentum. We determine this mass in subsection 2 of this appendix.

The scaled fields $\phi(x)$ are related to the unscaled fields $\phi^u(x)$ by an $a$-dependent noncanonical scaling transformation which corresponds to an a priori renormalization procedure. It reads (in \cite{bQCD}, a similar equation has a misprint in the exponent of $a$!)
\bequ\lb{s}
\phi(x)\,=\, s(a)\,\phi^u(x)\quad;\quad s(a)\,=\, \left(2d\kappa_u^2a^{d-2}\,+\,m_u^2a^d \right)^{1/2}\,\geq\,\left(2d\kappa_u^2\right)^{1/2}\,a^{(d-2)/2}\,,
\eequ
which leads to the scaled field partition function
\bequ\lb{scscaledpart}
Z_\Lambda\,=\,\dis\int\,e^{-A(\phi)}\,D\phi\quad;\quad A(\phi)\,=\,A^u(\phi^u\,=\,s^{-1}\phi)\,,
\eequ
where $D\phi$ is defined similarly to $D\phi^u$.

For \bequ\lb{scaledkappa}\kappa^2\,=\, (2d\,+\,r)^{-1}\qquad,\qquad r\,=\, m_u^2a^2/\kappa_u^2\,,\eequ the scaled field action of Eq. (\ref{scscaledpart}) is given by
\bequ \lb{scact}
\barr{lll}
A(\phi)&=&-\kappa^2\,\sum_{x,\mu}\,\phi(x)\,\phi(x_\mu^+)\,+\,\dfrac12\,\sum_x\,[\phi(x)]^2\vspace{2mm}\\
&=&\dfrac{\kappa^2}2\,\sum_{x,\mu}\,\left(\phi(x^+_\mu)\,-\,\phi(x)\right)^2\,+\,\dfrac{\kappa^2r}2\,\sum_x\,[\phi(x)]^2\,.
\earr
\eequ

Thus, by a change of variables, the unscaled and scaled finite lattice partition functions are related by
\bequ\lb{Zrel} Z_\Lambda\,=\,s^{\Lambda_s}\, Z^u_\Lambda\,,\eequ
and the corresponding finite lattice free energies satisfy
\bequ\lb{relation}
f_\Lambda(a)\,=\, \ln s\,+\,f^u_\Lambda(a)\,,
\eequ
where, as made precise above and omitting the $a$-dependence, $f_\Lambda\,=\,\dfrac1{\Lambda_s}\,\ln Z_\Lambda$ and $f^u_\Lambda\,=\,\dfrac1{N_s}\,\ln Z^u$. Eq. (\ref{relation}) above, tells us that the singularities of $f^u_\Lambda(a)$, when $\Lambda\nearrow a\mathbb Z^d$ and, subsequently, in the $a\searrow$ limit are killed by the $\ln s$ term. Thus, we have isolated the divergences of $f^u_\Lambda(a)$.

Indeed, as it is proved in Ref. \cite{bQCD}, we have the stability bounds for the scaled partition function of Eq. (\ref{Zrel})
\bequ\lb{stab1}e^{c_\ell  \Lambda_s}\leq Z_{\Lambda,a}\leq e^{c_u \Lambda_s}\,,
\eequ
so as, by the Bolzano-Weierstrass theorem \cite{Rudin}, the scaled free energy $f_\Lambda(a)$, in the thermodynamic limit $\Lambda\nearrow a\mathbb Z^d$ followed  by the continuum continuum limit $a\searrow 0$ is bounded. Letting $f$ denote the result of these two limits, with these considerations, we have
\bequ\lb{stabfscaledf}
c_\ell\,\leq\,f\,\leq\,c_u\,.\vspace{3mm}
\eequ

We now derive and sum a power series representation, in the scaled hopping parameter $\kappa^2$, for the free energy. The series is proved to converge absolutely up to and including the critical value $\kappa^2\,=\,\kappa_c^2\,=\,(1/2d)$.

From Appendix A of Ref. \cite{bQCD}, for $d=1,2,3,4$, we have the momentum representation
\bequ\lb{repf}
f(a)\,=\,-\,\dfrac1{2(2\pi)^d}\,\dis\int_{(-\pi,\pi]^d}\,\ln\left[ 1\,-\,2\kappa^2\sum_\mu\cos q^\mu\right]\,d^dq\,.
\eequ

Recalling that the scaled hopping parameter $\kappa^2$ depends on $a$ [see Eq. (\ref{scact})], the integral in Eq. (\ref{repf}) is proved to be uniformly bounded for $a\in(0,1]$ and also, by the Lebesgue dominated convergence theorem, that the limit $a\searrow 0$ exists and is equal to the $a=0$ value of the r.h.s. of Eq. (\ref{repf}).
\begin{rema}
For $d=1$, the integral in Eq. (\ref{repf}) gives the closed form expression $$f(a)\,=\,-\,\dfrac12\,\ln\left[\dfrac{1\,+\,\sqrt{1\,-\,4\,\kappa^4}}{2} \right]\,.$$ At the critical point, $\kappa^2\,=\,(1/2)$, $f(a)$ remains bounded and takes the value $\ln\sqrt{2}$.
\end{rema}

Expanding the logarithm in the integrand of Eq. (\ref{repf}), and using the multinomial theorem, we obtain
\bequ\lb{logexp}
f(a)\,=\,\dfrac12\,\sum_{r\geq 1}\,\dfrac {(2\kappa^2)^r}r \sum_{r_1,\ldots,r_d|\sum_j r_j=r}\;\dfrac{r!}{\,r_1!\ldots r_d!}\:\prod_{j=1,\ldots,d}\left[\dfrac1{2\pi}\,\dis\int_{(-\pi,\pi]}\,\cos^{r_j}q\,dq\right]\,.
\eequ

We claim that the series is absolutely convergent, for $\kappa^2\,<\,\kappa_c^2\,=\,(1/2d)$. To see this, bound the integral factors inside the product by one and then sum over the $r_j$ to get $d^r$. Next, summing over $r$ gives $[\,(-1/2)\,\ln (1\,-\,2 \kappa^2d)\,]$. This is bounded for $\kappa^2<(1/2d)$ which proves our claim.

Using Ref. \cite{grad}, Chap. 3, for the trigonometric integrals gives, for $0\leq r_j\leq r$, $j=1,\ldots,d$,
\bequ\lb{logexp2}
f(a)\,=\,\dfrac12\,\sum_{r\geq 2,\,{\mathrm even}}\,\dfrac {(2\kappa^2)^r}r \sum_{r_1,\ldots,r_d,\,{\mathrm even}|\sum_j r_j=r}\;\dfrac{r!}{r_1!\ldots r_d!}\:\prod_{j=1,\ldots,d}\dfrac{(r_j-1)!!}{r_j!!}\,,
\eequ
with the convention $0!!\,\equiv\,1$ and $(-1)!!\,\equiv\,1$.

Note that all the coefficients of the above series are positive. As $\kappa^2$ is a monotone increasing function of a monotonically decreasing $a$, by the monotone convergence in the space $\ell_1$ of the counting measure, and by the Lebesgue dominated convergence theorem, the series in Eq. (\ref{logexp2}) converges absolutely for all $0\,\leq\,\kappa^2\,\leq\,\kappa_c^2\,=\, (1/2d)$. Denoting by $f$ the $a\searrow 0$ limit of $f(a)$, we have the exact result
\bequ\lb{f}
f\,=\,\dfrac12\,\sum_{r\geq 2,{\mathrm even}}\,\dfrac {1}{r\,d^r} \sum_{r_1,\ldots,r_d,\,{\mathrm even}|\sum_j r_j=r}\;\dfrac{r!}{r_1!\ldots r_d!}\:\prod_{j=1,\ldots,d}\dfrac{(r_j-1)!!}{r_j!!}\,.\vspace{2mm}
\eequ

\subsection{Correlations:}\vspace{2mm}

We now consider correlations of the real free field scalar case. We obtain the relation between unscaled and scaled correlations and prove their thermodynamic and continuum limits exist. We also analyze derivative field correlations. Explicit momentum space representations are also derived. The previous treatment of the partition function is widely used here.

First, we obtain the relation between the physical, unscaled two-point correlation and its scaled counterpart. Recalling Eqs. (\ref{scpart}) and (\ref{Zrel}), we have, with $s\equiv s(a)$,
\bequ\barr{lll}\lb{correlphi}
\langle \phi^u(x)\phi^u(y)  \rangle^u&\equiv&\dfrac 1{Z^u}\,\dis\int\, \phi^u(x)\,\phi^u(y)\,\exp\left\{-A^u(\phi^u)\right\}\, d\phi^u\vspace{2mm}\\
&=&\dfrac 1{s^2\,Z^u}\,\dis\int s\phi^u(x)\,s\phi^u(y)\,\exp\left\{ -A^u(\phi^u) \right\} \, d\phi^u\vspace{2mm}\\
&=&\dfrac 1{s^2Z}\,\dis\int \phi(x)\phi(y)\,\exp\left\{-A(\phi)\right\}\, d\phi\vspace{2mm}\\
&=&\dfrac 1{s^2}\,\langle \phi(x)\phi(y)\rangle\,,
\earr\eequ
where we changed from $\phi^u(x)$ to $\phi(x)\,=\,s\phi^u(x)$ and also set, following Eqs. (\ref{scaction}) and (\ref{scact}), $$A(\phi)\,=\,A^u(\phi^u=s^{-1}\phi)\,.$$

Similarly, recalling that $x_\mu^+\,=\, x\,+\,ae^\mu$ and with $\delta_\mu(x)\,=\, \phi(x_\mu^{+})\,-\,\phi(x)$, we have,
\bequ\lb{correldphi}
\langle \partial^a_\rho\phi^u(x) \partial^a_\nu\phi^u(y)\rangle^u\,=\,\dfrac1{(as)^2}\,\langle \delta_\rho\phi(x) \delta_\nu\phi(y)\rangle\,.
\eequ

As proved in Ref. \cite{bQCD}, the scaled two-point correlation $\langle \phi(x)\phi(y)\rangle$ at the thermodynamic limit admits the momentum space representation given by
\bequ\lb{2pfscalar}\barr{lll}
\langle \phi(x)\phi(y) \rangle&=& \dfrac1{(2\pi)^d}\,\dis\int_{(-\pi,\pi]^d}\,
\dfrac{e^{iq(x-y)/a}}{ \left(1\,-\,2\kappa^2\,\sum_\mu\,\cos q^\mu\right)}\,d^dq\vspace{2mm}\\
&=& \dfrac1{2(2\pi)^d\kappa^2}\,\dis\int_{(-\pi,\pi]^d}\,
\dfrac{e^{iq(x-y)/a}}{ \sum_\mu\,\left(1\,-\,\cos q^\mu\right)\,+\,r/2}\,d^dq\,.\earr
\eequ
\begin{rema}\lb{remaa2}
	For $d=1$, the above integral gives the closed form expression $$\langle \phi^2(x)\rangle\,=\,\left(1\,-\,4\kappa^4\right)^{-1/2}\,=\,1\,+\,\sum_{r\geq 2,\, \mathrm{even}}\,\dfrac{(r-1)!!}{r!!}\,\left(2\kappa^2\right)^r\qquad,\qquad \kappa^2\,<\,1/2\,,$$
which is infinite at the critical point $\kappa^2\,=\,1/2$.
\end{rema}
\begin{rema}\lb{unifcs}
From Eq. (\ref{2pfscalar}), for $d\,=\,3,4$, we that for $x=y$, the two-point function is bounded from above and below, uniformly in $a\in(0,1]$. Indeed, 
a lower bound is obtained by replacing $\kappa^2$ by its maximum value $(1/2d)$ and putting $r=m_u^2/\kappa_u^2$ (the maximum of $r$) in the denominator of the integrand, resulting in a finite integral. For an upper bound, we set $r=0$ in the integrand and take $a=1$ in $\kappa^{-2}$ [see Eq. (\ref{scaledkappa})].
\end{rema}

From Eq. (\ref{2pfscalar}), expanding the denominator of the first equality and using the multinomial expansion on each term, for coincident points $x=y$, we obtain, with $0\leq r_j\leq r$, $j=1,\ldots,d$,
\bequ\lb{seriescoinc}\barr{lll}
	\langle [\phi(x)]^2 \rangle&=&1\,+\,\dis\sum_{r\geq 1}\,(2\kappa^2)^r\;\sum_{r_j,\;{\mathrm even}|\sum_{j=1}^d\,r_j=r}\:\dfrac{r!}{r_1!\ldots r_d!}\;
	\prod_{j=1}^d\,\left[\dfrac1{2\pi}\,\dis\int_{-\pi}^\pi\,\cos^{r_j}q\;dq    \right]\,.
	%\vspace{3mm}\\
	%&=&1\,+\,\dis\sum_{r,\,{\mathrm %even}}\,(2\kappa^2)^r\;\sum_{r_j,\;{\mathrm %even}|\sum_{j=1}^d\,r_j=r}\:\dfrac{r!}{r_1!\ldots r_d!}\;
	%\dfrac{(r_j-1)!!}{r_j!!}\,.
	\earr
	\eequ 
At this stage, we see that all coefficients of powers of $\kappa^2$ are positive. The series is bounded for $0\,<\,\kappa^2\,< (1/2d)$ as the integrals are bounded by $(2\pi)$. Using these bounds and resumming, we get
$$
\langle [\phi(x)]^2 \rangle\,\leq\,\sum_{r=0}^\infty\,(2\kappa^2d)^r\,=\,\dfrac1{1\,-\,2\kappa^2d}\,.
$$

Now, from (see e.g Ref. \cite{grad}, Chap. 3),
$$
\int_{-\pi}^\pi\, \cos^{2m}x\,dx\,=\,4\,\int_0^{\pi/2}\, \cos^{2m}x\,dx\,=\,2\pi\;\dfrac{(2m-1)!!}{(2m)!!}\,,
$$
so that letting $2m=r_j$, and recalling the convention $0!!\,\equiv\,0$ and $(-1)!!\,\equiv\,1$, we  have
\bequ\lb{2pfscalarcoinc}
\langle [\phi(x)]^2 \rangle\,=\,1\,+\,\sum_{r\geq 2,\,{\mathrm even}}\,(2\kappa^2)^r\;\sum_{r_j,\;{\mathrm even}|\sum_{j=1}^d\,r_j=r}\:\dfrac{r!}{r_1!\ldots r_d!}\;\,\,\prod_{j=1}^d\,
\dfrac{(r_j-1)!!}{r_j!!}\,.
\eequ
The series is monotone increasing in $\kappa^2$. In Eq. (\ref{2pfscalarcoinc}), if we bound the product by one, we see that the series converges absolutely for all $0\,\leq\,\kappa^2\,<\,\kappa^2_c\,\equiv (1/2d)$. By the upper bound given below, for $d>2$, we can extend the $\kappa^2$ convergence domain up to and including $\kappa^2_c$.
\begin{rema} 
By the upper bound below, the convergence of the series in Eq. (\ref{2pfscalarcoinc}) is uniform in the lattice spacing $a\in(0,1]$. The continuum limit exists, up to and including $\kappa^2_c$, for $\langle [\phi(x)]^2 \rangle$ setting $2\kappa^2=1/d$ in Eq. (\ref{2pfscalarcoinc}) and considering $d>2$. Also, the zero mass limit and the continuum limit are interchangeable. Recalling Eq. (\ref{s}) and the scaled/unscaled relation $\langle [\phi^u(x)]^2 \rangle^u\,=\,s^{-2}\,\langle [\phi(x)]^2 \rangle$, and taking $m_u^2\,=\,0$ in $s\,\equiv\,s(a)$, we have the zero mass limit of $\langle [\phi(x)]^2 \rangle$.
\end{rema}
\begin{rema}
Comparing Eq. (\ref{2pfscalarcoinc}) with Eq. (\ref{logexp2}), for the thermodynamic limit of free energy, we have that \bequ\lb{df}\kappa^2\frac{df}{d\kappa^2}\,=\,[\langle\phi^2(x)\rangle\,-\,1]/2\,.\eequ Hence it is enough to determine only one expansion, for $f$ or for $\langle\phi^2(x)\rangle$. We remark that this equation gives a relation between the derivative of a global quantity in terms of a local average. This is a more general result. Consider a finite lattice $\Lambda$ model partition function defined with a quadratic form $Z_\Lambda\,=\,\int\,\exp\{-\frac12\,(\phi,A\phi) \}\;D\phi$, with a symmetric matrix $A>0$, $(\phi,A\phi)\,=\,\sum_{x,y\in\Lambda} \phi(x) A_{xy}\phi(y)$ and a product Lebesgue measure $D\phi$, such that we have
$Z_\Lambda\,=\,\exp\{-\frac12\,\tr \ln A\}$. Hence, the associated finite free energy is
$$
f_\Lambda\,=\,\dfrac{\ln Z_\Lambda}{\Lambda_s}\,=\, -\dfrac1{2\Lambda_s}\,\tr\ln A\,.
$$
On the other hand, for $A\,=\,1\,-\,\lambda M$ and assuming translation invariance, even at finite volume, expanding the log for small enough $|\lambda|$,
\bequ\lb{Nr}f_\Lambda\,=\,\frac12\,\sum_{n\geq 1}\,\dfrac1n\;(\lambda M)^n_{00}\,,\eequ
and
\bequ\lb{Nrr}
\langle \phi^2(x)  \rangle_\Lambda \,=\, [1\,-\,\lambda M]^{-1}_{xx}\,= \,\sum_{n\geq 0}\,(\lambda M)^n_{00}\,.\eequ
From the last two equations, we obtain
$$\lambda\,\dfrac{df_\Lambda}{d\lambda}\,=\,\dfrac12\,\left[ \langle \phi^2(x)\rangle\,-\,1\right]\,.$$
This is an equality of the same type as in Eq. (\ref{df}). Likely, this result can be extended to the case when translation invariance is present only in the thermodynamic limit of a lattice model.
\end{rema}

We now prove existence of $\langle [\phi(x)]^2 \rangle$ at $\kappa^2\,=\,\kappa^2_c\,=\,(2d)^{-1}$. From the second equality in Eq. (\ref{2pfscalar}), using the basic lower bound $(1\,-\,\cos u)\,\geq\,2u^2/\pi^2$, $u\in(-\pi,\pi]$, the upper bound $\kappa^{-2}\,\leq\,2d\,+\,(m_u/\kappa_u)^2$ and spherical coordinates, we get, for $d>2$,
\bequ\lb{bbb}
\langle \phi(x)\phi(y) \rangle\,\leq\,\dfrac{\pi^2\Omega_d}{4(2\pi)^d}\; \left[ 2d\,+\,(m_u/\kappa_u)^2 \right]\;\dfrac{(\sqrt{d}\pi)^{d-2}}{d-2}\,,
\eequ
where $\Omega_d$ is the $d$-dimensional solid angle and $[\rho(q)]^2\,=\, \sum_\mu\,(q^\mu)^2$. The last numerical factor arises from the radial integral by expanding the integration domain to $0\,\leq\,\rho\,\leq\sqrt{d}\pi$. Notice that the bound of Eq. (\ref{bbb}) holds for all $a\in(0,1]$ and $0,\leq\,\kappa^2\,\leq \kappa_c^2$.

Coming back to the spectral representation of Eq. (\ref{2pfscalar}),
the integrand is dominated, for $a\in(0,1]$ and $d\,=\,3,\,4$, by the integrable function  $\{1/[\sum_{\mu=0,1,...,(d-1)}\,(1\,-\,\cos q^\mu)]\}$. Recalling that $\kappa^{-2}\,=\,[2d\,+\,(m_u\,a/\kappa_u)^2]$, we have
$$
\lim_{a\searrow 0}\,\langle  \,\phi(x)\,\phi(y)\,\rangle\,=\,\left\{\barr{c} \langle \phi^2(0)\rangle_0\quad,\quad x= y;\vspace{1mm}\\0\quad,\quad x\not= y\,.\earr \right.\,,
$$
by the dominated convergence theorem, for $x=y$, and the Riemann-Lebesgue lemma for $x\not= y$. Here, 
$$
\langle \phi^2(0)\rangle_0\,(a)=\, \dfrac d{(2\pi)^d}\,\dis\int_{(-\pi,\pi]^d}\,\dfrac1{\sum_{\mu=0,1,\ldots,(d-1)}\,\left(1\,-\,\cos q^\mu\right)}\,d^dq\,=\,2d\,\Delta_1^{-1}(0,0)\,,$$ denote the zero mass ($r\,=\,0$) coincident point value of $\langle \,\phi(y)\rangle_0$, in the thermodynamic limit, and $\Delta_1^{-1}(x,y)$ is the kernel of the inverse of minus the unit lattice Laplacian operator, and $\Delta_1$ acts on $f\,\in\,\ell_2(\mathbb Z^d)$ by
\bequ\lb{DDelta}[\Delta_1 f](x)\,=\,2df(x)\,-\,\sum_\mu[f(x^{\mu+}=x+e^\mu)\,+ \,f(x^{\mu-}=x-e^\mu)]\,,\eequ
where, here, the lattice site $x\in\mathbb Z^d$. The integrand is integrable. Thus, 
\bequ\lb{bdb}
\langle\,\phi(x)\,\phi(y)\rangle\,\leq\,\dfrac{2d\,+\,(m_ua/\kappa_u)^2}{2d}\,
\langle\,\phi^2(0)\,\rangle_0\,\leq\,\dfrac{2d\,+\,(m_u/\kappa_u)^2}{2d}\:\Delta^{-1}(0,0)\,,
\eequ
and is bounded uniformly for $a\in(0,1]$.

We now obtain a series expansion for scaled scalar field lattice two-point correlation  $\langle \phi(0)\phi(x)\rangle_\Lambda$, in the thermodynamic limit. 
Expanding Eq. (\ref{2pfscalar}) as before, we obtain, writing $x\,=\,n_xa$. $n_x\,=\,(n_{x,1},\ldots,n_{x,d})\in \mathbb Z^d$ and with a Kronecker delta,
\bequ\lb{x}
\langle \phi(0)\phi(x)\rangle\,=\, \delta(x)\,+\, \sum_{r\geq 1}\,(2\kappa^2)^r\,\sum_{r_1,\ldots,r_d|\sum_jr_j=r}\,\dfrac{r!}{r_1!\ldots r_d!}\,\prod_{j=1,\ldots,d}\,\left[\dfrac1{2\pi}\,\dis\int_{-\pi}^{\pi}\,\cos(n_jq_j) \,\cos^{r_j}q_j\;dq_j\right]\,.
\eequ
Here, the $j$th integral with $n_j=m$, $r_j=s$, is
$$\barr{lll}
\dfrac1{2\pi}\,\dis\int_{-\pi}^{\pi}\,\cos(mq) \,\cos^sq\;dq&=&
%\vspace{2mm}\\&=& 
\dfrac1{2^{s+1}\pi}\,\dis\int_{-\pi}^{\pi}\,\,\left\{\left[\dfrac{e^{imq}\,+\,e^{-imq}}2\right]\;\,\sum_{k=0,\ldots,s}\,\left[\dfrac{s!}{k!(s-k)!}\,e^{iq(2k-s)} \right]\right\}\,dq\,.
\earr
$$
Substituting this result in Eq. (\ref{x}) gives
\bequ\lb{series}\barr{lll}
\langle \phi(0)\phi(x)\rangle&=& \delta(x)\,+\,\sum_{r\geq 1}\:(2\kappa^2)^r\,\sum_{r_1,\ldots,r_d|\sum_jr_j=r}\,\dfrac{r!}{r_1!\ldots r_d!}\,\prod_{j=1,\ldots,d}\,\left\{\dfrac1{2^{r_j+1}}\,\sum_{k_j=0,\ldots,r_j}
\right.\vspace{2mm}\\&&\left.\times\;\;\dfrac{r_j!}{k_j!(r_j-k_j)!}\;\left[ \delta(2k_j-r_j+n_j)\,+\,\delta(2k_j-r_j-n_j) \right] \right\}\vspace{2mm}\\
&=&\delta(x)\,+\,\sum_{r\geq 1}\,\kappa^{2r}\,\sum_{r_1,\,\ldots,\,r_d\;|\;\sum r_j=r}\;\left(\barr{c}  r\vspace{2mm}\\r_1\;r_2\;\ldots\;r_d\earr\right) \vspace{2mm}\\&&\times\;\prod_{j=1,\ldots,d}\, \left\{\,\sum_{k_j=0,\ldots,r_j}\,N_{r_j,k_j}\;\dfrac12\,\left[\delta(2k_j-r_j+n_j)\,+\,\delta(2k_j-r_j-n_j)\right] \right\}\,.\earr
\eequ
The second equality above is obtained as follows. In $d=1$, the number of paths of length $n$ that start at the origin and end at the site $k$ is
$$
N_{n,k}\,\equiv\,\left( \barr{c}n\vspace{1mm}\\k\earr\right)\,=\,\dfrac{n!}{k!\,(n-k)!}\,.
$$ 
For $n_j=0$, for any $j=1,\,\ldots,\,d$, such as the path terminates at the origin, $r_j\,=\,2k_j$ is even and $$N_{r_j,k_j}\,=\,N_{r_j, r_j/2}\,=\,\dfrac{r_j!}{\left[(r_j/2)!\right]^2}\,=\,N_r\,.$$

The series in Eq. (\ref{series}) converges absolutely for $0\leq\kappa^2\leq\kappa_c^2=(1/2d)$. For the point $x=n_1a$, $n_1\not= 0$, the smallest exponent of $r$ is $|n_1|$; for the point $x=(n_1a,0,\ldots,0)$, $n_1=1$, the leading contribution is of order $\kappa^2$.

For the derivative field correlation $\langle \delta_\mu\phi(x)\delta_\nu\phi(y) \rangle$ a spectral representation can be obtained from $\langle \phi(x)\phi(y)\rangle$ by including an additional factor $[(e^{iq^\mu}-1)\,(e^{-iq^\nu}-1)]$ in the integrand, and we show that, with no sum on $\rho$,
$$
\langle \delta_\rho\phi(0)\,\delta_\rho\phi(0) \rangle_0 \,=\,\left(\dfrac1{\kappa^2d}\right)(r=0)\,=\,2\,,
$$
We have an exact value of the coincident point derivative physical field correlation 
$$
\langle \partial^a_\rho\phi^u(0)\partial^a_\rho\phi^u(0) \rangle_0^u \,=\,\dfrac{2}{a^2\,s^{2}(m_\mu=0)}\,=\,\dfrac{2}{a^d\left(2d\kappa_u^2\right)^{d-2}}\,.
$$
and the physical derivative field correlation verifies the bound
$$
|\langle \partial^a_\rho\phi^u(0)\partial^a_\rho\phi^u(0) \rangle_0^u |\,\leq\,\dfrac1{(as)^2}\,\langle \delta_\rho\phi(0)\,\delta_\sigma\phi(0) \rangle_0\,\leq\,2{(as)^{-2}}\,\leq\,\dfrac2{a^{d}\,\left(2d\kappa_ u^2 \right)^{d-2}}\,.
$$
which shows that the singular behavior is $a^{-d}$.

With similar arguments for the unscaled two-point correlation, we have the thermodynamic limit
$$
\langle \phi^u(x)\phi^u(y)\rangle^u_a\,=\,\dfrac1{2(2\pi)^d} \,\dis\int_{(-\pi/a,\pi/a]^d}\,
\dfrac{e^{ip(x-y)}}{\dfrac{\kappa_u^2}{a^2}\sum_\mu \left(1\,-\,\cos ap^\mu\right)\,+m_u^2/2}\,d^dp\,.
$$

Thus, from Eq. (\ref{2pfscalar}), even at coincident points, and for $d=3,4$, the scaled two-point correlation is bounded uniformly in $a\in(0,1])$. Using the scaling relation between the unscaled, physical field and the scaled field correlations, the physical two-point correlation behaves as $a^{2-d}$ at coincident points and the derivative field two-point correlation singular behavior is $a^{-d}$. As mentioned above, these behaviors characterize the free field and can be taken as a measure of UV asymptotic freedom on the lattice.

The mass $m$ associated with the $a\mathbb Z^d$ lattice QFT is given by the zero of the Fourier transform of the convolution inverse of the two-point correlation $\langle \phi^u(x)\phi^u(y)\rangle^u_a$ at zero spatial momentum, namely $\tilde \Gamma(p^0=im, \vec p=0)$. It corresponds to the solution of the equation
$$
2\,\dfrac{\kappa_u^2}{a^2}\,\left[ 1\,-\,\cosh (ma)\right] \,+\, m_u^2\,=\,0\,,
$$
or, using the equality $(\cosh u\,-\,1)\,=\,2\,\sinh^2 (u/2)$, $u\in\mathbb R$,
$$
\sinh^2 \dfrac {ma}{2}\,=\, \dfrac{m^2_u a^2}{4\kappa_u^2}\,.
$$

With this, we see the solution is, with $r\,=\,(m_ua/\kappa_u)^2$,
$$\barr{lll}
m&=& \dfrac2a\,\sinh^{-1}\left(\dfrac{m_ua}{2\kappa_u} \right)\,
=\,\dfrac2a\,\,\ln\left[\,\dfrac{\sqrt r}{2} \,+\,\dfrac{\sqrt {4+r}}{2}\,\right]\vspace{3mm}\\
&=& \dfrac{m_u}{\kappa_u}\,+\, \mathcal{O}\left(a^2 \left(\dfrac{m_u}{\kappa_u}\right)^3 \right)\,.
\earr
$$

The mass $m$ is real analytic in $a$, $m_u$ and $\kappa_u\not= 0$.

\subsection{Scaled Two-Point Correlation: Random Walk Expansions, Number of Paths and Spectral Representations:}\vspace{2mm}

We would be remiss if we did not establish the connection with random walk expansions for the scaled two-point correlation $\langle \phi(x)\,\phi(y)\rangle$. We give a description of some of our findings. For a simple random walk in dimension one, it is well-known that the number of paths starting and ending at the origin, and of length $n$, even, is \cite{LL}
\bequ\lb{Xi}
\Xi_n\,=\,\dfrac{n!}{[(n/2)!]^2}\,.
\eequ
As shown below in Eq. (\ref{doubles}), in dimension $d>1$, the number of paths is given by,
\bequ\lb{NR}
N_r\,=\,\sum_{r_1,\ldots,r_d\,|\,\sum r_i=r\,,\;|\,r\:{\mathrm even}}\,\left(\barr{c} r\vspace{1mm}\\r_1\;r_2\,\ldots\,r_d   \earr\right)\,\prod_{j=1,\ldots,d}\,\Xi_{r_j}\,.\eequ

Furthermore, in the thermodynamic limit, the free energy
\bequ\lb{fexp}
f\,=\,\dfrac12\,\sum_{r\geq 2\;,\;{\mathrm even}}\,N_r\,\dfrac{\kappa^{2r}}r\quad;\quad\kappa^2\leq(1/2d),\;d=1,2,\ldots,
\eequ
and the local correlation 
\bequ\lb{corexp}
\langle \phi^2 (x)\rangle\,=\,1\,+\,\sum_{r\geq 2,\: \mathrm{even}}\,N_r\,\kappa^{2r}\qquad ;\qquad\kappa^2\leq(1/2d),\;d=3,4,\ldots
\eequ
Furthermore, \bequ\lb{relll}\kappa^2\,\dfrac{df}{d\kappa^2}\,=\,\frac12\,\left[ \langle   \phi^2(x)\rangle\,-\,1 \right]\,.\eequ
Thus, it turns out that the only input needed for these numerical formulas is the $d=1$ combinatorial factor $\Xi_n$. Whether or not this property is a consequence of a more fundamental behavior is to be investigated.

The number $[N_r/(2d)^r]\,\equiv\,p_r$ is the fraction of the total number of paths of length $r$, with initial and endpoint at the origin. Denoting by $\langle\phi^2(x)\rangle_c$ the value of $\langle\phi^2(x)\rangle$ at the critical point $\kappa^2\,=\,\kappa_c^2\,=\,(1/2d)$, we have
$$
\langle\phi^2(x)\rangle_c\,=\,1\,+\,\sum_{r\;\mathrm{even}\geq 2}\,p_r\,.
$$

For simplicity of notation, we do drop the $s$ (scaled) upper indices in this section. We write the free b.c. scaled field action of Eq. (\ref{scact}) as, with inner product in $\ell_2(a\mathbb Z^d)$,
$$
\dfrac12\,\left( \phi,[1\,-\,\kappa^2S]\,\phi \right)\,,
$$
with
$$ \langle \phi(x)\,\phi(y)\rangle\,=\, \left(1\,-\,\kappa^2S \right)_{xy}^{-1}\,=\,\delta_{xy}\,+\, \dis\sum_{r=1}^\infty\,\kappa^{2r}\,\left[\sum_{x_1,x_2,\ldots,x_{r-1}}\, S_{x,x_1}\,S_{x_1,x_2}\ldots\,S_{x_{r-1}y}\right]\,.
$$
Here, the elements of the symmetric matrix $S$ are given by
\bequ\lb{SSS}
S_{y_j,y_k}\,=\,\left\{\barr{lll}1&,&|y_j-y_k|\,=\,1,
\vspace{3mm}\\0&,&\mathrm{otherwise}\,.\earr\right.\,.\eequ
The above Neumann series converges absolutely for $|\kappa^2|\,<\,(1/2d)$, as $|S|\,=\,2d$.

Comparing this series with the series obtained from spectral representation, we can equate the $\kappa^{2r}$ coefficients. For example, for coincident points $x=y$, and for $r$ and $r_j$ ($j=1,\ldots,d$) being even integers, we have
\bequ\lb{doubles}\barr{lll}
\sum_{x_1,x_2,\ldots,x_{r-1}}\, S_{x,x_1}\,S_{x_1,x_2}\ldots\,S_{x_{r-1}x}&=&2^r\; \sum_{r_1,\,\ldots,r_d\in\{0,1,\ldots,r\}}\;\dfrac{r!}{r_1!\,r_2!\ldots r_d!} \;\,\prod_{j=1}^d\,\dfrac{(r_j-1)!!}{r_j!!}\vspace{4mm}\\&=& \sum_{r_1,\ldots,r_d\in\{0,1,\ldots,r\}}\:\,\left(\barr{c} r\vspace{2mm}\\r_1\:r_2\:\ldots\:r_d \earr  \right)\:\,\,\,\prod_{j=1}^d\,\Xi_{r_j}\qquad;\qquad\sum_{j=1}^d r_j\,=\,r\:,\earr
\eequ
where we have used the multinomial coefficient and used the number of paths starting and ending at zero, according to Eq. (\ref{Xi}).

\begin{rema}
The second equality above follows on writing $r\,=\,\sum_{j=1,\ldots,d}\, r_j$ in the exponent of $2^r$, and using double factorial identities, namely, for even $n$,
$$
\dfrac{2^n(n-1)!!}{n!!}\,=\,\dfrac{2(n-1)!}{(n/2-1)!\;(n/2)!}\,=\,\dfrac{n!}{[(n/2)!]^2}\,.
$$
In Ref. \cite{LL}, the number $\Xi_n$ is deduced by a combinatorial argument. The same result is obtained here by an analytic argument in Remark \ref{remaa2}.
\end{rema}
\begin{rema}
	Since, using Eq. (\ref{SSS}), the product in the left-hand-side of Eq. (\ref{doubles}) is $0$ or $1$, the right-hand-side of this equation is the counting of paths of length $r$ (even) that start and end at site $x=0$.
\end{rema}
\begin{rema}
From Eq. (\ref{Nr}), we see that the free energy satisfies Eq. (\ref{fexp}), for $d=1,2,3,4$ and $\kappa^2<(1/2d)$. Likewise, Eq. (\ref{Nrr}) tell us that $\langle   \phi^2(x)\rangle$ obeys Eq. (\ref{corexp}). Hence, we have the relation of Eq. (\ref{relll}). Note the only input for all this results is the combinatorial factor $\Xi_n$, for $d=1$.

On the other hand, for $d=1$, we also have $$f\,=\,-\frac12\,\ln\dfrac{1+\sqrt{1-4\kappa^4}}2\,,$$ $\kappa^2\leq\frac12$, so that $0\leq f\leq \ln\sqrt{2}$. In Appendix C, we show the inequality $f\,\leq\,\ln\sqrt{2}$, holds for any dimension $d$, provided that $\kappa^2\,\leq\,(1/2d)$. We also have the correlation inequality
$$
\langle   \phi^2(x)\rangle\,=\,\dfrac1{\sqrt{1-4\kappa^4}}\,\geq\,1\qquad;\qquad d=1\;\;,\;\;\kappa^2<1/2\,.
$$
 Whether or not there is a deeper reason for this bound to be true needs further investigation.
\end{rema}

Eq. (\ref{doubles}) is an explicit formula for the number of paths of length $r$ that begin at $x$ and terminate at the same point $x$. Note that if there are no restrictions on endpoints of the paths, then the r.h.s. is $[(2d)^r]$ upon replacing the $j$ product by $1$, and removing the $r_j$ even restriction, as it should be.

Next, we consider free b.c. and follow closely the treatment given in Ref. \cite{bQCD}, for periodic b.c. We derive the spectral representation for the partition function and free energy. Similar considerations apply for the two-point correlation. For the free b.c. action, we write
$$
A\,=\,\sum_{x,y\in\Lambda}\, \phi(x) M(x,y)\phi(y)\,,
$$
where $M(x,y)\,=\, \dfrac12\,\delta_{xy}\,-\,\dfrac{\kappa^2}2\,\sum_\mu\,H_\mu(x,y)\,
$, and where $H_\mu(x,y)$ is associated with a $L\times L$ tridiagonal symmetric matrix with unit elements on the nearest neighbor of the diagonal, and zero otherwise. $M$ is diagonalized by a product over $\mu$ of eigenvectors of $H_\mu$. The orthogonal eigenvectors are $\sin p^\mu x^\mu$ with  $x^\mu\,=\,a,2a,\ldots La$ (recall we chose the number $L$ of sites on the hypercubic lattice side to be $L$ even). The orthogonal eigenvectors of $H$ are
$$
\prod_\mu\,\sin\dfrac{\pi p^\mu x^\mu}{L+1}\,.
$$
Here, we identify the $L$ points along a line in the $\mu\,=\,0,1,\ldots,(d-1)$ coordinate direction with $a$, $2a$,...,$La$. The momenta $p^\mu$ lie in the set
$$
{\cal P}\,=\,\left\{ p\,=\,(p^0,\ldots,p^{d-1})\:;\:p^\mu\,=\,\dfrac{\pi r_\mu}{a(L+1)},\: \mu\,=\,0,1,\ldots,(d-1)\right\}  \,,
$$
with $r_\mu\,=\,-L,-L+2,\ldots,-2,1,3,\ldots,(L-1)$. There are $L^d$ elements in the set ${\cal P}$. We have chosen the $p$'s so that ${\cal P}$ lies in $(-\pi/a,\pi/a]^d$. The $p^\mu$ spacing is $\{2\pi/[a(L+1)]\}$, for $r_\mu$ positive and $r_\mu$ negative.

The eigenvalues of $M$ are
$$
\lambda(p)\,=\,\dfrac12\,\left[ 1\,-\,2\kappa^2\,\sum_{\mu}\,\cos (p^\mu a)\right]\,,
$$
and the spectral representation for the free b.c. partition function, in terms of the spectral parameter $q\,=\,pa$, $q\in a{\cal P}$, now follows as in Ref. \cite{bQCD} for the periodic b.c. case. Namely,  
$$
Z_\Lambda\,=\, \exp\left\{-\dfrac12\,
%\left( \dfrac{L+1}{2\pi}\right)^d\,
\sum_q\,\ln\left( 2\lambda(p=q/a)\right)\right\}\,.
$$
Here, the $q^\mu$ spacing is $[2\pi/(L+1)]$ and $|q^\mu|\,<\,\pi$.

From this spectral representation and bounds on the Riemann sum approximation, the thermodynamic limit of the finite lattice free energy 
$$
f_\Lambda\,=\,\dfrac1{L^d}\,\ln Z_\Lambda\,=\, -\,\dfrac{(L+1)^d}{2(2\pi)^dL^d}\,
\;\dis\sum_q\,\left(\dfrac{2\pi}{L+1} \right)^d\,\ln\left( 1\,-\,2\kappa^2\dis\sum_\mu\,\cos q^\mu\right)\,,$$
is the same as the thermodynamic limit of the periodic b.c. free energy given in Eq. (\ref{repf}).

We now show the relation between $\langle\phi(x)\phi(y)\rangle$, the scaled free field two-point correlation and the much analyzed resolvent of minus the unit lattice Laplacian $\Delta_1$ (see e.g. \cite{DJ,PLM,LL,Zucker,MS} and references therein).

Letting $x\equiv n_x a$, $y\equiv n_y a$ and $n_x-n_y\equiv n$, we have 
$$\langle\phi(n_xa)\phi(n_ya)\rangle\,=\,\dfrac1{(2\pi)^d}\,\dis\int_{(-\pi,\pi]^d}\;\dfrac{e^{inq}}{1\,-\,2\kappa^2\sum_\mu\,\cos q^\mu}\;d^dq\,.$$

On the other hand, noting that we have, in Fourier space, the unit lattice Laplacian $\tilde\Delta_1(q)\,=\,\sum_\mu\,2\left( 1\,-\,\cos q^\mu \right)$, we have the resolvent
$$
\left( \Delta_1\,-\,z \right)^{-1}\,=\,\dfrac1{(2\pi)^d}\,\dis\int_{(-\pi,\pi]^d}\;\dfrac{e^{inq}}{\sum_\mu\,2\,\left(1\,-\,\cos q^\mu\right)\,-\,z}\;d^dq\,,
$$
for $z\notin [0,4d]$. Thus, we have the desired relation
$$\langle\phi(n_xa)\phi(n_ya)\rangle\,=\,\dfrac1{\kappa^2}\,\left( \Delta_1\,-\,z \right)^{-1}(n_x,n_y)\,,$$
where 
$$z\,=\,-\,\left(\dfrac1{\kappa^2}\,-\,2d \right)\,=\,-\dfrac{\,m_u^2a^2}{\kappa_u^2}\,.
$$
%==================================================================================
%==================================================================================
%%%%%%%%%%%%%%%%%%%%%%%%%%%%%%%%%%%%%%%%%%%%%%%%%%%%%%%%%%%%%%%%%%%%%%%%%%%%%
%%%%%%%%%%%%%%%%%%%%%%%%%%%%%%%%%%%%%%%%%%%%%%%%%%%%%%%%%%%%%%%%%%%%%%%%%%%%%
\appendix{\begin{center}{\bf APPENDIX B: Truncated Model of Real Scalar Fields}\end{center}}
%%%%%%%%%%%%%%%%%%%%%%%%%%%%%%%%%%%%%%%%%%%%%%%%%%%%%%%%%%%%%%%%%%%%%%%%%%%%%
%%%%%%%%%%%%%%%%%%%%%%%%%%%%%%%%%%%%%%%%%%%%%%%%%%%%%%%%%%%%%%%%%%%%%%%%%%%%%
%%%%%%%%%%%%%%%%%%%%%%%%%%%%%%%%%%%%%%%%%%%%%%%%%%%%%%%%%%%%%%%%%%%%%%%%%%%%%%
\lb{appB}
\setcounter{equation}{0}
\setcounter{lemma}{0}
\setcounter{thm}{0}
\renewcommand{\theequation}{B\arabic{equation}}
\renewcommand{\thethm}{B\arabic{thm}}
\renewcommand{\thelemma}{B\arabic{lemma}$\:$}
\renewcommand{\therema}{{\em{B\arabic{rema}$\:$}}}\vspace{.5cm}
%%%%%%%%%%%%%%%%%%%%%%%%%%%%%%%%%%%%%%%%%%%%%%%%%%%%%%%%%%%%%%%%%%%%%%%%%%%%%
%*************************************************************************
%%%%%%%%%%%%%%%%%%%%%%%%%%%%%%%%
In this Appendix, we analyze the scalar scaled field truncated model. The ingredients for proving the results in this Appendix are the same used in the text or involve well known results, and will only be indicated. This lattice model, non-Gaussian in perturbation theory, illustrates that a TUV stability bound where the exponent is proportional to the number of lattice sites $\Lambda_s\,=\,L^d$ and {\em not} the volume $(aL)^d$ in $\mathbb R^d$, with a constant of proportionality which is independent of $L\in\mathbb N$ (even) and $a\in(0,1]$ is sufficient to bound correlations uniformly in $a\in(0,1]$.

Starting from the scalar scaled field model of Appendix A, recalling that $\kappa^2>0$ is the (squared) hopping parameter and $x^+_\mu=x+ae^\mu$, the truncated model is obtained by replacing the bond factor $\exp[\kappa^2\,\phi(x)\phi(x^+_\mu)]$,by the truncation $\left[ 1\,+\,\alpha \,\phi(x)\phi(x^+_\mu)\right]$. Its finite lattice $\Lambda\,\subset\,a\mathbb Z^d$, $a\in(0,1] $, partition function reads
\bequ\lb{parttrunc}
Z^t_\Lambda\,=\,\dis\int\, \prod_{x,\mu}\,\left[1\,+\,\alpha\,\phi(x)\phi(x^+_\mu) \right]\,d\mu_I(\phi)\,,
\eequ
where the measure is a product of normalized Gaussian measures, with identity covariances $d\mu_I(\phi)\,=\,\prod_x\,\left[\exp\left(-\,\phi^2(x)/2\right)\;d\phi(x) /\sqrt{2\pi}\;\right]$ and the parameter $\alpha$ verifies the condition $\alpha>0$.

We notice that the model satisfies Osterwalder-Schrader positivity and Griffiths I inequality (see e.g. Ref. \cite{GJ,Gini}). Concerning the truncated model, our first result is a stability bound.
\begin{propo}Letting $\Lambda_s=L^d$ denote the number of sites in $\Lambda$, we have the stability bound
\bequ\lb{stabtrunc}\barr{ll}
e^{c_\ell\,\Lambda_s}\,=\,1\,\leq\,Z^t_\Lambda&\leq \dis\int\,\prod_{x,\mu}\,\left[1\,+\,\sqrt{\alpha}\,|\phi(x)| \right]\,\left[1\,+\,\sqrt{\alpha}\,|\phi(x^+_\mu)| \right]\,d\mu_I(\phi)\vspace{2mm}\\&\leq
\dis\int\,\prod_{x}\,\left[1\,+\,\sqrt{\alpha}\,|\phi(x)|\right]^{2d}\,d\mu_I(\phi)\,\leq\,e^{c_u\Lambda_s}\earr
\,,\eequ where $c_\ell\,=\,0$ by Griffiths I inequality and $c_u\,=\,\ln\left[\dis\int\,\left(  1\,+\,\sqrt{\alpha}|\phi|\right)^{2d} \,e^{-\phi^2/2}\,d\phi/\sqrt{2\pi}\right]$.
\end{propo}
\noindent{\bf Proof}: For the upper bound, use $|1+\alpha \phi(x)\phi(x_\mu^+)|\leq 1+\alpha |\phi(x)|\,|\phi(x^+_\mu)|\leq \left[1\,+\,\sqrt{\alpha}\;|\phi(x)|\, \right]\,\left[1\,+\,\sqrt{\alpha}\,|\phi(x^+_\mu)| \,\right]$. For the lower bound, use Griffiths I inequality \cite{GJ,Gini}.\qed

Consider the finite lattice generating function with a single point source
\bequ\lb{genfcttrunc}
\langle e^{J\phi(y)}\rangle_\Lambda\,=\,\dfrac{Z^t_\Lambda(J)}{Z^t_\Lambda}\,,
\eequ
where $Z^t_\Lambda(J)\,=\,\dis\int \exp[J\phi(y)]\,\prod_{x,\mu}\,\left[1\,+\,\alpha\,\phi(x)\phi(x^+_\mu)\right]\,d\mu_I(\phi)$, verifying $Z^t_\Lambda(J=0)\,=\,Z^t_\Lambda$.
Concerning this quantity, we have the following upper bound.
\begin{propo}
	The generating function 
$\langle e^{J\phi(y)}\rangle_\Lambda$ satisfies the bound
\bequ\lb{bdgenertrunc}
\langle e^{J\phi(y)}\rangle_\Lambda\,\leq\,e^{-c_\ell+J^2+c_{u,2}}
\eequ
where $c_{u,2}\,=\,\ln\left[\dis\int\,\left( 1\,+\,\sqrt{\alpha} |\phi| \right)^{4d} \, e^{-\phi^2/2}\,d\phi/\sqrt{2\pi}\;\right]$, with a single point normalized Gaussian measure with unit covariance (see Eq. (\ref{parttrunc})).
\end{propo}
\noindent{\bf Proof}: The proof uses the multireflection bound \cite{GJ,bQCD} and Cauchy-Schwarz inequality multiple times in the underlying quantum Hilbert space. Finally, the Cauchy-Schwarz inequality is used to factorize the $J$ dependence. An analogous proof for the YM model is given in the text. With these methods, we obtain
$$\barr{lll} \langle e^{J\phi(y)} \rangle_\Lambda & \leq&\left[ \dfrac1{Z^t_\Lambda}\, \dis\int\, \exp\left\{J\sum_x\,\phi(x)\right\}\;\prod_{x,\mu}\,\left[1+\alpha\,\phi(x)\phi(x^+_\mu)\right]\,d\mu_I(\phi) \right]^{1/\Lambda_s}\vspace{2mm}\\&\leq&e^{-c_\ell}\,\left\{\dis\int\, \exp\left[2J\sum_x\,\phi(x) \right] \,d\mu_I(\phi) \right\}^{1/2\Lambda_s}\;\left\{\dis\int\,\prod_{x,\mu}\,\left[ 1\,+\,\sqrt{\alpha}\,|\phi(x)| \right]^2\;\left[ 1\,+\,\sqrt{\alpha}\,|\phi(x^+_\mu)| \right]^2\,d\mu_I(\phi)\right\}^{1/2\Lambda_s}\vspace{2mm}\\
&\leq&e^{-c_\ell}\,e^{J^2}\,e^{c_{u,2}}\,.
\earr
$$
Above, in passing from the first to the second equality, we have written $e^{-\phi^2/2}\,=\,e^{-\phi^2/4}\,e^{-\phi^2/4}$ and considered each of the exponential terms in the r.h.s. in each factor of $d\mu_I(\phi)$. \qed

As usual [see Eq. (\ref{scaledcorrel})], correlations are defined by the zero source value of the source derivatives of the generating function 
$\langle e^{J\phi(y)}\rangle_\Lambda$. For the truncated model, we have the coincident point correlation bound
\begin{propo}
	We have the bound
\bequ\lb{boundcoinctrunc}	\left| \langle \phi^r(y)\rangle_\Lambda \right|\,\leq c_r\,\equiv\,e^{-c_\ell+c_{u,2}+1}\,r!
\eequ
\end{propo}
\noindent{\bf Proof}: The coefficients of the Taylor series of $Z_\Lambda(J)$ in $J$ are all positive by Griffiths I inequality \cite{GJ,Gini}. The bound extends to complex source $J$ and $\langle e^{J\phi(y)} \rangle_\Lambda$ is an entire function of $J$. The upper bound results from using Cauchy estimates on the source derivatives,namely,
\bequ\lb{boundcoinctruncexplanation}	\left| \langle \phi^r(y)\rangle_\Lambda \right|\,=\,\left|\left[ \dfrac {d^r}{dJ^r}\,\langle \exp[J\phi(y)]\rangle_\Lambda \right]_{J=0}\right|\,\leq r!\,{\mathrm max}_{|J|=1}\left[ e^{c_\ell}\,e^{|J|^2}\,e^{c_{u,2}}\right]\,\equiv c_r\,.
\eequ
\qed

For $d=1$, we can obtain exact explicit results for the above correlations. They emerge, by inspection, just by expanding the product of bond factors and controlling the terms individually. In doing so, we note that if a lattice site is intersected by only one bond, then the integral for that site vanishes.\vspace{3mm}

\noindent $\mathbf{d=1}$ {\bf and free b.c.:}\vspace{2mm}

We have $Z^t_\Lambda\,=\,1$. We also have TUV stability for all $\alpha>0$ and we have $c_\ell\,=\,c_u\,=\,0$. Also, we have
$$
\langle \phi(x)\phi(y)\rangle\,=\, \alpha^{|x-y|/a}\,=\,\exp\{(|x-y|/a)\ln\alpha\}\,.
$$
This behavior gives a decay for $0<\alpha<1$ and a blowup for $\alpha>1$. The $\alpha=1$ value is the critical point. We perform a finite renormalization for $a\in(0,1]$ setting $\alpha=e^{-ma}$, for the mass $m>0$. Thus,
$$
\langle \phi(x)\phi(y)\rangle\,=\,e^{-m|x-y|}\,.
$$

For a power of the field, at a single point, we obtain a Gaussian behavior, namely (recalling we are dealing with a unit covariance),
$$
\langle \phi^r(y)\rangle\,=\,(r-1)!!\,.
$$
However, if $x\not= y$, we have 
$$
\langle \phi^2(x)\phi^2(y)\rangle\,=\,1\qquad,\qquad x\not= y\,
$$
so that the truncated four-point correlation satisfies
\bequ\lb{nonG}\barr{lll}
\langle \phi^2(x)\phi^2(y)\rangle^{{\mathrm tr}}&=&
\langle \phi^2(x)\phi^2(y)\rangle\,-\,
\langle \phi^2(x)\rangle\;\langle\phi^2(y)\rangle\,-\,2\,
\langle \phi(x)\phi(y)\rangle^2\vspace{2mm}\\&=&-\,2\,
\langle \phi(x)\phi(y)\rangle^2\vspace{2mm}\\&=&-2\,e^{-2m|x-y|}\qquad,\qquad x\not=y\,,
\earr\eequ
which implies that our truncated scalar field model is non-Gaussian and triviality does not hold. This is in contrast with the recent result of Ref. \cite{AizDC} for the complete scalar field model with a quartic interaction, in $d=4$.\vspace{3mm}

\noindent{\bf Scaling Limit} (denoted by ${\mathit scl}$)\vspace{2mm}

The scaling limit is obtained by fixing the mass as $m_s>0$ taking $a\searrow 0$, $n_x,\,n_y\nearrow \infty$, where $x=n_x\,a$ and $y=n_y\,a$ so that $x,\,y\,\rightarrow\,x_s,\,y_s\,\in\mathbb R$. Thus, 
$$\barr{lll}
{\mathit scl} \langle \phi(x)\,\phi(y)\rangle&=& e^{-m_s|x_s-y_s|}\,,\vspace{2mm}\\
{\mathit scl} \langle \phi^2(x)\,\phi^2(y)\rangle^{tr}&=& -2\,e^{-2m_s|x_s-y_s|}\,.\earr
$$

We now consider periodic b.c. \vspace{3mm}

\noindent{$\mathbf{d=1}$ {\bf and periodic b.c}:\vspace{2mm}

Using periodic b.c., with similar methods, the following list results is derived for $d=1$.\vspace{2mm}

\noindent Partition Function: $Z^t_\Lambda\,=\,1\,+\,\alpha^{L}$.\vspace{2mm}

\noindent Free Energy: $f_\Lambda\,=\,\dfrac1L\,\ln(1\,+\,\alpha^{L})\;
\substack{L\nearrow\infty\\\longrightarrow}\;\left\{ \barr{lll} 0&;&\alpha\leq 1\vspace{1mm}\\\ln\alpha&;&\alpha>1\,.\earr  \right.$\vspace{2mm}

\noindent In the thermodynamic limit, the two-point correlation presents a discontinuous $\alpha$ derivative at $\alpha=1$. Indeed, we have
$$
\langle \phi(x)\phi(y)\rangle \,=\,\left\{ \barr{lll}\alpha^{|x-y|/a}&;&\alpha\leq 1\vspace{1mm}\\ \alpha^{-|x-y|/a}&;&\alpha> 1\,.\earr \right.\vspace{1mm}
$$

$\alpha\,=\,\alpha_c\,=\,1$ is a critical point.\vspace{2mm}

We can perform a finite renormalization setting $\alpha\,=\,e^{-ma}$, for $m>0$ and $\alpha<1$, and $\alpha\,=\,e^{ma}$, for $\alpha>1$, and $m$ is the scalar field mass. Thus,
$$
\langle \phi(x)\phi(y)\rangle \,=\,e^{-m|x-y|}\,.\vspace{1mm}
$$

Also, we can take the scaling limit $\mathit{scl}$ and obtain non-triviality of the model like in the $d=1$ free b.c. case.\vspace{3mm}

Finally, for the free or periodic b.c., we can obtain the critical exponents and, also, we can take the limit $m_s\rightarrow 0$.\vspace{2mm}

Now, taking $d\,=\,3,4$, we carry out perturbation theory to order $\alpha^2$ and obtain
$$
\alpha_c\,=\,\dfrac12\,\left[1\,-\,\sqrt{1\,-\,2/d}\right]\,=\,
\dfrac1{2d}\,+\,\dfrac1{4d^2}\,+\,\ldots\,,
$$
which is to be compared with the free scaled field value $\alpha_c\,=\,1/(2d)$. Here, $\alpha_c$ is small but not very small (not $\alpha_c\ll 1$).

For the configuration $x_1=x_2=0$ and $x_3=x_4=ae^\mu$, the truncated four-point correlation $\langle \phi(x_1)\phi(x_2)\phi(x_3)\phi(x_4)\rangle^{\mathrm tr} $ verifies the non-Gaussian (nontrivial) behavior
$$
\langle \phi(x_1)\phi(x_2)\phi(x_3)\phi(x_4)\rangle^{\mathrm tr}\,=\,-2\,\alpha^2\,\not=\,0
$$

With the negative sign and considering the ladder, leading order approximation in a lattice Bethe-Salpeter equation (see Refs. \cite{Sp,BS,BS2}, the negative sign of the above truncated four-point correlation implies that potential between the interacting particles is repulsive and does not favor the formation of bound states. 
%%%%%%%%%%%%%%%%%%%%%%%%%%%%%%%%%%%%%%%%%%%%%%%%%%%%%%%%%%%%%%%%%%%%%%%%%%%%%
%%%%%%%%%%%%%%%%%%%%%%%%%%%%%%%%%%%%%%%%%%%%%%%%%%%%%%%%%%%%%%%%%%%%%%%%%%%%%
\appendix{\begin{center}{\bf APPENDIX C: Factorized Bounds for Bose Field Model Partition Functions}\end{center}}
%%%%%%%%%%%%%%%%%%%%%%%%%%%%%%%%%%%%%%%%%%%%%%%%%%%%%%%%%%%%%%%%%%%%%%%%%%%%%
%%%%%%%%%%%%%%%%%%%%%%%%%%%%%%%%%%%%%%%%%%%%%%%%%%%%%%%%%%%%%%%%%%%%%%%%%%%%%
%%%%%%%%%%%%%%%%%%%%%%%%%%%%%%%%%%%%%%%%%%%%%%%%%%%%%%%%%%%%%%%%%%%%%%%%%%%%%%
\lb{appC}
\setcounter{equation}{0}
\setcounter{lemma}{0}
\setcounter{thm}{0}
\setcounter{rema}{0}
\renewcommand{\theequation}{C\arabic{equation}}
\renewcommand{\thethm}{C\arabic{thm}}
\renewcommand{\thelemma}{C\arabic{lemma}$\:$}
\renewcommand{\therema}{{\em{C\arabic{rema}$\:$}}}\vspace{.5cm}
%%%%%%%%%%%%%%%%%%%%%%%%%%%%%%%%%%%%%%%%%%%%%%%%%%%%%%%%%%%%%%%%%%%%%%%%%%%%%
%*************************************************************************
%%%%%%%%%%%%%%%%%%%%%%%%%%%%%%%%
In the text, bounds are obtained on the YM partition function which admit a factorization into local quantities, i.e. single plaquette partition functions of a single bond variable. This factorization seems to be peculiar to local gauge-invariant YM models. 

In  this Appendix, we show how to obtain factorized local bounds for Bose field models. Each factor involves only the `transfer matrix´ of a single bond. For the continuum limit $a\searrow 0$, it is important that the constants appearing in the bounds are uniform in the lattice spacing $a\in(0,1]$, and include the critical values of the model parameters. In particular, for the free field free energy, the bounds on the partition function allow us to show, in dimension $d$, that $f(a)\,\leq\,\ln\sqrt{2}$, in the thermodynamic limit, as long as $\kappa^2\,\leq\,(1/2d)$. Our bounds also  allow the bond factor coupling parameters to be space dependent where, as other bounds (e.g. those using the momentum representation) require the couplings to be space independent.

Variants of the method apply to numerous Bosonic models, as for $\lambda\phi^4_3$ with small $\lambda>0$, the truncated model of Appendix B, etc. For simplicity, we treat the free scalar scaled field model partition function with free b.c. and the local bond factors given by
$$
F_{x,\mu}\,=\,e^{\kappa^2\,\phi(x)\phi(x^+_\mu)}\,.
$$

For this model with free b.c., we have the  finite lattice partition function, given in terms of the bond factors, which reads
\bequ\lb{partc}
\barr{lll}
Z_\Lambda&=&\dis\int\,\prod_{\mu,x}\,F_{x,\mu}\,d\mu_I(\phi)\,,
\earr\eequ
where $d\mu_I(\phi)$ is a product measure of normalized Gaussian measures with unit covariance.

By the generalized H\"older inequality, we have
$$\barr{lll}
Z_\Lambda&\leq&\prod_\mu\,\dis\int\,\left[\prod_{x}\,F^d_{x,\mu}\,d\mu_I(\phi)\right]^{1/d}\,,\vspace{2mm}\\
&\leq&\prod_\mu\, \left[ \prod_c\,Z^\mu_{\Lambda,c}\right]^{1/d}\,,
\earr
$$
where $Z^\mu_{\Lambda,c}$ is the partition function of a one-dimensional chain which we shall now define.

For a fixed spacetime direction $\mu=0,1,\ldots,(d-1)$, we index each sequence of $L^{d-1}$ points (lattice sites) in the $x^\mu=1$ hyperplane by $c$. $Z^\mu_{\Lambda,c}$ of a chain with $L$ sites parallel to the $\mu$ direction, starting at $x^\mu=1$ denotes the partition function and ending at $x^\mu=L$. The couplings in the chain $c$ are denoted by $\kappa_{c_j}^2$. For the generic chain $c$, we denote the $L$ sites by $1,2,\ldots,L$ and the $L-1$ bound coupling parameters by $\kappa_{1\mu}^2$, $\kappa_{2\mu}2$, ...,$\kappa_{(L-1)\mu}^2$, suppressing the $c$ dependence. The field variables are denoted by $\phi_1$, $\phi_2$, ..., $\phi_L$. Having done this, we now write
$$
Z^\mu_c\,=\, \left( f\,,\,\prod_{j=1,\ldots,(L-1)}\,T(\kappa_{c_j}^2,\mu)\,f  \right)_{L^2({\mathbb R})}\,.
$$
The product is the composition of integral operators $T(\kappa_{c_j}^2,\mu)$ with the operator kernel $K(\phi,\phi^\prime, \kappa_{c_j}^{2})$ given by
$$
K(\phi,\phi^\prime,\beta)\,=\, \dfrac1{\sqrt{2\pi}}\; \exp\left(-\dfrac14\,\phi^2  \right)\;\exp\left(d\beta\phi\phi^\prime\right)\;\exp\left(-\dfrac14\,(\phi^\prime)^2  \right)\,,
$$
and $f(\phi)\,=\, (2\pi)^{-1/4}\,e^{-\phi^2/4}$. Thus,
$$
Z^\mu_c\,\leq\,\|f\|^2\;\prod_{j=1,\ldots,(L-1)}\,\|T(\kappa_{c_j}^2,\mu)\|\,,
$$
and
$$\barr{lll}
Z_\Lambda&\leq&\prod_\mu\,\left\{ \prod_c \,\left[\|f\|^2\, \prod_{j=1,\ldots,(L-1)}\,\|T(\kappa_{c_j}^2,\mu)\|\right]\right\}^{1/d}\vspace{2mm}\\
&=&\|f\|^{2L^{d-1}}\,\prod_b\, \|T(\kappa_{c_j}^2,\mu)\|^{1/d}\,,\earr
$$
where we have indexed the $d(L-1)L^{d-1}$ bonds of the lattice $\Lambda$ by $b$.

This is our factorized bound!  If all the  couplings are equal, say $\kappa_b^2\,=\,\kappa^2$ for any bond $b$, the product over the bonds is $\|T(\kappa^2)\|^{(L-1)L^{d-1}}$ and
$$
Z_\Lambda\,\leq\,e^{2L^{d-1}\ln\|f\|}\;\exp\left[L^{d-1}(L-1)\ln\|T(\kappa^2)\| \right]\,\equiv\, e^{c_u\,L^d}\,.
$$
Of course, here, $$c_u\,=\, \dfrac2L\,\ln\|f\|\,+\, (1\,-L^{-1})\,\ln\|T(\kappa^2)\|\,,$$
which is our upper stability bound.

We obtain a convenient bound for $\|T(\kappa^2)\|$ using a Holmgren bound  $\|T(\kappa^2)\|_H$ for the norm (see e.g. Ref. \cite{Simon3} and Chap. 4 of Ref. \cite{Kato}). We have
$$\barr{lll}\|T(\kappa^2)\|\,\leq\,\|K\|_H&=&{\mathrm sup}_\phi\,\left[\dis\int_R\,K(\phi,\phi^\prime,\kappa^2)\,d\phi^\prime\,\right]\vspace{3mm}\\&=&{\mathrm  sup}_\phi\,\left[\sqrt{2}\, \exp\left(\dfrac14\,(1\,-\,4d^2\kappa^4)\,\phi^2 \right)\right]\vspace{3mm}\\
&=&\sqrt{2}\qquad;\qquad \kappa^2\,\leq\,(1/2d)\,.
\earr$$

Since $\|f\|\,=\,1$, the thermodynamic limit $f(a)$ of the finite lattice model free energy $f(\Lambda,a)$ satisfies
$$
f(a)\,=\,\lim_{L^d\nearrow\infty}\,\dfrac{\ln Z_\Lambda}{L^d}\,\leq\,\dfrac12\,\ln 2\qquad;\qquad\kappa^2\,\leq\,(1/2d)\,.
$$
The bound is independent of the coupling parameters.

Alternatively, for an upper bound on $Z_c$, we set the maximum $\kappa$ condition, $d\,\kappa^2\,=\,1/2$, in $Z_c$. For free b.c., we evaluate $Z_c$ by successive integration. The result is expressed in the following Lemma. We have,
\begin{lemma} We have
\bequ\lb{Zcc} 
Z_c\,=\,\prod_{j=1}^{L-1}\,\dfrac1{\sqrt{1\,-\,b_j}}\,,\eequ
where $b_j$ satisfy the recursion relation
$$b_{j+1}\,=\,\frac1{4(1\,-\,b_j)}\quad,\quad b_1\,=\,\dfrac14\,.$$
	
The solution to this recursion is $$b_j\,=\,\dfrac j{2(j+1)}\,,$$
so that we have
\bequ\lb{Zccc}
Z_c\,=\,\prod_{j=1}^{L-1}\,\sqrt{\dfrac{2(j+1)}{j+2}}\,=\,\dfrac{2^{(L-1)/2}}{\sqrt{L+1}}\,.
\eequ	
\end{lemma}

\noindent{\bf Proof:} By induction, we obtain the $b_j$ recursion. To solve the recursion, we note that $b^*\,=\,1/2$ is a fixed point and pass to the variable $c_j$ where $b_j\,=\,b^*\,+\,c_j$. The  recursion for $c_n$ is
$$
c_{n+1}\,=\,\dfrac{c_n}{1-2c_n}\quad , \quad c_1\,=\,-\,\dfrac14\,,
$$
or 
$$
\dfrac1{c_{n+1}}\,=\,\dfrac1{c_n}\,-\,2\,.
$$

Telescoping $(1/c_{n+1})$ gives 
$$
\dfrac1{c_{n}}\,=\,\dfrac1{c_{1}}\,+\, \sum_{k=1}^{n-1}\, \left[\dfrac1{c_{k+1}}\,-\,\dfrac1{c_{k}} \right]\,=\,\dfrac1{c_{1}}\,+\, \sum_{k=1}^{n-1}\,(-2)\,=\,\dfrac1{c_{1}}\,-\,2(n-1)\,.
$$
or
$$
{c_{n}}\,=\,-\,\dfrac1{2(n+1)}\,.
$$
so that $b_n\,=\,\dfrac n{2(n+1)}$, and $1\,-\,b_n\,=\,\dfrac{n+2}{2(n+1)}\,$.

% As seen by Fourier transform methods, and periodic b.c., the thermodynamic limit of one-dimensional chain's free energy is
%\bequ\lb{faa}\barr{lll}
%f(a)&=& -\dfrac1{4\pi}\,\dis\int_{-\pi}^{\pi}\,\ln\left( 1\,-\,2d\kappa^2\,\cos q \right)\,dq\vspace{2mm}\\
%&=&\-\dfrac12\,\ln\left[ \dfrac{1\,+\,\sqrt{1\,-\,4d^2\kappa^4}}2\right]\,,\earr
%\eequ
%where Ref. \cite{grad}, Chap. 3, was used to evaluate the integral.
%
%From the first line of Eq. (\ref{faa}), a power series in $\kappa^2$ can be obtained by expanding the logarithm and performing the integral over $q$. The series coefficients are all positive and, again, the coupling along the chain is constant.
%
%We obtain a new representation for the chain partition function $Z_c$ by successive integration and allowing the couplings to vary along the chain. We have
%$$
%Z_c\,=\,\prod_{k=1,\ldots,(L=1)}\,\dfrac1{\sqrt{1\,-\,B_k}}\,,
%$$
%where $\{B_j\}$ satisfy the recursion relation
%$$
%B_{j+1}\,=\,\dfrac{\kappa^4_j}{1\,-\,B_j}\qquad;\qquad j\geq1\quad,\quad B_1\,=\,\kappa^4_1\,\leq\,(1/4)\,.
%$$
%Notice that $Z_c$ factorizes. If $\kappa^4\leq (1/2)$, {\bf cannot read comment, is out of pic}
%%%%%%%%%%%%%%%%%%%%%%%%%%%%%%%%%%%%%%%%%%%%%%%%%%%%%%%%%%%%%%%%%%%%%%%%%%%%
%%%%%%%%%%%%%%%%%%%%%%%%%%%%%%%%%%%%%%%%%%%%%%%%%%%%%%%%%%%%%%%%%%%%%%%%%%
\begin{acknowledgements}
We would like to thank the anonymous referees for suggestions. We also acknowledge the partial support of FAPESP and the Conselho Nacional de Desenvolvimento Cient\'\i fico e Tecnol\'ogico (CNPq) during the early stage of this work.\vspace{3mm}
\end{acknowledgements}
%%%%%%%%%%%%%%%%%%%%%%%%%%%%%%%%%%%%%%%%%%%%%%%%%%%%%%%%%%%%%%%%%%%%%%%%%%%%
%%%%%%%%%%%%%%%%%%%%%%%%%%%%%%%%%%%%%%%%%%%%%%%%%%%%%%%%%%%%%%%%%%%%%%%%%%
%%%%%%%%%%%%%%%%%%%%%%%%%%%%%%%%
%%%%%%%%%%%%%%%%%%%%%%%%%%%%%%%%%%%%%%%%%%%%%%%%%%%%%%%%%%%%%%%%%%%%%%%%%%%%
%%%%%%%%%%%%%%%%%%%%%%%%%%%%%%%%%%%%%%%%%%%%%%%%%%%%%%%%%%%%%%%%%%%%%%%%%%

\end{document}